\documentclass[a4paper, amsfonts, amssymb, amsmath, reprint, showkeys, superscriptaddress, nofootinbib, twoside, floatfix, aps, pra]{revtex4-1}
\usepackage[english]{babel}
\usepackage[utf8]{inputenc}
\usepackage[colorinlistoftodos, color=green!40, prependcaption]{todonotes}
\usepackage{amsthm}
\usepackage{mathtools} 
\usepackage{blkarray}
\usepackage{amsmath}
\usepackage{physics}
\usepackage{xcolor}
\usepackage{makecell}
\usepackage{colortbl}
\usepackage{graphicx}
\usepackage{subcaption}
\usepackage[left=23mm,right=13mm,top=35mm,columnsep=15pt]{geometry} 
\usepackage{adjustbox}
\usepackage{placeins}
\usepackage[T1]{fontenc}
\usepackage{lipsum}
\usepackage{csquotes}
\usepackage{qcircuit}
\usepackage{array}
\usepackage[unicode=true, colorlinks=true]{hyperref} 
\usepackage{cleveref}
\usepackage{multirow}
\usepackage{ragged2e}
\begin{document}

\title{Quantum Resource Assay for the Grid-Based Simulation of the Photodynamics of Pyrazine}

\author{Xiaoning Feng}
    \email{xiaoning.feng@chem.ox.ac.uk}
    \affiliation{Department of Chemistry, University of Oxford, Oxford OX1 3TA, United Kingdom}
\author{Hans Hon Sang Chan}
    \affiliation{Department of Materials, University of Oxford, Oxford OX1 3PH, United Kingdom}
\author{David P. Tew}
    \email{david.tew@chem.ox.ac.uk}
    \affiliation{Department of Chemistry, University of Oxford, Oxford OX1 3TA, United Kingdom}
\date{\today} 

\begin{abstract}
We establish and analyse the performance and resource requirements of an end-to-end fault-tolerant quantum algorithm for computing the absorption spectrum and population dynamics of photoexcited pyrazine. The quantum circuit construction consists of initial state preparation using uniformly controlled rotations, the time-dependent Hamiltonian propagation based on the grid-based Split Operator Quantum Fourier Transform (SO-QFT) method, and cost-effective measurements including statistical and canonical phase estimation. We use classical emulations to validate the quantum resources required for the task, and propose generalised formulae for the qubit count and gate depth calculation. Simulating the vibronic dynamics of pyrazine in a low-dimensional abstraction requires 17-qubit circuits with a gate depth of $\mathcal{O}(10^5)$, whereas a full-dimensional simulation of pyrazine in 24 modes requires at least 97-qubit circuits with a gate depth of $\mathcal{O}(10^6)$. Our work provides a foundational framework for understanding high-dimensional wavepacket-based quantum simulations of photo-induced dynamics and vibronic spectra, anticipating future applications in the simulation of even larger molecular systems on fault-tolerant quantum computers.
\end{abstract}

\maketitle

\section{Introduction}
Quantum computing promises significant time and resource advantage over its classical counterpart for simulating molecular quantum dynamics of large and complex chemical systems. Its potential stems mainly from the ability to encode and manipulate exponentially-scaling representations of many-body systems using only a linearly-scaling number of qubits and polynomially-scaling operations. Given an initial state of a chemical system, one can compute and reveal dynamical properties of interest by propagating the system with the corresponding time-dependent Hamiltonian. While it is usually extremely expensive on classical computers due to the exponential complexity of the Hilbert space, practical quantum algorithms may present exponential runtime reduction that outperform the best classical approaches~\cite{Montanaro2016,brown2010}. 

The development of quantum algorithms for studying the time evolution of quantum systems has attracted considerable interest\cite{Zalka1998,Shen2023,Pauline2023,Katherine2022,Nishi2023,Barison2021}, and researchers seek greater clarity on exactly which simulation tasks can be ported to quantum computers to deliver a compelling advantage. Considerable progress has been made in the design of quantum algorithms at the quantum gate or circuit level for preparing initial states, applying time-evolution operators and extracting observable quantities from simulations of interacting quantum particles~\cite{Abrams1997,hastings2014,Jordan2012,george2014,Berry_2015,Zhang2022,Kokcu2022,Wada2022,Chan2023,Astrakhantsev2023,Tepaske2023,Haah2023,Yuan2023}. 

In this work, we numerically investigate an end-to-end error-corrected, fault-tolerant digital quantum simulation of molecular photo-electron absorption spectra and the associated state population dynamics after photo-excitation, with a focus on the corresponding quantum circuit construction, feasibility assessments and cost estimations. We choose the pyrazine molecule as a case study, which is notable as one of the essential building blocks of many organic compounds~\cite{BREDA2006}. The photoinduced dynamics of pyrazine is a prototype for diazine rings due to its well-characterised inter-state transition via the conical intersection, and is thus a representative system for studying molecular vibronic-coupling. 

Different abstractions of the system have been explored and the \textit{ab initio} Potential Energy Surfaces (PES) characterised using Complete Active Space Self Consistent Field (CASSCF), Multi-Reference Configuration Interaction (MRCI), Second Order Møller-Plesset Perturbation Theory (MP2) and Configuration Interaction Singles (CIS) techniques~\cite{Woywod1994,HE2009}. Many standard techniques for modelling molecular quantum dynamics on conventional computers have been applied to this system, including the Path-Integral (PI) approach~\cite{Krempl1994,Krempl1995},  the Multiconfiguration Time-Dependent Hartree (MCTDH) method~\cite{Worth1996,Raab1999,Sala2014}, Semiclassical descriptions~\cite{Thoss2000}, Time-Dependent Discrete Variable Representation (TDDVR) approach~\cite{Puzari2005}, Matching-Pursuit/Split Operator Fourier Transform (MP/SOFT) methods~\cite{Chen2006}, and also Short-Time Trajectories techniques~\cite{Saller2015,Saller2017}. 

Among these aforementioned approaches, the grid-based Split Operator Fourier Transform (SO-FT) stands out because it can be straightforwardly translated into a quantum algorithm by performing the Quantum Fourier Transform (QFT) in place of the discrete Fourier transform. QFT was first implemented in Shor’s Algorithm~\cite{Shor1994}, and nowadays forms the basis of many quantum simulation algorithms due to its exponential speed-up over the discrete Fourier Transform~\cite{OLIVEIRA2007,BAND2013}. Based on the first-order Lie–Trotter product formula~\cite{Trotter1959}, many further studies were conducted to develop high-order variations~\cite{Suzuki1976,SUZUKI1990,Suzuki1991,Andre1991,Andre1992,Hatano2005} and characterise the Trotter error~\cite{tannor2007,Childs2021,roulet2021}. With timestep-wise unitary operations and the use of the Fourier transform, the Trotterized SO-QFT method is particularly appropriate for quantum simulation, and has been investigated in many previous works~\cite{Kassal2008,jones2019,Damian2019}. This approach is well-suited to vibrational dynamics, because, in contrast to electronic wavefunctions, the quantum states do not exhibit derivative discontinuities and accurate grid-based representations can be realised with relatively few qubits per degree of freedom.

While pyrazine in a reduced dimensional representation is considered a classically `solved' problem, understanding the requirements for performing a quantum simulation of this model helps us to extrapolate to more meaningful systems, where classical computers are expected to struggle. Previous phenomenological representations of a 2D vibronic model of pyrazine have been studied in an analog setting~\cite{Navickas2025,Ryan2021}. In this work, we focus on the digital representation of a fault-tolerant quantum algorithm that encodes the simulation through qubits and gates, enabling systematic resource estimation and scalability analysis for higher-dimensional vibronic models.
We first assess the performance of the grid-based SO-QFT method for the quality of the pyrazine spectrum and the accuracy of the population dynamics. Since grid-based modelling is one of the most promising use cases for fault-tolerant quantum computers, the focus of this work is on illustrating the quantum implementation of this algorithm and evaluating the quantum resources required. Specifically, we examine initial state preparation, circuit design for the time-evolution process, appropriate measurement techniques of both absorption spectra and population dynamics, and present the corresponding calculations of relevant gate depths.

\section{Background}
\subsection{The Pyrazine Photosystem}
The three lowest-energy electronic states of pyrazine are the ground state $S_{0}$, the dark $S_{1}$ $n\pi^{*}$ state, and the bright $S_{2}$ $\pi \pi^{*}$ state. 
Photo-excitation from the ground $S_{0}$ state initially populates the $S_{2}$ state, after which the vibronic coupling between the $S_{2}$ and $S_{1}$ states triggers an extremely fast and efficient $S_{2} \rightarrow S_{1}$ internal conversion process resulting in a broadband $S_{2}$ absorption spectrum and significant population transfer from the $S_{2}$ to the $S_{1}$ state~\cite{Thoss2000,Sala2014,Saller2015}.

The conical intersection of the $S_{1}$ and $S_{2}$ states of pyrazine has been thoroughly characterised in previous studies~\cite{SCHNEIDER1988,Woywod1994,Kanno2015,Schile2019,Gu2020,Neville2022}. A full 24-dimensional vibronic model Hamiltonian has also been developed based on \textit{ab initio} calculations~\cite{Köuppel1984,Woywod1994,Raab1999,Puzari2005}.
Of the 24 vibrations, only 4 of the modes $\nu_{6a},\nu_{1},\nu_{9a},\nu_{10a}$ are directly active in the dynamics and are sufficient for a low-dimensional, qualitative description of the vibronic spectrum~\cite{Innes1988,Woywod1994,Worth1996,Thoss2000}.
According to group-theoretic considerations, the $\nu_{10a}$ mode is responsible for the coupling between the $S_{1}$ and $S_{2}$ states~\cite{Worth1996}. The remaining three symmetric modes act as tuning modes~\cite{Krempl1994}. 
The terms of the 4D linear vibronic model Hamiltonian of pyrazine are collected into diagonal kinetic and potential terms, and off-diagonal coupling terms:
\begin{equation}
\begin{split}
H &= \underbrace{- \displaystyle\sum_{k} \frac{\omega_{k}}{2} \frac{\partial^2}{\partial Q_{k}^{2}}}_{K} +
    \underbrace{\begin{pmatrix}
    V_{S1}  & 0  \\
    0 & V_{S2}
    \end{pmatrix}}_{V_\text{diag}} +
    \underbrace{Q_{10a} \begin{pmatrix}
    0 & \lambda   \\
    \lambda  & 0
    \end{pmatrix}}_{V_\text{c}} 
    \end{split}
\label{eq:hamil}
\end{equation}
where
\begin{align}
V_{S1} &=  -\Delta +
    \displaystyle\sum_{j}
     \kappa_{j}^{(1)} Q_{j}+\frac{1}{2}\displaystyle\sum_{k} \omega_{k}Q_{k}^{2} \\
     V_{S2} &=  \Delta +
    \displaystyle\sum_{j}
     \kappa_{j}^{(2)} Q_{j}+\frac{1}{2} \displaystyle\sum_{k} \omega_{k}Q_{k}^{2} \end{align}
and $k = \{6a,1,9a,10a\}, j = \{6a,1,9a\}$. 
Here $\omega$ denotes the vibrational frequencies, $\kappa$ the intra-state coupling constants, $\lambda$ the inter-state coupling constants and 2$\Delta$ the energy gap between $S_{1}$ and $S_{2}$ states. The model parameters are taken from \textcite{Krempl1994,Worth1996} and are listed in Table~\ref{table parameter}.

\begin{table}[!htbp]
    \centering 
    \renewcommand{\arraystretch}{1.3}
    \begin{tabular}{|p{1cm}<{\centering}|p{1.5cm}<{\centering}|p{1.5cm}<{\centering}|p{1.5cm}<{\centering}|} 
        \hline 
        \multicolumn{4}{|c|}{$\lambda$ = 0.1825, $\Delta$ = 0.4617} \\
        \hline 
        & $\omega$ & $\kappa^{(1)}$ & $\kappa^{(2)}$ \\ 
        \hline 
        $\nu_{6a}$  & 0.0740  & -0.0964 &  0.1194\\ 
        $\nu_{1}$  & 0.1273  & 0.0470 &  0.2012\\ 
        $\nu_{9a}$  & 0.1568  & 0.1594 & 0.0484\\ 
        $\nu_{10a}$ & 0.0936 & \textbackslash & \textbackslash \\
        \hline 
    \end{tabular}
    \caption{\justifying Values of parameters $\{ \omega, \kappa, \lambda, \Delta \}$ used in the 4D model Hamiltonian (units: eV).} 
    \label{table parameter} 
\end{table}

The coordinates $Q_k$ are the dimensionless ground-state normal-mode coordinates and the electronic degree of freedom is represented as the state vector in the basis of the two excited states:
\begin{equation}
\ket{S_{1}}:
\begin{pmatrix}
   1\\0 
\end{pmatrix}, \hspace{0.2cm} 
\ket{S_{2}}:
\begin{pmatrix}
   0\\1 
\end{pmatrix}.   
\end{equation}
In this model, the two excited states are represented in the diabatic basis, and the vibrational modes are approximated as displaced harmonic oscillators with the same characteristic frequencies as the normal modes of the ground state. 



\subsection{The Split Operator Approach}

While classical grid-based representations are hindered by the exponential scaling of the memory required to build and store the states, the number of qubits needed to store and represent the same state on a quantum computer scales only linearly with the dimension of the system. The grid-based SO-QFT method is one of the most promising algorithms suitable for implementing Hamiltonian evolution on fault-tolerant quantum devices.
Competing grid-based alternatives such as linear combination of unitaries (LCU)~\cite{Zeng2025,Childs2022,Berry2024,An2023} and qubitization~\cite{Berry2021,Mukhop2024,Low2019,vonBurg2021}, achieve asymptotically optimal error scaling, but are typically associated with nontrivial block encodings or quantum signal processing sequences. 
In this regard, SO-QFT offers a more straightforward framework for practical contexts where intuitive and accessible circuit design and moderate resource demands are of primary importance.

The initial wavepacket of this work is the vibrational ground state of the harmonic $S_0$ potential, projected onto the $S_2$ state. In the position space, it is represented as a tensor product state on a uniform five-dimensional rectangular grid, consisting of four normal mode coordinates $\mathbf{Q}$ and one electronic state coordinate $s$:
\begin{equation}  
\psi(\mathbf{Q}s,t=0) = \left(\bigotimes_k \psi(Q_k)\right) \otimes \ket{S_2},
\label{init state}
\end{equation}
where each $\psi(Q_k)$ is a Gaussian ground state.
The SO-FT method simulates Hamiltonian evolution by Trotterizing the Hamiltonian evolution operator into small time increments $dt$ and by performing time evolution of the potential and kinetic operators separately as
$$\psi(\mathbf{Q}s,t+dt) \approx e^{-\frac{iV dt}{2}}\text{FT}^{-1} e^{-iKdt} \text{FT} e^{-\frac{iVdt}{2}} \psi(\mathbf{Q}s,t) ,$$
where $V=V_{\text{diag}}+V_\text{c}$ and $\text{FT}^{-1}$ is the inverse Fourier transform.
The kinetic and potential components are applied in the momentum and position representations respectively, where they are local and diagonal-dominant matrices and thus significantly easier to evaluate. Conversion between position space and momentum space representations is effected via forward and inverse Fourier Transforms.  Since $V$ and $K$ do not commute, this introduces an error, which scales as $\mathcal{O}(dt^{3})$ for the second-order Split Operator approximation shown here \cite{FEIT1982,kosloff1988}.  

The SO-QFT method benefits from further advantages compared to its classical counterpart above in two aspects. First, because the computational basis states of a quantum computer increase exponentially with the overall qubit number, it becomes possible to use high-dimensional grids with fine spatial resolution. Second, while the classical Fast Fourier Transform (FFT) normally requires $\mathcal{O}(n2^{n})$ elementary operations over $2^n$ grid points, the QFT operates with only $\mathcal{O}(n^2)$ quantum gates~\cite{Benenti2008}, leading to an exponential compression in resource scaling.

\subsection{Observable Properties}
\label{observable}

\begin{figure*}[!htbp]
    \centering
    \includegraphics[scale=0.13]{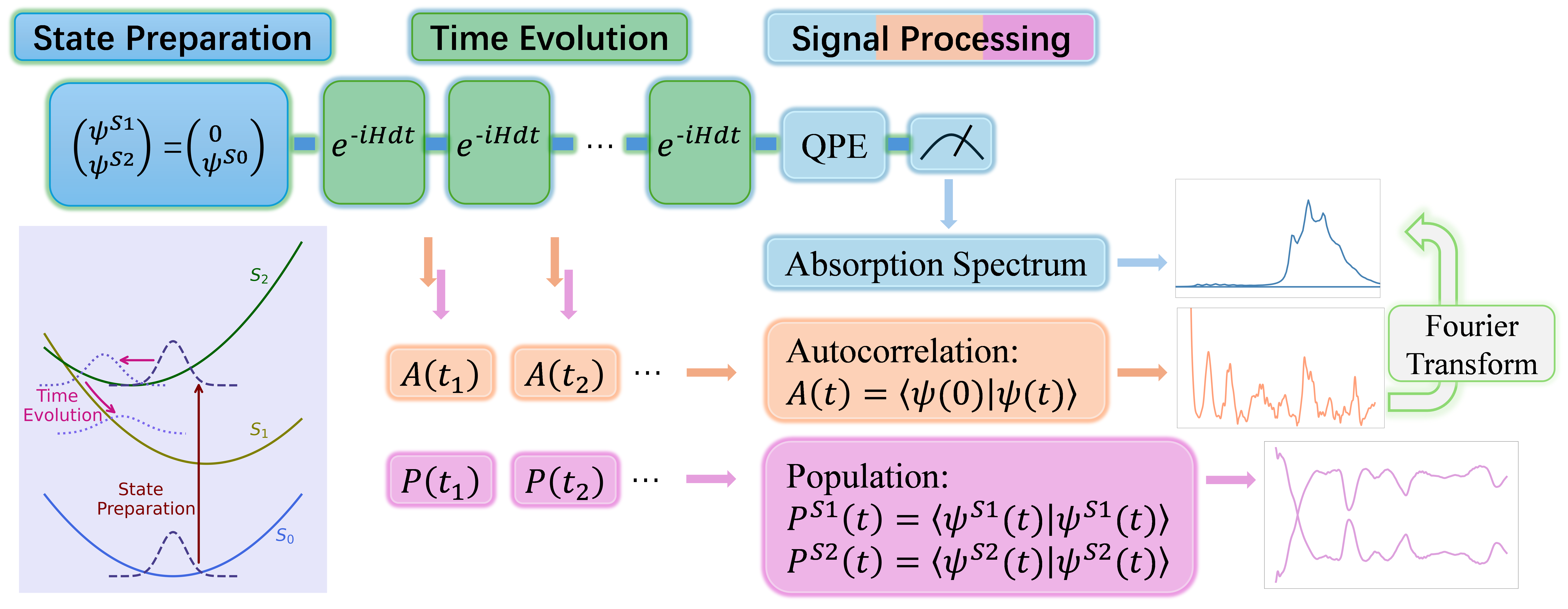}
    \caption{\justifying Schematic overview of the proposed end-to-end quantum algorithmic framework, comprising three main stages: state preparation, time evolution, and signal processing. Each component is discussed in detail in Sections~\ref{State preparation}-~\ref{signal processing}.
    }
    \label{fig:framework}
\end{figure*}

Under the Franck--Condon approximation~\cite{Franck1926,Condon1926,Birge1926,Luis2004}, the intensities of vibronic spectra are given by the vibronic transition moments, which are proportional to the overlap of the Born--Oppenheimer vibrational states in the two electronic PESs~\cite{IUPAC2025}. In this framework, vertical excitation occurs under the sudden approximation, where the initial state for time-evolution is the zero point vibrational wavefunction of the $S_0$ PES projected into the $S_2$ electronic PES~\cite{Kundu2022}. The initial wavefunction is propagated with the 4D Hamiltonian for the $S_1$-$S_2$ system and the absorption spectrum is given by the Fourier transform of the autocorrelation function of the wavefunction~\cite{Worth1996}:
\begin{equation}
    I(\omega) \propto \int_{-\infty}^{\infty} \braket{\psi(0)}{\psi(t)} e^{i\omega t} \mathrm{d}t 
\label{eq:absorption}
\end{equation}
Population dynamics are extracted directly from the probability distribution across the $S_1$ and $S_2$ states.


Lines in the experimental spectra are broadened due to spectrometer resolution and the inherent ultrafast photorelaxation of the system~\cite{Worth1996}. Since the 4D model neglects the other 20 normal modes, which are responsible for fast vibrational relaxation, homogeneous phenomenological broadening is accounted for by multiplying the autocorrelation function in the integrand of Eq.~\ref{eq:absorption} by a time-dependent broadening function:
\begin{equation}
    B(t) = \exp(-\frac{t}{\tau}).  
\label{eq:damp_b}
\end{equation} 
Following previous work on this system, the short effective relaxation time is selected to be $\tau = 30$~fs~\cite{Worth1996,Thoss2000,Chen2006}. In practical simulations the state is evolved for a finite time $T$, which gives rise to artificial features in the spectrum resulting from the Fourier Transform in Eq.~\ref{eq:absorption}. Another damping function: 
\begin{equation}
     D(t) = \cos(\frac{\pi t}{2T}) 
\label{eq:damp_d}   
\end{equation}
that simultaneously adds broadening due to the spectral resolution, and reduces artefacts due to the Gibbs phenomenon is therefore also applied.

\section{Quantum Implementation}

Our goal is to establish and analyse the performance characteristics of the SO-QFT quantum algorithm for computing the absorption spectrum and population dynamics of photo-chemical systems, using the 4D model for pyrazine as a prototypical test case where full classical emulation is feasible. The end-to-end quantum algorithm consists of three steps: preparation of a relevant initial state on a register of qubits; application of the SO-QFT time-propagation circuit; qubit measurement for the extraction of time-dependent system properties. This is summarised in Figure~\ref{fig:framework}.

\subsection{State Preparation}
\label{State preparation}

As shown in Eq.~\ref{init state}, the initial state for the model system is the tensor product of independent Gaussian distributions in four normal modes, each centred at $Q=0$, and the electronic state vector in the $S_2$ state. Several methods have been proposed for preparing Gaussian distributions across the amplitudes of a qubit register, including the uniformly-controlled rotation method~\cite{Mttnen2004}, the Matrix Product States method~\cite{Iaconis_2024}, the Fourier Series Loader method~\cite{moosa2023} and the discrete random walk method~\cite{Rattew_2021}. Here we choose the uniformly-controlled rotation approach, which uses only $\mathbf{R_{y}}$ and $\mathbf{CNOT}$ gates~\cite{moosa2023,Mttnen2004}, and where all the angles of $\mathbf{R_{y}}$ gates are conveniently calculated from the wavefunction amplitudes.

We allocate $2^n$ grid points to each normal mode; this is achieved by assigning $n$-qubit sub-registers to the representation of each normal mode, and therefore $dn$ qubits in total to represent $d$ spatial degrees of freedom: 
\begin{equation}
\overbrace{\underbrace{\ket{k_{0}k_{1}k_{2}\cdots k_{n-1}}}_{n\text{-qubit} \medspace \text{sub-register}}
\otimes\underbrace{\ket{k_{0}k_{1}k_{2}\cdots k_{n-1}}}_{n\text{-qubit} \medspace \text{sub-register}}
\otimes \cdots }^{d \medspace \text{sub-registers}},
\label{eq:register}    
\end{equation}
 One further qubit is allocated to represent the electronic degree of freedom. The quantum resource for encoding the molecular system counts to $dn + 1$. For our numerical emulations, we select $n=4$ (i.e. 16 grid points per normal mode, determined through systematic analyses in Section~\ref{spatial test}), and the 4D pyrazine model requires overall 17 qubits for the molecular state register. 

The quantum circuit to prepare the initial state consists of 4 identical uniform-rotation circuits applied in parallel to each of the normal mode sub-registers. The corresponding gate depth is given by $2^{n+1}-3$, which is 29 for $n=4$.  For the details of the quantum circuit and how the angles of the gates relate to the wavefunction amplitudes, we refer the reader to Appendix~\ref{App state prep}, where we also demonstrate the veracity of the approach through emulated random measurement outcomes of the tuning mode state prepared using Qiskit~\cite{moosa2023,Qiskit}. This state preparation algorithm is straightforwardly applicable to higher dimensional harmonic models without extra gate depth or additional time costs, since the states of different sub-registers are prepared in parallel.

\subsection{Time Evolution}
\label{Time evolution}

\begin{figure*}[!htbp]
\centering
\hspace*{1.28cm}
\scalebox{1.1}{
\Qcircuit @C=1em @R=1.5em {
\lstick{\text{\normalfont Electronic state}\ket{q}} &\qw & \ctrlo{1} & \ctrl{1} & \multigate{1}{U_\text{c}} & \qw & \qw &\qw&\ctrlo{1} & \ctrl{1}&\multigate{1}{U_\text{c}} &  \qw \\
\lstick{\nu_{10a}} &\qw {/} & \multigate{1}{U_{S1}} & \multigate{1}{U_{S2}}& \ghost{U_\text{c}} &\multigate{1}{U_\text{QFT}} & \multigate{1}{U_{K}} & \multigate{1}{U_\text{QFT}^{-1}} & \multigate{1}{U_{S1}} & \multigate{1}{U_{S2}}& \ghost{U_\text{c}}&  \qw   \\
\lstick{\text{\normalfont Tuning modes}} &\qw {/} & \ghost{U_{S1}} & \ghost{U_{S2}}& \qw &\ghost{U_\text{QFT}} & \ghost{U_{K}} & \ghost{U_\text{QFT}^{-1}} & \ghost{U_{S2}} & \ghost{U_{S1}}& \qw&  \qw  
}}
\caption{\justifying Quantum circuit of time evolution for one time step $dt$.}
\label{fig:1}
\end{figure*}
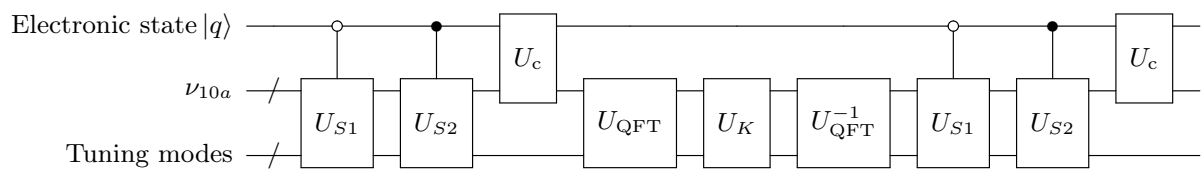

The SO-QFT method has previously been applied to  first-quantized real-space quantum simulation of a simplified Marcus model with two coupled 1D harmonic PESs \cite{Pauline2020}. We extend their time evolution decomposition to second order to obtain 
\begin{equation}
e^{-iHdt}\approx  U_\text{diag}  U_\text{c} U^{-1}_\text{QFT} U_K U_\text{QFT}U_\text{diag}U_\text{c}
\label{eq:1}
\end{equation}
where
$$U_\text{diag} = e^{-\frac{iV_\text{diag}dt}{2}},
\quad U_\text{c} = e^{-\frac{iV_\text{c}dt}{2}}, 
 \quad  U_{K} = e^{-iKdt}.$$

Figure~\ref{fig:1} summarises the quantum circuit schematically for one time step. This sequence is repeated $n_t$ time steps to evolve for a total time of $T=n_tdt$. $U_\text{diag}$ is decomposed into two controlled unitary operators $U_{S1}$ and $U_{S2}$. As shown in Eq.~\ref{eq:hamil}, $V_\text{diag}$ and $K$ involve $0^\text{th}$-, $1^\text{st}$- and $2^\text{nd}$-order polynomial terms in either position or momentum coordinates. Quantum circuits that apply the exponential of polynomials with these orders are well established and commonly used in quantum algorithms~\cite{Pauline2020}. Re-writing the time evolution operators by expressing $Q$ as a binary expansion is equivalent to applying various combinations of elementary gates including $\mathbf{U_{1}}$, Pauli $\mathbf{X}$ and controlled $\mathbf{R_{x}}$. The rotation angles are functions of the time resolution and the parameters used in the Hamiltonian. Details of the specific circuits for $U_\text{diag}$, $U_\text{c}$ and $U_K$ are described in Appendix~\ref{App evolve} and Appendix~\ref{App qft} gives details of $U_\text{QFT}$. 

The operations within $U_\text{diag}$ or $U_{K}$ block permit concurrent execution across different normal modes if they were not subject to global control. Since they are now collectively controlled by either the same electronic state register or the same ancilla qubit required by measurement techniques, this enforces sequential executions and causes the circuit depth to scale with the number of normal mode registers $d$. 
 
When allocating $n$ qubits to each normal mode sub-register and combing the resulting blocks, each $U_{S1}$ or $U_{S2}$ consist of $d(n^2+n)+4$ gates, with $d=4$ corresponding to four normal modes of the 4D model Hamiltonian. Together, the two $U_{S1}$ and $U_{S2}$ blocks contribute $2d(n^2+n)+8$ gates to the entire $U_\text{diag}$ section. The $U_\text{c}$ and $U_{K}$ blocks add further $n$ and $dn^{2}$ gates, respectively. For one forward or inverse QFT block, its gate depth is calculated to be $n^{2}/2+n$ for even $n$, according to the circuit shown in the Appendix~\ref{App qft}. For odd $n$, we don't have to swap the middle qubit in the following SWAP procedure, then the gate count reduces to $n^{2}/2+n-1/2$.

Consequently, the time evolving operations of one single time step illustrated in Figure~\ref{fig:1}, including all the $U_\text{diag}$, $U_\text{c}$, $U_{K}$ and QFTs, generate $6dn^{2}+(6d+2)n+16$ gates for even $n$ (or $6dn^{2}+(6d+2)n+16-d$ in the case of odd $n$). The 4D model with $n=4$ and $d=4$ presents a total gate count of 504 for the time propagation within one $dt$. The total circuit depth then depends on the number of discrete time steps $n_t$. 

We also note that we can select an alternative second-order split with comparable accuracy by performing a half time step of the kinetic operator and a full time step of the potential operator. As the circuit for $U_K$ is shorter than $U_\text{diag}U_\text{c}$, this reversed split saves $dn^2+(2d+1)n+8$ gates per time step, but requires additional QFTs at the beginning of the time evolution and at the end, before each measurement.


\subsection{Signal Processing}
\label{signal processing}

\subsubsection{Absorption Spectrum Measurement}
\label{spec theory}

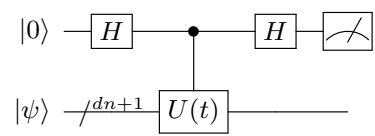
\begin{figure}[!htbp]
\centering
\scalebox{1.1}{
\Qcircuit @C=1em @R=1.5em {
\lstick{\ket{0}} & \gate{H} & \ctrl{1} & \gate{H}  & \meter  \\
\lstick{\ket{\psi}} & \qw {/}^{dn+1} & \gate{U(t)} & \qw & \qw  
}}
\caption{\justifying Autocorrelation measurement at a single time step using an ancillary qubit.}
\label{fig:6}
\end{figure}

The absorption spectrum is obtained from the Fourier Transform of the autocorrelation time function, as shown in Eq.~\ref{eq:absorption}.
The autocorrelation signal will be extracted from digital quantum time propagation experiments from single-ancilla-qubit measurements via techniques such as the iterative phase estimation or the Hadamard test~\cite{jensen2023,aharonov2006,Cleve1998,Ryan2023,lorenzo2024,dantong2024},
which requires repeated measurement `shots' sampling at multiple time steps to get an accurate signal.
We present the circuit of the Hadamard test in Figure \ref{fig:6}; given a time-evolution unitary $U(t)$ controlled by a single ancillary qubit, the phase acquired by the state during the evolution is encoded in the ancillary qubit. The difference in the probability of measuring the ancilla in state $\ket{0}$ and state $\ket{1}$ indicates us this phase -- specifically the real part of the autocorrelation signal (see details in Appendix~\ref{App hadamard}).
To obtain the imaginary counterpart, we need only add an $\mathbf{S}$ gate on the ancilla after the controlled operation and before the final Hadamard gate (illustrated in Appendix~\ref{App hadamard}).

This procedure is suitable when the qubit budget is limited, since it only requires a single ancilla qubit for autocorrelation extraction. With this approach, the final frequency spectrum is obtained by classical post-processing of the measured autocorrelation signal in the time domain. Advanced post-processing techniques used in related algorithms, such as statistical phase estimation, can also be used here to extract the absorption spectrum using fewer measurements~\cite{Wan2022,Lin2022,Blunt2023}.

\begin{figure}[!htbp]
\flushright
\scalebox{1}{
\Qcircuit @C=0.8em @R=0.8em {
\lstick{\ket{0}} & \gate{H} & \qw & \qw & \qw  & \cdots & & \ctrl{4} &\multigate{3}{\text{QFT}^{-1}} & \meter  \\
\lstick{\vdots} & & & & & \vdots & & & \\
\lstick{\ket{0}} & \gate{H} & \qw & \ctrl{2}& \qw  & \cdots & & \qw &\ghost{QFT^{-1}} & \meter  \\
\lstick{\ket{0}} & \gate{H} & \ctrl{1} & \qw& \qw  & \cdots & & \qw &\ghost{QFT^{-1}} & \meter  \\
\lstick{\ket{\psi}} & \qw {/}^{dn+1} & \gate{U^{2^{0}}} & \gate{U^{2^{1}}} & \qw &\cdots&& \gate{U^{2^{m-1}}}& \qw
}}
\caption{\justifying Spectrum signal measurements using canonical QPE techniques, with $m$ qubits in the time register.}
\label{fig:7}
\end{figure}
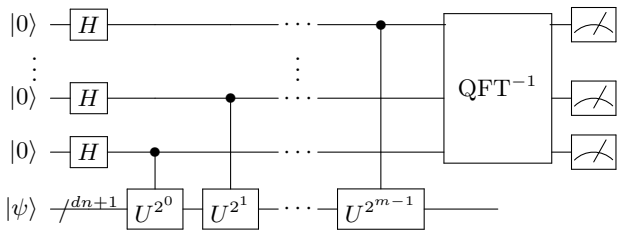

An alternative approach to extracting the absorption spectrum, shown in Figure~\ref{fig:7}, is to simply apply the canonical Quantum Phase Estimation (QPE) protocol~\cite{kitaev1995, Cleve1998,nielsen2010}. It uses multiple ancilla qubits to directly sample the spectrum, bypassing the measurement of the autocorrelation signal.
This approach generalises the single ancilla Hadamard test and replicates that across multiple ancilla qubits, each controlling time evolution to different time steps, capturing the autocorrelation signal in the time domain across the ancilla qubit register. After the full time evolution, the inverse QFT is applied to the time register, which retrieves the phase corresponding to eigenvalues of the unitary time evolution operator (see details in Appendix~\ref{General QPE}). As the eigenvalues are easily deduced from the measured phases, the probability of occurrences of one specific phase is proportional to the intensity of the corresponding eigenvalue. The frequency range and spectra intensity are accumulated until convergence after sufficiently many repeated measurements. There have been many studies detailing specific implementations and error analyses of QPE~\cite{Cleve1998,miroslav2008,Cao2019,OBrien2019,Ni2023}.

Note that canonical phase estimation cannot be used to obtain the absorption spectrum directly if the damping functions $D(t)$ and $B(t)$ are required; classical post-processing of the autocorrelation signal will have to take place before computing the spectrum via Fourier transform, implying the use of techniques like single-qubit statistical phase estimation.
It is likely that neither damping function would be necessary in at-scale, fault-tolerant applications, where all modes are explicitly represented in long time propagations and multi-ancilla canonical QPE is relevant -- the function $B(t)$ is needed to account for the neglected modes in low-dimensional models, and $D(t)$ to remove artefacts arising from short time propagation.

\subsubsection{Population Dynamics Measurement}

Throughout the photorelaxation process, the populations of the two states fluctuates due to the transfer through the conical intersection. The instantaneous populations are simply extracted by sampling the electronic qubit in the $Z$ basis. The probability of measuring the $\ket{0}$ and $\ket{1}$ states reflects the population of $S_{1}$ and $S_{2}$ states: $$P_{S1}=\frac{C_{0}}{C_{0}+C_{1}} \text{ and } P_{S2}=\frac{C_{1}}{C_{0}+C_{1}},$$ where $C_{0}$ and $C_{1}$ are the frequencies with which the $\ket{0}$ and $\ket{1}$ states are measured, respectively.

\section{Cost Evaluation}

In this section, we analyse the minimum resources required to execute the SO-QFT algorithm for simulating the absorption spectrum and population dynamics of the 4D model of pyrazine. We will first characterise the number of qubits, entangling gates and repeated measurements needed to obtain expected vibronic properties, before generalising to the case of a model Hamiltonian in higher dimensions. 

To quantify the resource cost of implementing the algorithm, we need to ascertain the spatial- and time-resolution necessary for obtaining accurate results with controlled simulation errors; these parameters will determine the number of qubits required to represent the state and the number of time steps required for the time-evolution.
Furthermore, the resolution of the resulting spectrum is limited by the number of sampled time points in the autocorrelation signal, which need not have the same density as the SO-QFT propagated time steps. A higher sampling density means higher spectrum measurement costs in both the statistical and canonical phase estimation.

\begin{table*}[!htbp]
    \centering 
    \renewcommand{\arraystretch}{1.3}
    \begin{tabular}{|p{1cm}<{\centering}|p{1.7cm}<{\centering}|p{1.5cm}<{\centering}|p{1.5cm}<{\centering}|p{1.5cm}<{\centering}|p{1.5cm}<{\centering}|p{1.5cm}<{\centering}|p{1.5cm}<{\centering}|} 
        \hline 
        $N$ &  Range & $E_0$ & $E_1$ & $E_2$ & $E_3$ & $E_4$ & $E_5$ \\ 
        \hline 
        8 & $[-2.5:2.5]$ & 0.22548  & 0.68594  & 1.09920  & 1.79509  & 1.86961  & 3.29204  \\ 
        16 &$[-5: 5]$ & 0.22585  & 0.67755  & 1.12925  & 1.58094 & 2.03270  & 2.48402   \\ 
        32  &$[-10:10]$ & 0.22585  & 0.67755  & 1.12925  & 1.58095  & 2.03267   & 2.48423   \\ 
        64  &$[-20: 20]$ & 0.22585  & 0.67755  & 1.12925  & 1.58095  & 2.03266  & 2.48428  \\ 
        \hline 
    \end{tabular}
    \caption{\justifying Numerical energies of the six lowest vibrational states, computed using $N$ grid points over dimensionless normal coordinates that span $[Q_\text{min}:Q_\text{max}]$.} 
    \label{table spatial ran} 
\end{table*}

\begin{table*}[!htbp]
    \centering  
    \renewcommand{\arraystretch}{1.3}
    \begin{tabular}{|p{1cm}<{\centering}|p{1.7cm}<{\centering}|p{1.5cm}<{\centering}|p{1.5cm}<{\centering}|p{1.5cm}<{\centering}|p{1.5cm}<{\centering}|p{1.5cm}<{\centering}|p{1.5cm}<{\centering}|} 
        \hline 
        $N$ &  Spacing & $E_0$ & $E_1$ & $E_2$ & $E_3$ & $E_4$ & $E_5$ \\ 
        \hline 
        4 & 3.33  & 0.65244  & 0.75274  & 0.65243  & 0.75262  & 0.66857  & 0.75458  \\ 
        8 & 1.43 & 0.23403  & 0.61630  & 1.35052  & 1.42281 & 1.29920  & 1.46433 \\ 
        16  & 0.67  & 0.22585  & 0.67755  & 1.12925  & 1.58094   & 2.03270  & 2.48402  \\ 
        32  & 0.32  & 0.22585  & 0.67755  & 1.12925  & 1.58095  & 2.03265  & 2.48437  \\ 
        64 & 0.16 & 0.22585  & 0.67755  & 1.12925  & 1.58095 & 2.03265  & 2.48442 \\
        \hline 
    \end{tabular}
    \caption{\justifying Numerical energies of the six lowest vibrational states, computed using $N$ grid points between $[-5, 5]$ in dimensionless normal mode coordinates.} 
    \label{table spatial res} 
\end{table*}

We therefore start this section with a series of numerical experiments aimed at identifying optimal simulation spatial size, spatial resolution, total simulation time and time resolution, such that the computed autocorrelation signal, population dynamics and absorption spectra closely reproduce benchmark results reported in Ref.~\cite{Krempl1994, Worth1996}, which were in agreement with experimental data~\cite{Yamazaki1983}.
Given the absence of large-scale fault-tolerant quantum hardware, our numerical experiments reported here were performed via high-performance classical emulations.

\subsection{Spatial Size and Spatial Resolution}
\label{spatial test}

In grid-based methods, the size of the simulation environment and the density of grid points over this space must be selected such that the wavefunction is accurately represented throughout the simulated time window without using an excessively dense grid. We also need to make sure that the simulation environment is large enough to model the physics without introducing any significant artificial reflections at the boundaries.

In Tables~\ref{table spatial ran} and~\ref{table spatial res}, we report the numerical energies of the six lowest vibrational states as a function of the spatial range and grid density in dimensionless normal coordinates; the energies were calculated as
\begin{equation}
    \langle E_\text{numerical} \rangle = \bra{\psi} V\ket{\psi} + \bra{\psi}  K\ket{\psi},   
\end{equation}
with each term evaluated in real and momentum space, respectively. We conducted two sets of systematic convergence tests: first, by fixing the spatial resolution and varying the spatial range (Table~\ref{table spatial ran}) to determine the smallest domain ensuring energy convergence, and second, by fixing this optimal size and adjusting the spatial resolution (Table~\ref{table spatial res}) to identify the minimal number of grid points that retains accuracy. While the higher excited states require finer grids over larger ranges for full convergence, the grid of 16 points within the range of $Q \in [-5,5]$ was considered sufficient for the present analysis, which provides a satisfactory level of accuracy in the eigenenergies of the six states examined.

\begin{figure}[!htbp]
    \centering
    \includegraphics[scale=0.15]{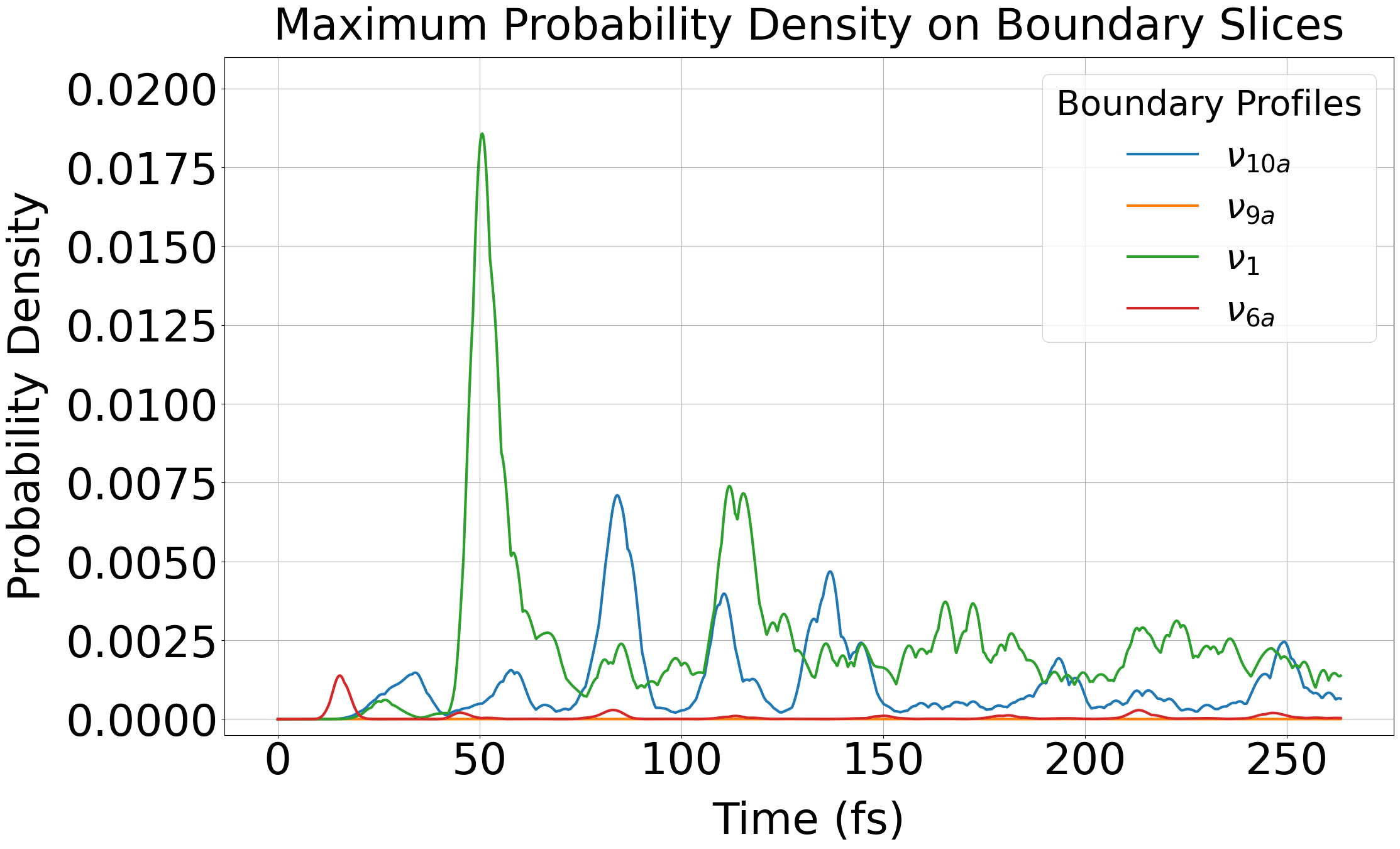}
    \caption{\justifying Maximum values of the probability density at the boundaries of the simulation box for each normal mode as a function of time. The persistently low amplitudes indicate that the chosen simulation size is sufficient to contain the evolving wavefunction.
    }
    \label{fig:spatial}
\end{figure}

In addition, we validate that the size of the simulation environment is large enough to capture the essential time-dependent physics. Taking the simulation coordinates of $Q=[-5,5]$ and 16 grid points for each coordinate, we performed a time evolution of the system using 2048 time steps within a total time of around 264~fs. Figure~\ref{fig:spatial} shows the maximum values of the probability density at the extreme ends of the simulation range during that simulated time, evaluated as: 
\begin{equation}
    P_i(t) = \max \bra{ \psi(t)}  \delta(Q_i \pm 5) \ket{\psi(t)}.
\end{equation}
We see that the probability amplitudes of the wavefunction at edges of the simulation range remain persistently low throughout the time evolution, confirming that the size of the simulation environment is sufficient.

\subsection{Total Time and Time Resolution}
\label{time resolution}

To obtain an accurate absorption spectrum, the dynamics must be simulated for a sufficiently long time to record the features of the vibronic dynamics.
Furthermore, the granularity of the SO-QFT decomposition (time resolution) must be sufficiently small to control the Trotter error. Finally, the autocorrelation sampling must also be frequent enough to resolve the relevant features of the spectrum.
The overall quantum resource, gate depth and measurement requirements will depend on these algorithmic parameters.

\begin{figure} [!htbp]
\centering
\includegraphics[scale=0.15]{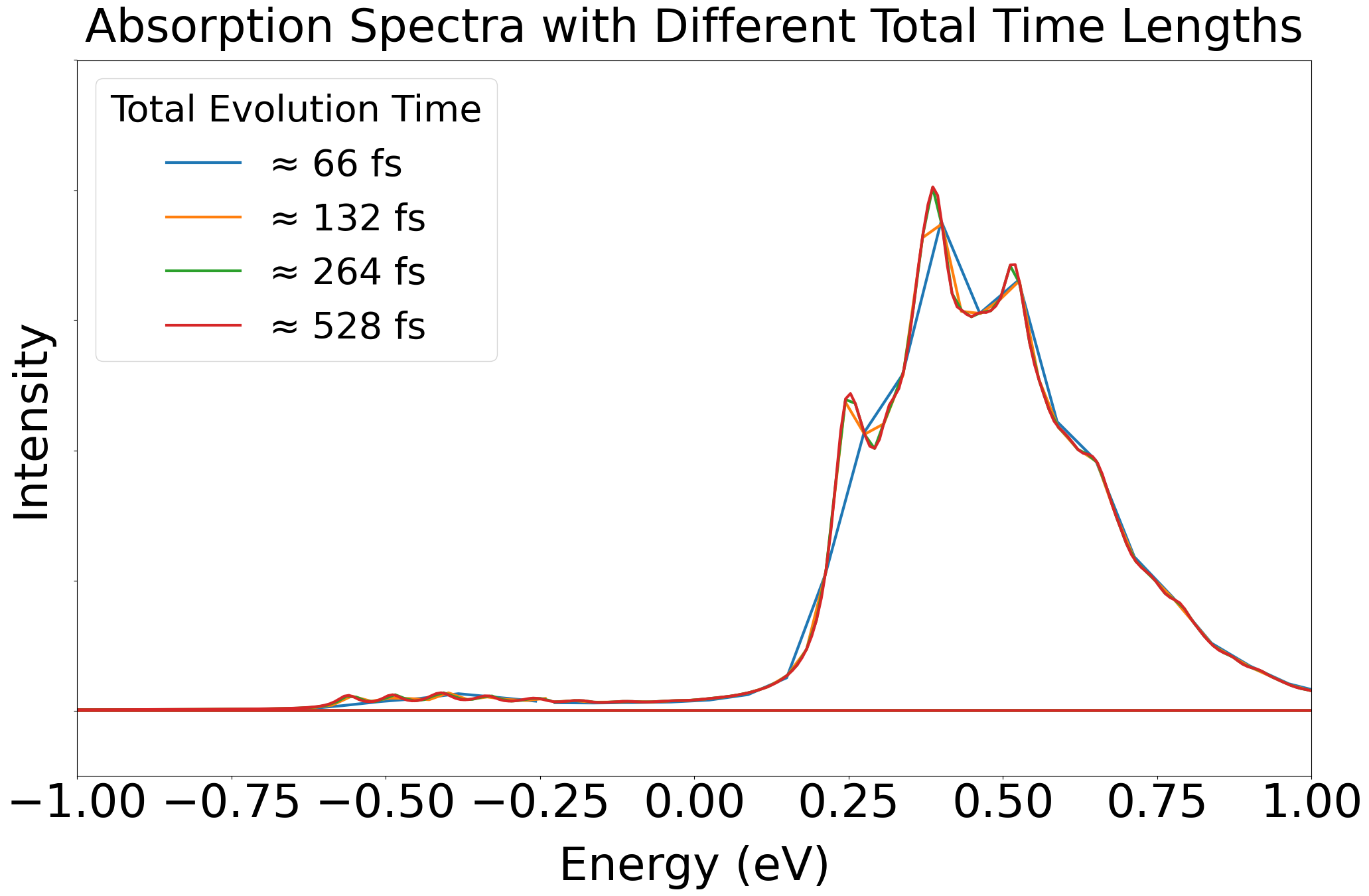}
\caption{\justifying Comparison of absorption spectra across various time propagation lengths, with the number of time steps set to 512, 1024, 2048 and 4096 for total time lengths of 66, 132, 264, 528~fs, respectively. This setup ensures a time resolution of $dt=0.13$~fs in all cases.}
\label{fig:auto spec timelen} 
\end{figure}

\begin{figure} [!htbp]
\centering
\subfloat[\label{fig:auto diffres}]{
\includegraphics[scale=0.15]{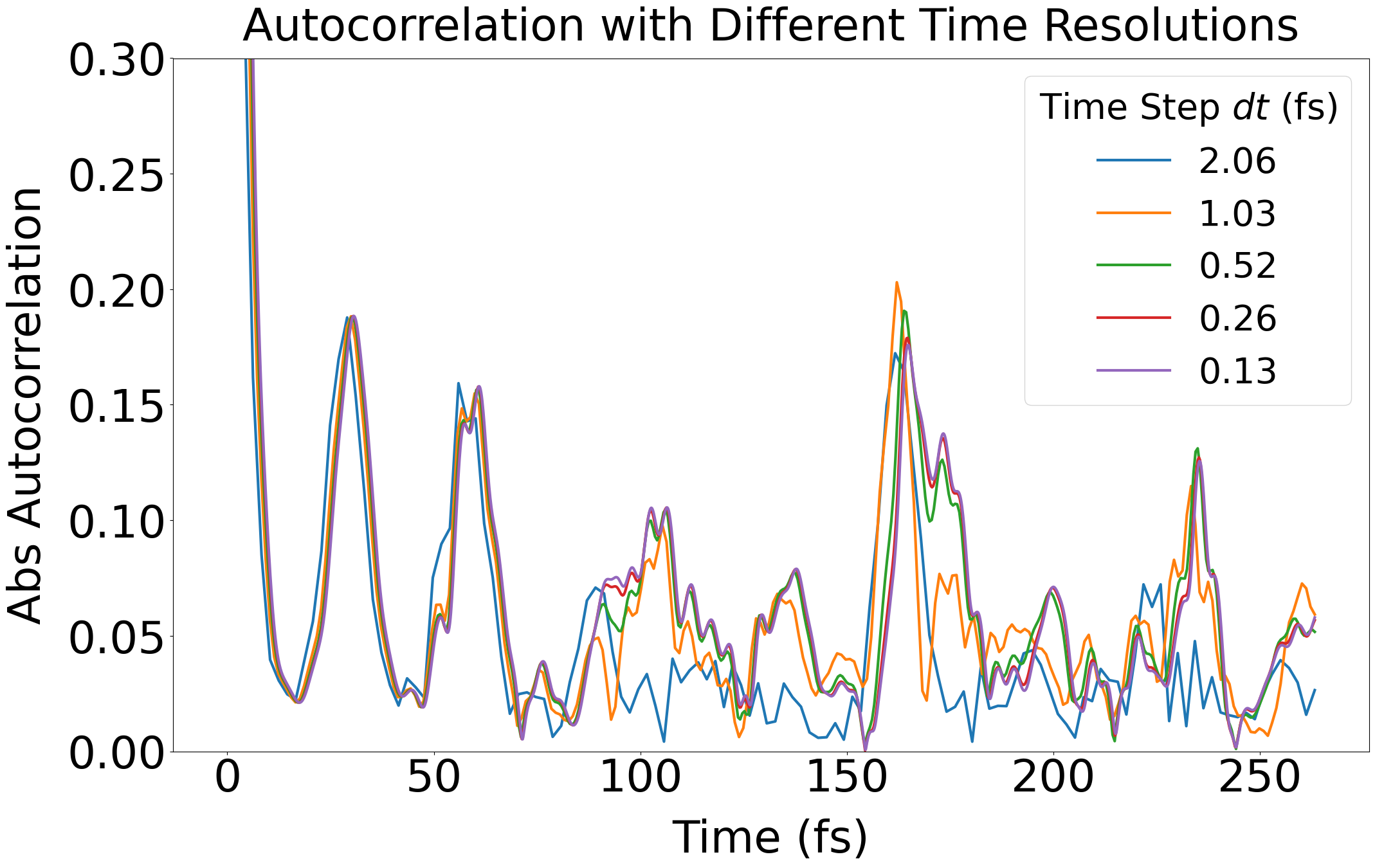}}
\\
\subfloat[\label{fig:spec diffres}]{
\includegraphics[scale=0.15]{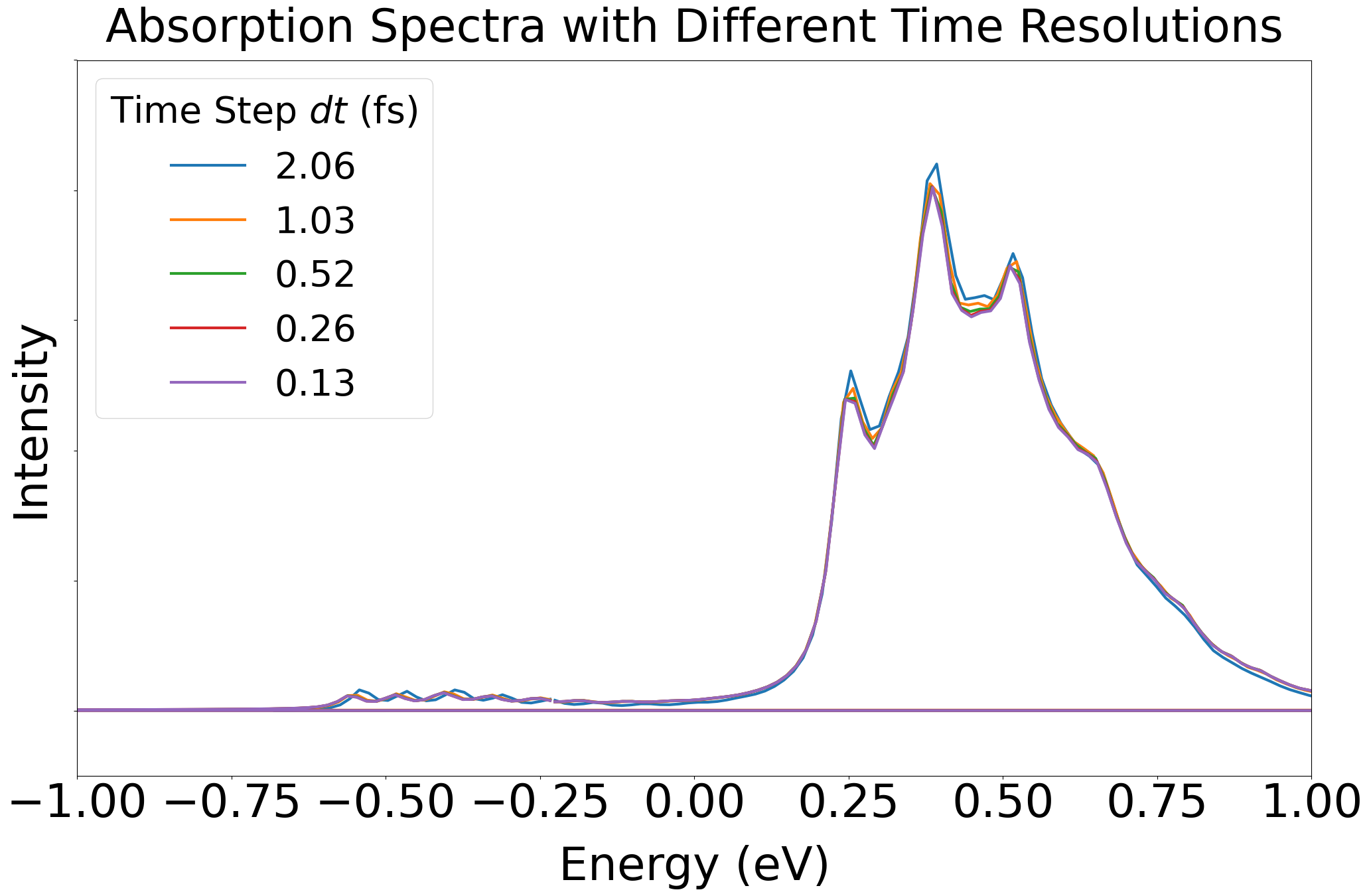}}
\\
\subfloat[\label{fig:popu diffres}]{
\includegraphics[scale=0.15]{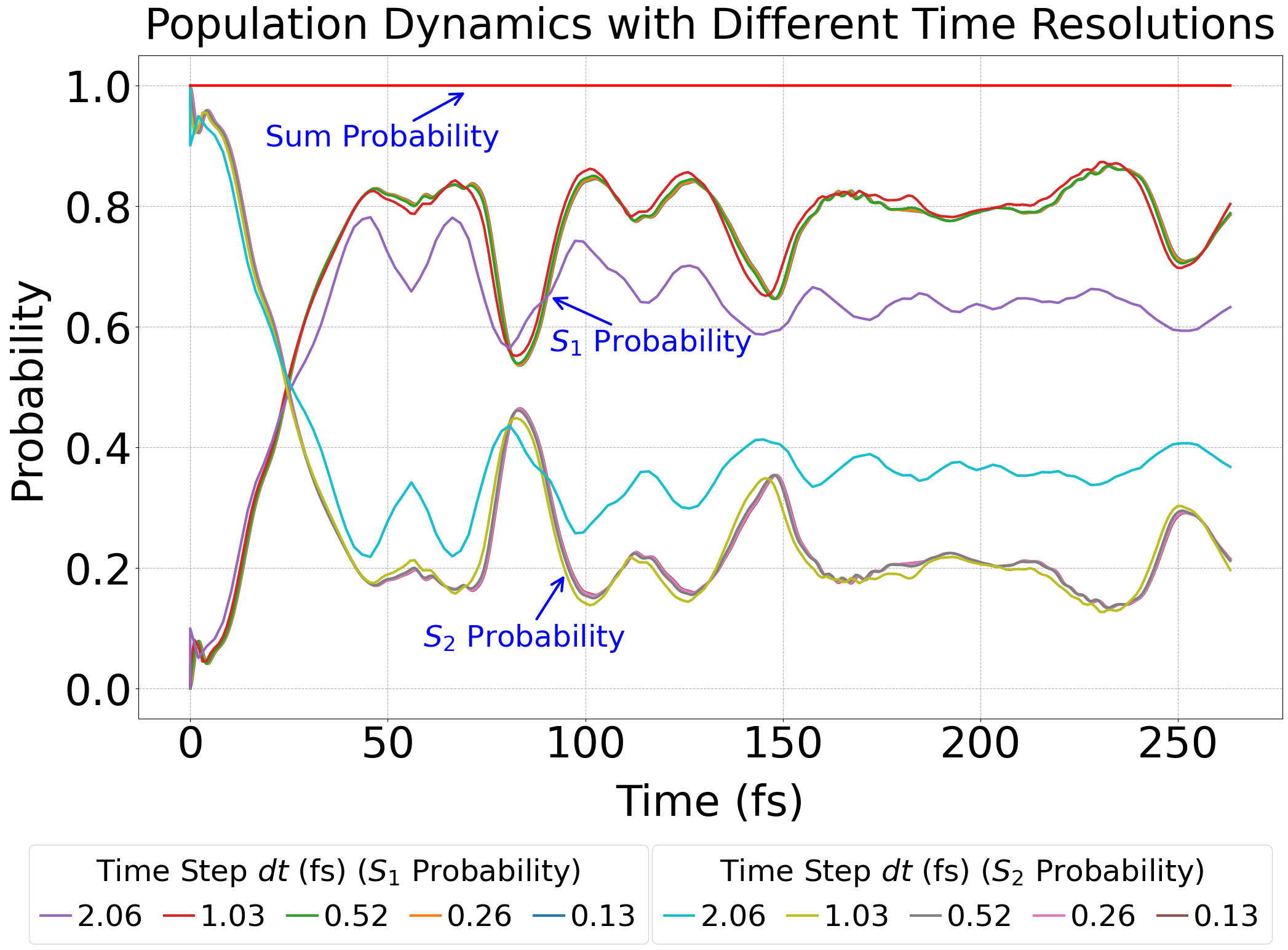}}
\caption{\justifying Comparison of (a) absolute autocorrelation, (b) absorption spectra, and (c) population dynamics with 128, 256, 512, 1024 and 2048 time steps used within a fixed total evolution time of around 264~fs.}
\label{fig:diff timeres} 
\end{figure}

\begin{figure}[!htbp]
\centering
\subfloat[\label{fig:auto diffsam}]{
\includegraphics[scale=0.15]{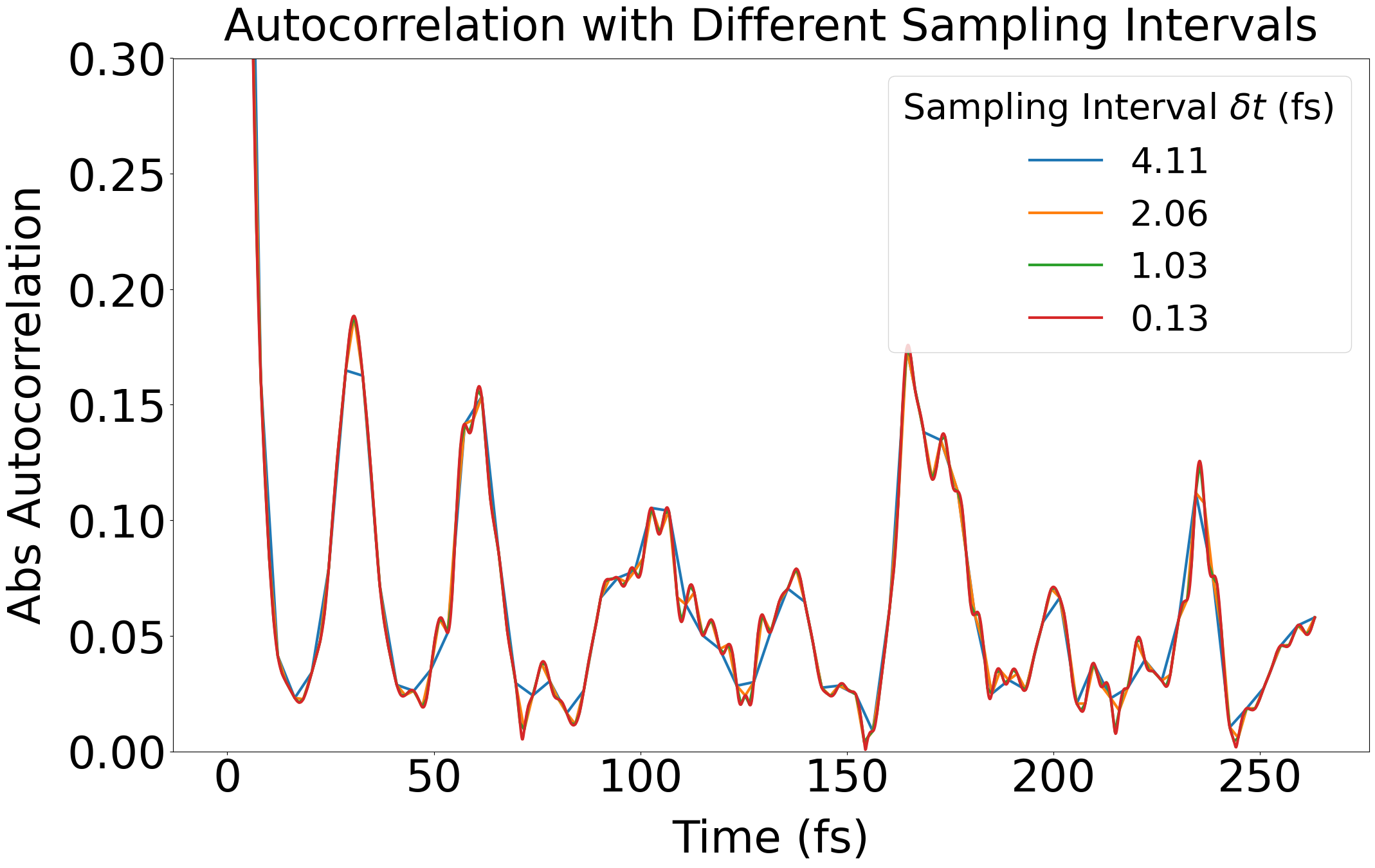}}
\\
\subfloat[\label{fig:spec diffsam}]{
\includegraphics[scale=0.15]{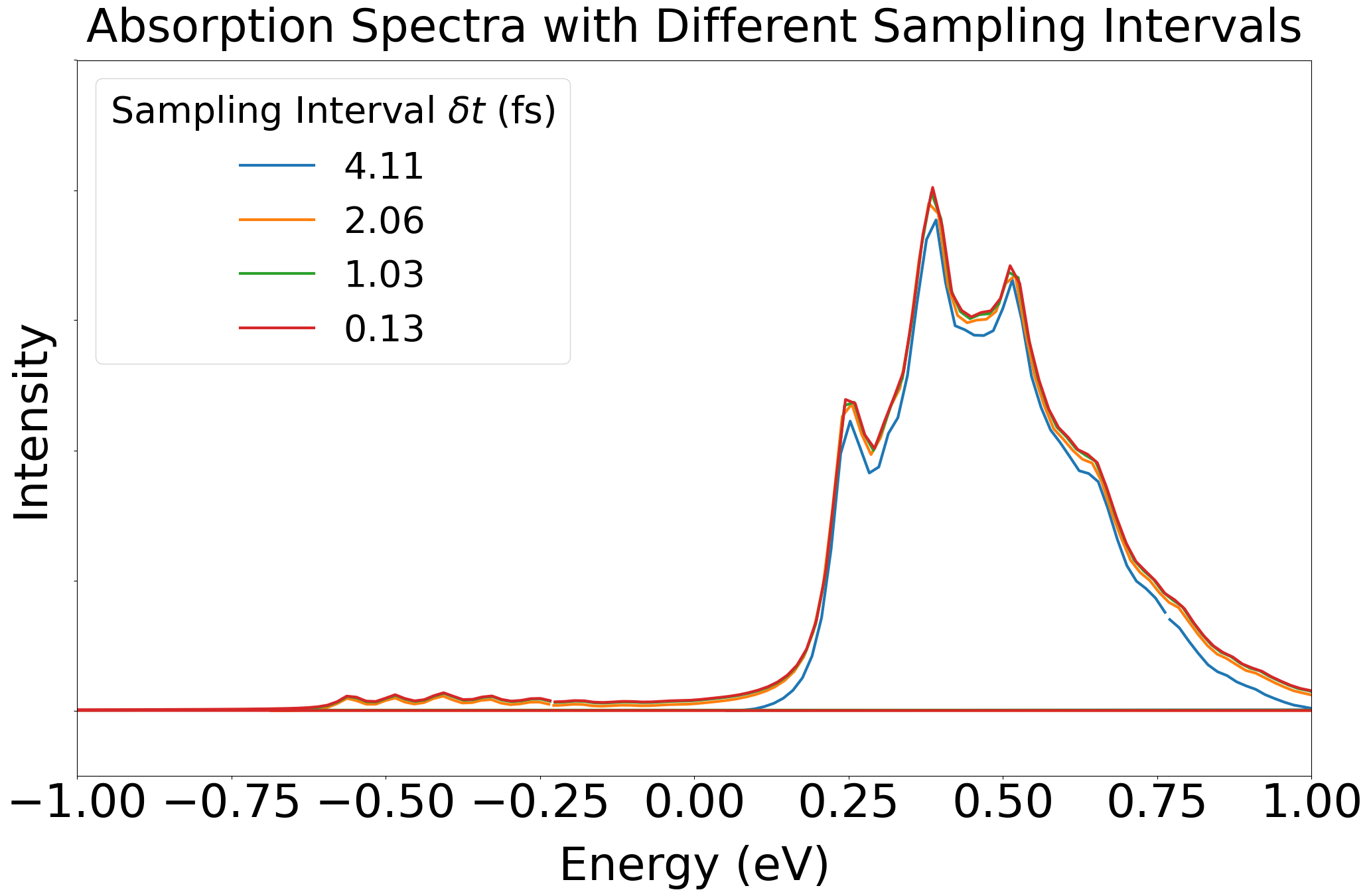}}
\\
\subfloat[\label{fig:popu diffsam}]{
\includegraphics[scale=0.15]{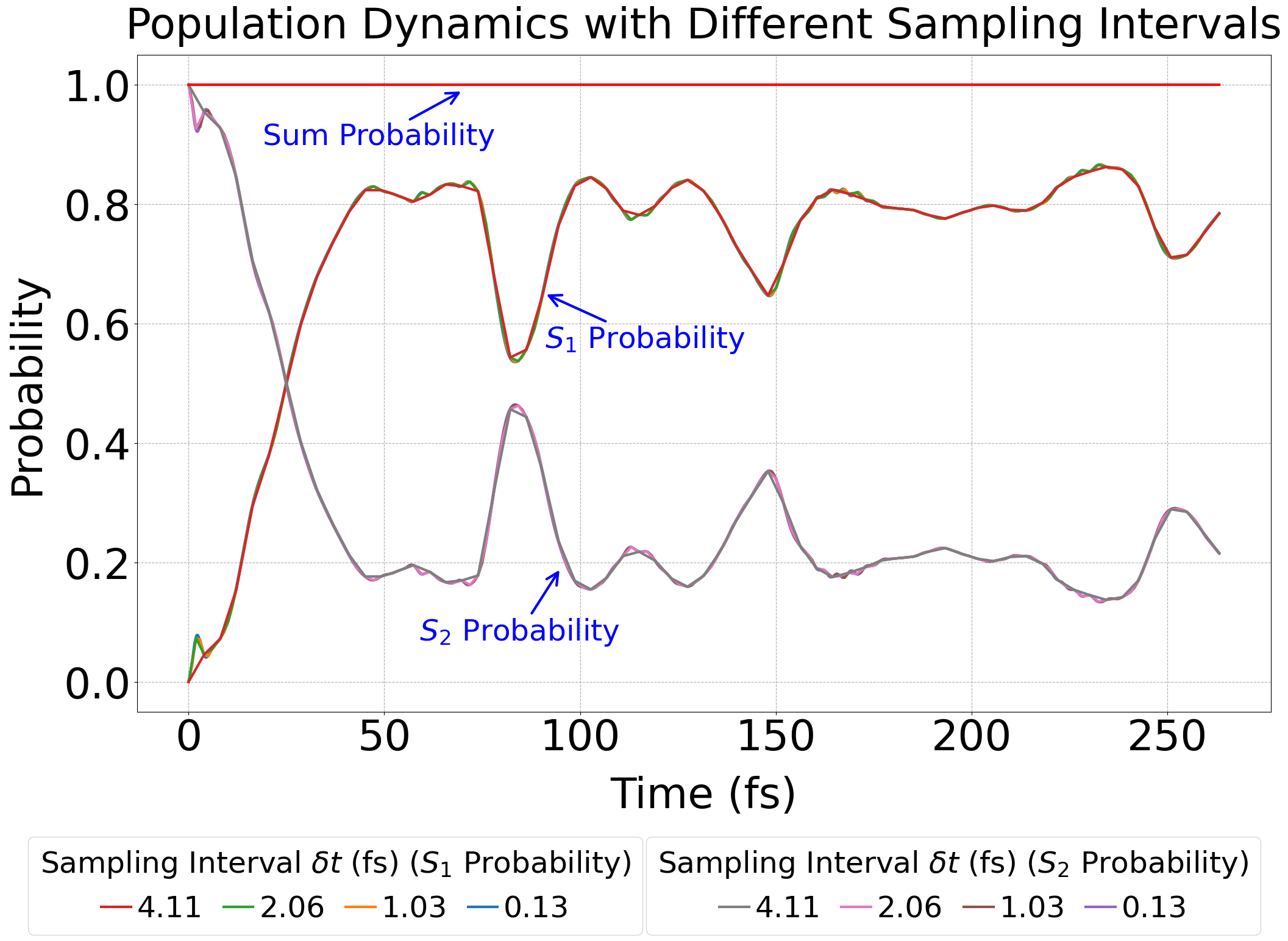}}
\caption{\justifying Comparison of (a) absolute autocorrelation, (b) Absorption spectra, and (c) population dynamics with different sampling time intervals, while keeping the time evolution setting fixed at 2048 steps within 264~fs.
}
\label{fig:sampling} 
\end{figure}

We first performed simulations to determine the necessary total simulation time for capturing the correct dynamics by
simulating increasing periods of total evolution time $T$ but with the same time resolution $dt=0.13$~fs.
The autocorrelation function was recorded at every time step.
Figure~\ref{fig:auto spec timelen} shows the resulting absorption spectra.
We see that a total simulated time of 264~fs suffices to produce an accurate absorption spectrum.
For simulations of this length, we found the inclusion of the damping function $D(t)$ has negligible impact on the resulting spectrum; we therefore omit its use for the remainder of this work.

We then explored, using a total simulation time of $T\approx264$~fs, how the time step affects the time propagation errors. Figure~\ref{fig:diff timeres} reports the autocorrelation function, absorption spectra, and population dynamics resulting from simulations that used between 128-2048 time steps. The autocorrelation function was again sampled at the same frequency as the time resolution $dt$.
The minimum number of time steps that yields a qualitative spectrum and population dynamics is 512 (corresponding to a time resolution of $dt=0.52$~fs). Further increasing the number of time steps to 1024 (i.e. $dt=0.26$~fs) ensures a converged quantitative reproduction of the vibronic dynamics.

As discussed, it is not necessary to sample the autocorrelation signal at every time step in the Trotter evolution.
To investigate this, we computed autocorrelation signals obtained from simulating $T\approx264$~fs with a Trotter step of $0.13$~fs, but sampled at longer intervals: every 32, 16 and 8 time steps. The resulting autocorrelation functions and corresponding absorption spectra are compared in Figure~\ref{fig:sampling}, where we visualise the severity of under-sampling with a smooth interpolation between the recorded data points. Similarly, the coherence of population dynamics curvatures across $S_1$ and $S_2$ states is also presented to assess the reliability and fidelity of the chosen sampling strategy. To obtain an accurate spectral function, it is only necessary to sample the autocorrelation function through quantum circuit measurements at a frequency of once every $2.06$~fs, which can significantly reduce the measurement overhead.

\subsection{Gate Depth Calculation}
\label{gate depth cal}

\begin{table*}[!htbp]
    \centering 
    \renewcommand{\arraystretch}{1.3}
    \begin{tabular}{|c<{\centering\arraybackslash}|c<{\centering\arraybackslash}|c<{\centering\arraybackslash}|c<{\centering\arraybackslash}|} 
        \hline 
           &Gate depth & $n=4$, $n_{t} = 512$& $n=5$, $n_{t} = 1024$ \\ 
        \hline 
        \hline 
        State Preparation &  $N_\text{i} = 2^{n+1} - 3$  & 29  &  61 \\ 
        \hline
        \multirow{2}{*}{Time Evolution}    & $N_\text{t} = (6dn^{2}+(6d+2)n+16)(n_t-1)$ for even $n$ & \multirow{2}{*}{$257{,}544$} & \multirow{2}{*}{$759{,}066$} \\ 
            & $N_\text{t} = (6dn^{2}+(6d+2)n+16-d)(n_t-1)$ for odd $n$ &  &  \\
        \hline
        \multirow{2}{*}{Signal processing} & $N_\text{m} = (\log_{2}n_t)^{2}/2+\log_{2}n_t$ for even $\log_{2}n_t$  &\multirow{2}{*}{49} &\multirow{2}{*}{60}\\ 
        & $N_\text{m} = (\log_{2}n_t)^{2}/2+\log_{2}n_t-1/2$ for odd $\log_{2}n_t$ &  &  \\
        \hline
        Full algorithm A &  $N_\text{i} +N_\text{t} + 2$ & $257{,}575$ & $759{,}129$ \\
        \hline 
        Full algorithm B & $N_\text{i} +N_\text{t} + N_\text{m}$ & $257{,}622$ & $759{,}187$ \\
        \hline 
    \end{tabular}
    \caption{\justifying Cost evaluation through gate depth for 4D ($d=4$) simulations of pyrazine photodynamics in terms of the number of qubits per mode $n$ and the number of time steps $n_t$. Algorithm A is the classical post-processing approach where the autocorrelation function is obtained through measurement, and B is the fully quantum approach which directly obtains spectral signals through QPE.
    } 
    \label{table circuit depth} 
\end{table*}

We now present a quantitative estimate of the gate depth required for a simulation using the spatial and time resolutions established in the preceding sections. 
The circuit depth is evaluated as the number of sequential gate layers, with gates acting on disjoint qubit registers assumed to be parallelisable and therefore grouped into the same layer. Accordingly, multiple controlled unitaries are treated as simultaneous only if they share neither control nor target qubits.


From Section~\ref{State preparation} we concluded that the initial state preparation using uniformly controlled rotations has a gate depth of $2^{n+1}-3$, where $n$ is the number of qubits in each normal mode register. Section \ref{Time evolution} concludes that the time evolution adds $6dn^{2}+(6d+2)n+16$ gates per time step.
As mentioned in Section~\ref{spec theory}, an additional ancillary qubit is necessary to accomplish Hadamard tests when extracting the autocorrelation signal statistically. The initial Hadamard gate can be performed in parallel to the state preparation, but the Hadamard gate at the end of the test contributes one additional gate. The $\mathbf{S}$ gate required to measure the imaginary part of the signal contributes one further gate.

We now consider the gate depth for performing time evolution to obtain the autocorrelation signal and population dynamics.
Recall that we must first propagate the initial state to a particular point in `simulated' time, then measure the Hadamard test ancillary qubit for deriving the autocorrelation signal, or the qubit representing the electronic states to get the population dynamics. To obtain the next time point, the workflow must be restarted from the very beginning. The circuit contributing the greatest depth is therefore that for extracting the imaginary part of the autocorrelation at the last time point. Note that the population measurement requires two fewer gates, as it does not involve ancilla manipulation.

In Table~\ref{table circuit depth}, we present the full cost in terms of quantum gate depth. $n$ represents the number of qubits per vibrational mode and $n_t$ represents the number of time steps evolved within 264~fs. The circuit associated with preparing the initial $t=0$ as an individual time point always contributes. To illustrate the estimation concretely, we provide representative costs in two regimes: a simulation with $n=4$ and $n_t=512$, and a more expensive simulation with $n=5$ and $n_t=1024$. The deepest circuits involve 257,575 and 759,129 gates, respectively.

In the case where no explicit broadening or damping functions for post-processing the autocorrelation function is necessary, a fully quantum canonical QPE approach provides one of the most direct ways to derive the spectrum.
This approach requires an auxiliary register of $m$ qubits instead of just one, where the number of time points sampled is $2^{m}$.
The initial Hadamard gates applied to qubits on this time register are applied in parallel with the initial state preparation on the state registers encoding the normal modes. However, much like the single-ancilla, statistical version of phase estimation, the final inverse QFT on the ancilla register adds another $m^{2}/2+m$ gate counts after the time evolution. 
Table~\ref{table circuit depth} summarises the generalised formula calculating the circuit depth required for the canonical QPE approach. For the exploratory run with $n=4$ and $n_t=512$, the deepest circuits involve 257,622 gates and for the production run with $n=5$ and $n_t=1024$ the gate depth becomes 759,187.


\subsection{Measurement Shot Estimation}
\label{measure shot}

We now give an assessment of the minimum number of measurement shots necessary to reproduce the absorption spectrum, relevant to both the statistical autocorrelation-based and direct spectrum sampling scenarios, using a total simulation time of $T\approx264$~fs, a time step of $dt=0.26$~fs and autocorrelation sampling of $\delta t=0.26$~fs.
Our target is for the empirical spectrum to achieve a Total Variation Distance (TVD) below a specified threshold.
Here the TVD is defined as a unitless relative error measure: 
\begin{equation}
    \frac{1}{2}\sum_{i}\left|\frac{I(\omega_i)^\text{empirical}}{\sum_j I(\omega_j)^\text{empirical}}-\frac{I(\omega_i)^\text{target}}{\sum_j I(\omega_j)^\text{target}}\right|.
\end{equation}

Our first investigation separately sampled real and imaginary parts of the autocorrelation function $A(t)$, as would be done on a real quantum hardware. We classically mimicked this process by drawing binary outcomes from the target binomial distributions: 
\begin{equation}
    \begin{aligned}
    &\text{sampling of the real part follows}  \\
    &P(\ket{0})=\frac{1+\operatorname{\mathbb{R}e}\{A(t)\}}{2}, \ P(\ket{1})=1-P(\ket{0}); \\
    &\text{sampling of the imaginary part follows}  \\
    &P(\ket{1})=\frac{1+\operatorname{\mathbb{I}m}\{A(t)\}}{2}, \ P(\ket{0})=1-P(\ket{1}).    
    \end{aligned}
\end{equation}
For each discrete time point, sampling counts ranged from 1,000 to 300,000, with increments of 1,000. After combining real and imaginary parts, we performed the Fourier Transform to obtain the estimated spectrum and evaluated its fidelity via TVD against the target spectrum.

In the second assessment of the full QPE approach, we treated the target spectrum as a normalized probability distribution and drew samples directly from it using a discrete random variable model. At each iteration, we simulated $M$ measurements and the empirical spectrum was constructed and compared to the target spectrum via TVD, with $M$ increasing from 1,000 to 200,000 in steps of 1,000.  

\begin{figure*}[!htbp]
    \centering

    \begin{subfigure}[b]{0.48\textwidth}
        \includegraphics[scale=0.15]{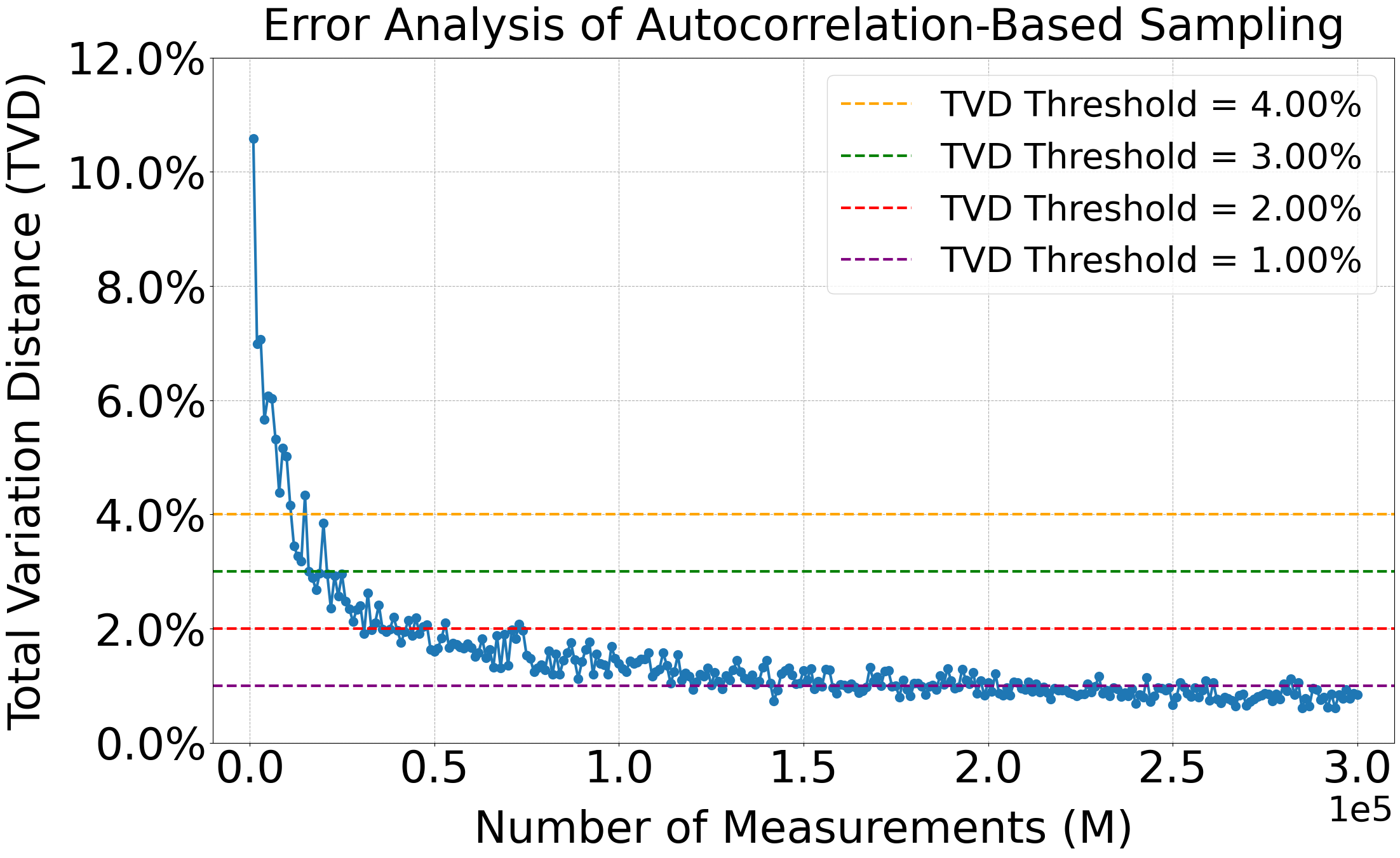}
        \caption{}
        \label{fig:autocorr_tvd}
    \end{subfigure}
    \hfill
    \begin{subfigure}[b]{0.48\textwidth}
        \includegraphics[scale=0.15]{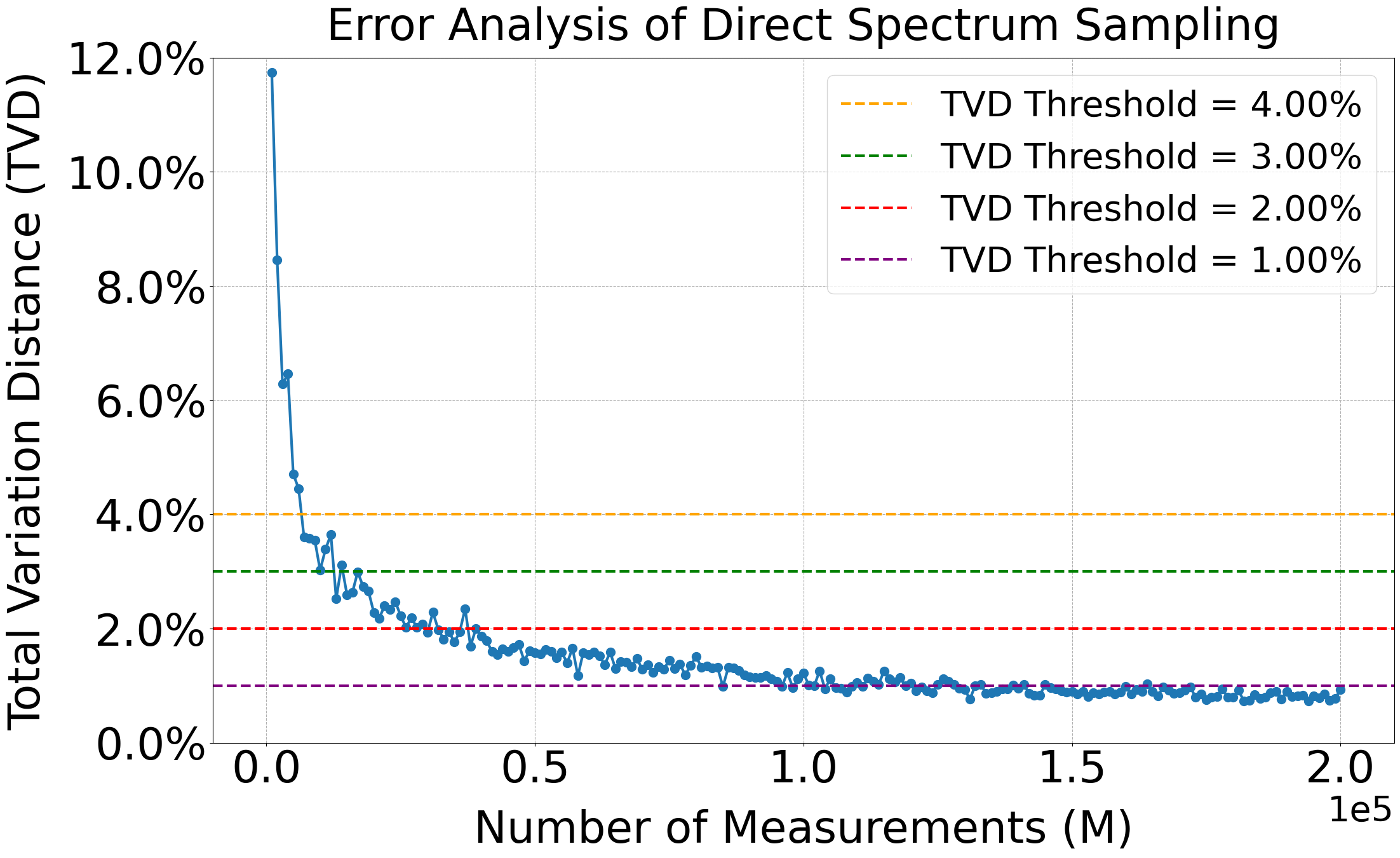}
        \caption{}
        \label{fig:direct_tvd}
    \end{subfigure}

    \vspace{0.5cm}

    \begin{subfigure}[b]{0.48\textwidth}
        \includegraphics[scale=0.15]{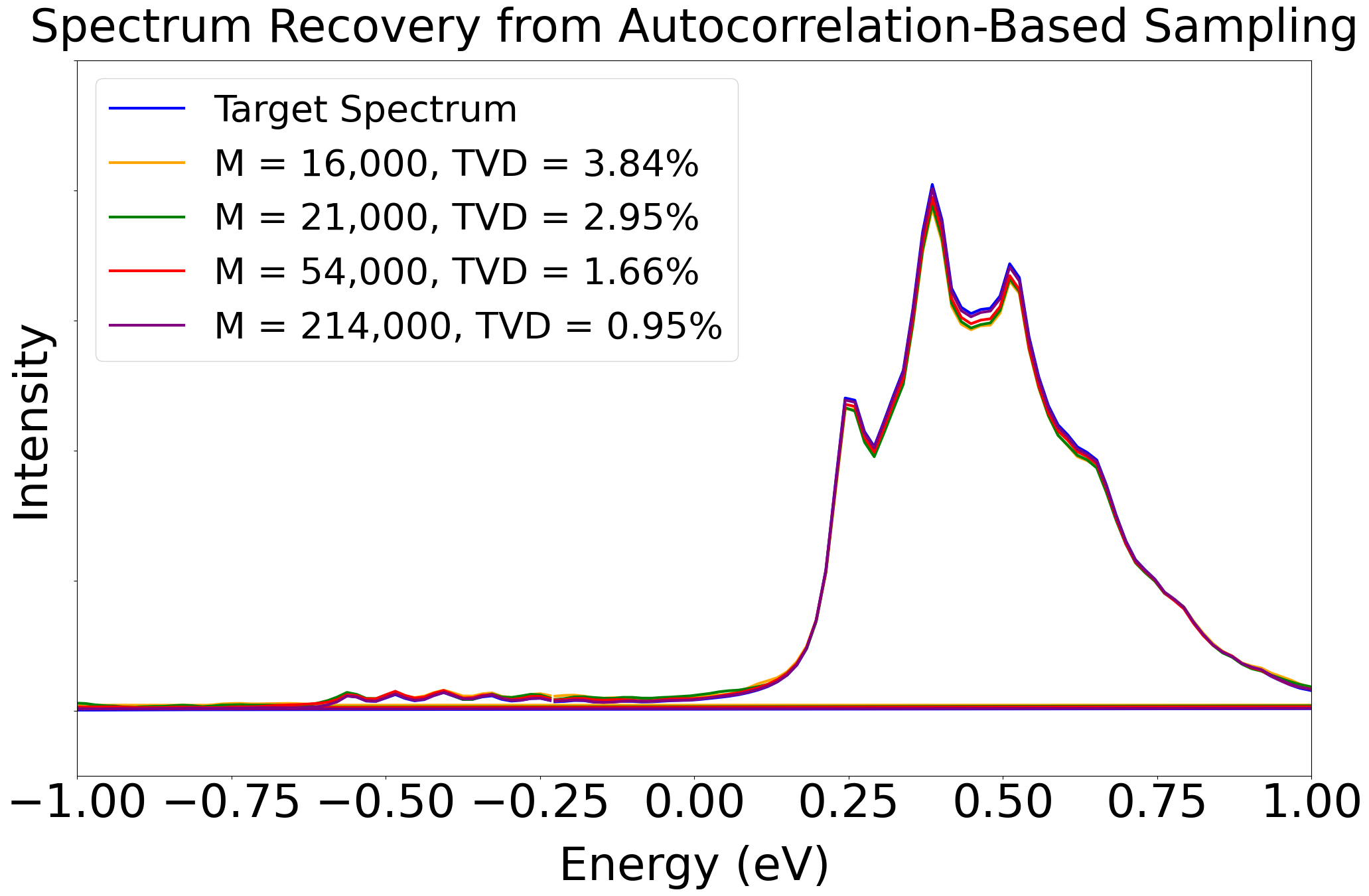}
        \caption{}
        \label{fig:autocorr_spec}
    \end{subfigure}
    \hfill
    \begin{subfigure}[b]{0.48\textwidth}
        \includegraphics[scale=0.15]{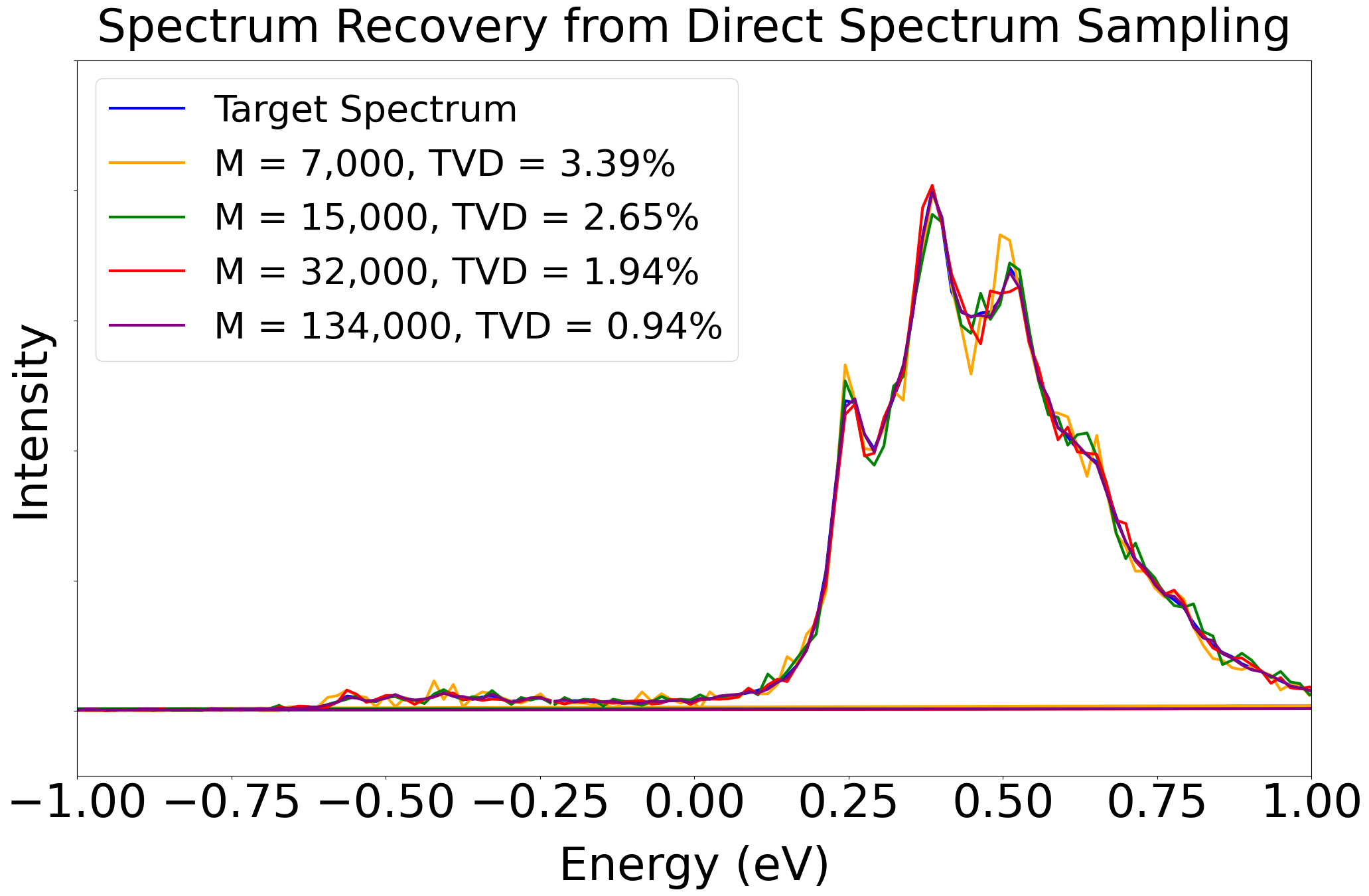}
        \caption{}
        \label{fig:direct_spec}
    \end{subfigure}

    \caption{
        \justifying Estimating measurement requirements for two different sampling strategies: left panels depict results from autocorrelation-based sampling, while the right panels correspond to direct spectrum sampling. Panels (a) and (b) plot the TVD between the simulated and target spectra as a function of the number of measurements. Panels (c) and (d) show the recovered spectra under four selected TVD thresholds (4\%, 3\%, 2\%, and 1\%). These results serve to calibrate the measurement cost required for each approach to reach practical TVD thresholds.
    }
    \label{fig:four_panel_comparison}
\end{figure*}

For both methods, we track the evolution of TVD as a function of measurement counts in Figure~\ref{fig:autocorr_tvd} and~\ref{fig:direct_tvd}, report the measurement counts required to achieve TVD thresholds of 4\%, 3\%, 2\%, and 1\% in Table~\ref{table TVD}, and visualize the recovered spectra at selected measurement counts in Figure~\ref{fig:autocorr_spec} and~\ref{fig:direct_spec}. 

The differing ranges of the horizontal axes in Figures~\ref{fig:autocorr_tvd} and~\ref{fig:direct_tvd} reveal the variations in sampling efficiencies: the direct spectrum sampling exhibits a steeper decline in TVD with increasing samples, compared to the autocorrelation-based sampling. In other words, reaching high accuracy levels requires fewer measurements for the full QPE approach.

As illustrated in Figure~\ref{fig:autocorr_spec} and~\ref{fig:direct_spec}, both methods demonstrate systematic improvement of the reconstructed spectrum as the sample size grows. Table~\ref{table TVD} lists the measurement counts at which each TVD threshold is first met and then sustained across five consecutive evaluations. This criterion ensures robust convergence, reducing the likelihood of reaching the threshold by chance fluctuation. Notably, a measurement count of 134,000 is sufficient to bring the TVD below 1.00\% in the direct spectrum sampling case, whereas the autocorrelation-based method demands 214,000 shots at each time point to reach the same threshold. 


\begin{table}[!htbp]
    \centering 
    \renewcommand{\arraystretch}{1.3}
    \begin{tabular}{|c<{\centering}|c<{\centering}|c<{\centering}|c<{\centering}|} 
        \hline
        \multicolumn{2}{|c|}{Autocorrelation Sampling} & \multicolumn{2}{|c|}{Direct Spectrum Sampling}\\
        \hline
        $M$ & TVD & $M$ & TVD \\
        \hline
        16,000  & $3.84\% $ & 7,000  & $3.39\% $ \\ 
        \hline 
        22,000  & $2.95\% $ & 15,000  & $2.65\% $ \\ 
        \hline 
        54,000  & $1.66\% $ & 32,000  & $1.94\% $\\ 
        \hline 
        214,000  & $0.95\% $ & 134,000  & $0.94\% $ \\ 
        \hline 
    \end{tabular}
    \caption{\justifying Measurement counts $M$ required for achieving TVD thresholds of 4\%, 3\%, 2\% and 1\% using autocorrelation-based sampling or direct spectrum sampling schemes.} 
    \label{table TVD} 
\end{table}

\section{Extension To Higher Dimensions}
\label{Dimension Extension}

Quantum algorithms are attractive because the exponential growth in the computational space provided by linear addition of qubits makes it possible to embed the many coupled degrees of freedom in a molecule as a large state vector on a quantum computer. We now  extend our cost evaluation to the full 24D treatment of pyrazine, quantifying both the qubit requirements and the circuit depth estimates.

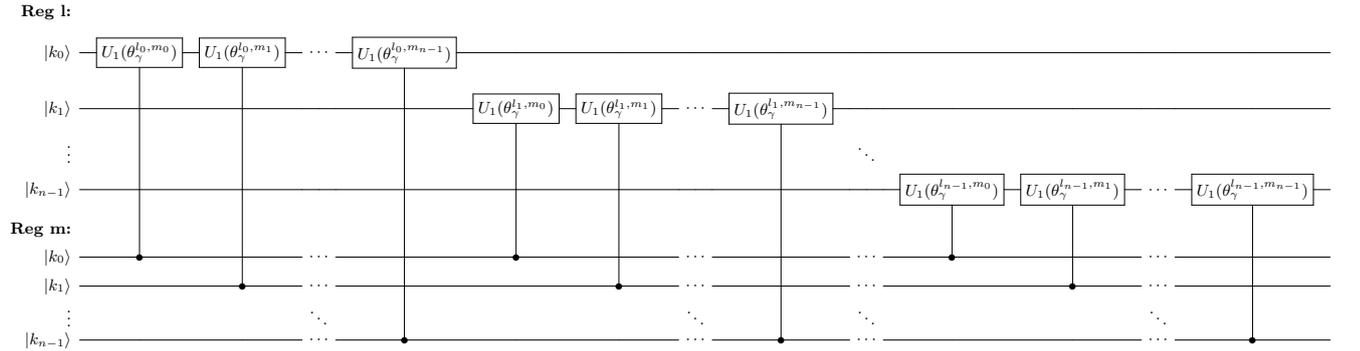
\begin{figure*}[!htbp]
\flushright
\scalebox{0.68}{
\Qcircuit @C=1em @R=1.5em {
\lstick{\textbf{Reg l:}}\\
\lstick{\ket{k_{0}}} & \gate{U_1(\theta_{\gamma}^{l_0,m_0})} &\gate{U_1(\theta_{\gamma}^{l_0,m_1})}&\qw &\cdots & & \gate{U_1(\theta_{\gamma}^{l_0,m_{n-1}})}& \qw &\qw &\qw&\qw  &\qw & \qw &\qw &\qw&\qw  &\qw & \qw&\qw &\qw&\qw &\qw&\qw \\
\lstick{\ket{k_{1}}} & \qw &\qw &\qw&\qw  &\qw & \qw & \gate{U_1(\theta_{\gamma}^{l_1,m_0})} &\gate{U_1(\theta_{\gamma}^{l_1,m_1})}&\qw &\cdots & & \gate{U_1(\theta_{\gamma}^{l_1,m_{n-1}})}&\qw&\qw&\qw &\qw&\qw  &\qw & \qw&\qw &\qw&\qw \\
\lstick{\vdots} & & &  & & &&  &  & & & & & & \ddots \\
\lstick{\ket{k_{n-1}}}   & \qw  & \qw &\qw&\qw& \qw  &\qw& \qw & \qw  & \qw &\qw&\qw& \qw  &\qw& \qw&\qw& \gate{U_1(\theta_{\gamma}^{l_{n-1},m_0})} &\gate{U_1(\theta_{\gamma}^{l_{n-1},m_1})}&\qw &\cdots & & \gate{U_1(\theta_{\gamma}^{l_{n-1},m_{n-1}})}&\qw\\
\lstick{\textbf{Reg m:}}\\
\lstick{\ket{k_{0}}} & \ctrl{-5} &\qw  &\qw&\cdots & & \qw  & \ctrl{-4} &\qw  &\qw&\cdots & & \qw & \qw  &\cdots & &\ctrl{-2} &\qw  &\qw&\cdots & & \qw&\qw \\
\lstick{\ket{k_{1}}} & \qw & \ctrl{-6}   &\qw&\cdots  &  &\qw & \qw & \ctrl{-5}   &\qw&\cdots  &  & \qw &\qw &\cdots & &\qw & \ctrl{-3}   &\qw&\cdots  &  &\qw & \qw\\
\lstick{\vdots} & & &  & \ddots &  & & & & & \ddots  & & && \ddots& & & & &  \ddots  \\
\lstick{\ket{k_{n-1}}} & \qw & \qw& \qw& \cdots&  & \ctrl{-8}& \qw & \qw& \qw& \cdots&  & \ctrl{-7}& \qw&\cdots & & \qw & \qw& \qw& \cdots&  & \ctrl{-5}& \qw
}}
\caption{\justifying Quantum circuit for $e^{-i\gamma_{l,m}Q_{l}Q_{m}dt}$, the time evolution operators of on-diagonal bilinear terms.}
\label{fig:24Dond}
\end{figure*}

\begin{figure*}[!htbp]
\flushright
\scalebox{0.68}{
\Qcircuit @C=1em @R=1.5em {
\hspace{3em}\textbf{Electronic State Reg:}\\
\lstick{\ket{q}} & \gate{R_{x}(\theta_{\mu}^{l_0,m_0})} & \gate{R_{x}(\theta_{\mu}^{l_0,m_1})}&\qw &\cdots && \gate{R_{x}(\theta_{\mu}^{l_0,m_{n-1}})}& \gate{R_{x}(\theta_{\mu}^{l_1,m_0})} & \gate{R_{x}(\theta_{\mu}^{l_1,m_1})}&\qw &\cdots && \gate{R_{x}(\theta_{\mu}^{l_1,m_{n-1}})}& \qw &\cdots&& \gate{R_{x}(\theta_{\mu}^{l_{n-1},m_0})} & \gate{R_{x}(\theta_{\mu}^{l_{n-1},m_1})}&\qw &\cdots && \gate{R_{x}(\theta_{\mu}^{l_{n-1},m_{n-1}})}&\qw \\
\lstick{\textbf{Reg l:}}\\
\lstick{\ket{k_{0}}} & \ctrl{-2} &\ctrl{-2}&\qw &\cdots & & \ctrl{-2}& \qw &\qw &\qw&\qw  &\qw & \qw &\qw &\qw&\qw  &\qw & \qw&\qw &\qw&\qw &\qw&\qw \\
\lstick{\ket{k_{1}}} & \qw &\qw &\qw&\qw  &\qw & \qw & \ctrl{-3} &\ctrl{-3}&\qw &\cdots & & \ctrl{-3}&\qw&\qw&\qw &\qw&\qw  &\qw & \qw&\qw &\qw&\qw \\
\lstick{\vdots} & & &  & & &&  &  & & & & & & \ddots \\
\lstick{\ket{k_{n-1}}}   & \qw  & \qw &\qw&\qw& \qw  &\qw& \qw & \qw  & \qw &\qw&\qw& \qw  &\qw& \qw&\qw& \ctrl{-5} &\ctrl{-5}&\qw &\cdots & & \ctrl{-5}&\qw\\
\lstick{\textbf{Reg m:}}\\
\lstick{\ket{k_{0}}} & \ctrl{-7} &\qw  &\qw&\cdots & & \qw  & \ctrl{-7} &\qw  &\qw&\cdots & & \qw & \qw  &\cdots & &\ctrl{-7} &\qw  &\qw&\cdots & & \qw&\qw \\
\lstick{\ket{k_{1}}} & \qw & \ctrl{-8}   &\qw&\cdots  &  &\qw & \qw & \ctrl{-8}   &\qw&\cdots  &  & \qw &\qw &\cdots & &\qw & \ctrl{-8}   &\qw&\cdots  &  &\qw & \qw\\
\lstick{\vdots} & & &  & \ddots &  & & & & & \ddots  & & && \ddots& & & & &  \ddots  \\
\lstick{\ket{k_{n-1}}} & \qw & \qw& \qw& \cdots&  & \ctrl{-10}& \qw & \qw& \qw& \cdots&  & \ctrl{-10}& \qw&\cdots & & \qw & \qw& \qw& \cdots&  & \ctrl{-10}& \qw
}}
\caption{\justifying Quantum circuit for $e^{-i\mu_{l,m}Q_{l}Q_{m}dt}$, the time evolution operators of off-diagonal bilinear terms.}
\label{fig:24Doffd}
\end{figure*}

The simplest 24D model treats the remaining 20 modes as an external bath represented by displaced harmonic oscillators~\cite{Raab1999,Krempl1995,Puzari2005,Thoss2000}. Since the Hamiltonian has the same functional structure as the 4D treatment in Eq.~\ref{eq:hamil}, the depth of the circuit incorporating bath-mode operations can be evaluated by substituting $d=24$ into the scaling relations in Section~\ref{Time evolution}. 
Extra costs also stem from the additional number of qubits necessary for building expanded normal mode register, which increases the qubit requirement to 97 when 4 qubits are used for each vibrational mode. 


A more complete 24D model Hamiltonian of pyrazine includes all terms up to second order:
\begin{equation}
\begin{split}
H &= 
    \begin{pmatrix}
    K+V_\text{diag}^{S1}  & 0  \\
    0 & K+V_\text{diag}^{S2}
    \end{pmatrix} +
    \begin{pmatrix}
    0 & V_\text{off}   \\
    V_\text{off}  & 0
    \end{pmatrix}
\end{split}
\label{eq:hamil24}
\end{equation}
where
\begin{align}
K =& -\displaystyle\sum_{k} \frac{\omega_{k}}{2} \frac{\partial^2}{\partial Q_{k}^{2}} \\
V_\text{diag}^{S1} =&~\Delta +
    \displaystyle\sum_{j\in G_1}
     \kappa_{j}^{(1)} Q_{j}+\frac{1}{2}\displaystyle\sum_{k} \omega_{k}Q_{k}^{2} \\& + \displaystyle\sum_{(l,m)\in G_2} \gamma_{l,m}^{(1)}Q_{l}Q_{m} \nonumber \\
V_\text{diag}^{S2} =& -\Delta +
    \displaystyle\sum_{j\in G_1}
     \kappa_{j}^{(2)} Q_{j}+\frac{1}{2} \displaystyle\sum_{k} \omega_{k}Q_{k}^{2} \\& + \displaystyle\sum_{(l,m)\in G_2} \gamma_{l,m}^{(2)}Q_{l}Q_{m} \nonumber \\
V_\text{off} =& \displaystyle\sum_{j\in G_3}
     \lambda_{j} Q_{j}+\displaystyle\sum_{(l,m)\in G_4} \mu_{l,m}Q_{l}Q_{m}.    
\end{align}
In this expression, $k$ spans 24 normal modes, all of which are represented as displaced harmonic oscillators. 
The linear on-diagonal expansion terms involve modes in the $G_1$ set characterised by $A_g$ symmetry, and pairs of modes with identical symmetry, collectively labeled as $G_2$ set. Meanwhile, off-diagonal components capture interactions from the set of $G_3={\nu_{10a}}$ with $B_{1g}$ symmetry and the $G_4$ set containing all pairs of modes whose product exhibit $B_{1g}$ symmetry, i.e., pairs formed by combinations of $A_g\times B_{1g}$, $B_{2g}\times B_{3g}$, $A_u\times B_{1u}$ and $B_{2u}\times B_{3u}$~\cite{Raab1999}. The number of modes within each symmetry class is listed in Table~\ref{tab:symmetry}. From these counts, and applying the corresponding symmetry-combination rules, we obtain the number of terms in each set to be $d_{G_1}=5$, $d_{G_2}=31$, $d_{G_3}=1$, and $d_{G_4}=29$.

\begin{table}[!htbp]
    \centering 
    \renewcommand{\arraystretch}{1.3}
    \begin{tabular}{|c<{\centering}|c<{\centering}|c<{\centering}|c<{\centering}|c<{\centering}|c<{\centering}|c<{\centering}|c<{\centering}|c<{\centering}|} 
        \hline 
         Symmetry & $A_{g}$  &$B_{1g}$  &$B_{2g}$  &$B_{3g}$  &$A_{u}$  &$B_{1u}$  &$B_{2u}$  &$B_{3u}$  \\ 
        \hline
        Number of Modes & 5 & 1& 2 & 4 & 2 & 4 & 4 & 2\\
        \hline 
    \end{tabular}
    \caption{\justifying Number of modes in different symmetry groups.} 
    \label{tab:symmetry} 
\end{table} 

To estimate the quantum gate depth we adopt the same range of time and spatial discretisation parameters determined from our 4D model of pyrazine. While this extrapolation does not account for dimension-dependent increases in Trotter error due to the growing number of Hamiltonian terms, we note that discretisation schemes with similar parameters have been successfully employed in classical full-dimensional simulations of 24D pyrazine using SOFT-based approaches with either adaptive coherent-state expansions (Matching Pursuit-SOFT)~\cite{Chen2006} or tensor-train compression techniques (Tensor Train-SOFT)~\cite{Greene2017}. Assuming that the time resolution demands do not vary significantly with dimensionality allows us to derive more generalised resource projections of the quantum implementation.

\begin{table*}[!htbp]
    \centering 
    \fontsize{8pt}{9.6pt}\selectfont
    \renewcommand{\arraystretch}{1.3}
    \begin{tabular}{|c<{\centering\arraybackslash}|c<{\centering\arraybackslash}|c<{\centering\arraybackslash}|c<{\centering\arraybackslash}|} 
        \hline 
           &Gate depth & \makecell{$n=4$, \\ $n_{t} = 512$} & \makecell{$n=5$, \\ $n_{t} = 1024$} \\ 
        \hline 
        \hline 
        State Preparation &  $N_\text{i} = 2^{n+1} - 3$  & 29  &  61 \\ 
        \hline
        \multirow{2}{*}{Register Preparation} & $N_\text{p} = n^2/2 + n$ for even $n$  & \multirow{2}{*}{12} & \multirow{2}{*}{17} \\
        &$N_\text{p} = n^2/2 + n-1/2$ for odd $n$ & & \\
        \hline
        \multirow{2}{*}{Time Evolution}    & $N_\text{t} = 2N_\text{p} + ((5d+5d_{G_4}+2d_{G_2})n^2 + (2d+2d_{G_1}+d_{G_3})n + 8)(n_t-1)$ for even $n$ & \multirow{2}{*}{$2{,}798{,}260$} & \multirow{2}{*}{$8{,}648{,}476$} \\ 
            & $N_\text{t} = 2N_\text{p} + ((5d+5d_{G_4}+2d_{G_2})n^2 + (2d+2d_{G_1}+d_{G_3})n + 8-d)(n_t-1)$ for odd $n$ &  &  \\ 
        \hline
        \multirow{2}{*}{Signal processing} & $N_\text{m} = (\log_{2}n_t)^{2}/2+\log_{2}n_t$ for even $\log_{2}n_t$  &\multirow{2}{*}{49} &\multirow{2}{*}{60}\\ 
        & $N_\text{m} = (\log_{2}n_t)^{2}/2+\log_{2}n_t-1/2$ for odd $\log_{2}n_t$ &  &  \\
        \hline
        Full algorithm A &  $N_\text{i} +N_\text{t} + 2$ & $2{,}798{,}291$ & $8{,}648{,}539$ \\
        \hline 
        Full algorithm B & $N_\text{i} +N_\text{t} + N_\text{m}$ & $2{,}798{,}338$ & $8{,}648{,}597$ \\
        \hline 
    \end{tabular}
    \caption{\justifying Cost evaluation through gate depth for 24D ($d=24$) simulations of pyrazine photodynamics in terms of the number of qubits per mode $n$ and the number of time steps $n_t$. Algorithm A is the classical post-processing approach where the autocorrelation function is obtained through measurement, and B is the fully quantum approach which directly obtains spectral signals through QPE.
    } 
    \label{table QPE24D} 
\end{table*}


In adapting the quantum circuits to the 24D model Hamiltonian, we observe that the kinetic terms as well as the constant, linear and quadratic potential terms mirror those used in the 4D simulation. However, new circuits are required to implement the polynomials for the bilinear on-diagonal elements associated with $G_2$ modes and the bilinear off-diagonal elements stemming from $G_4$ modes. The quantum circuits to handle these terms are depicted in Figure~\ref{fig:24Dond} and Figure~\ref{fig:24Doffd}. The phase angles of $\mathbf{U_1}$ gates in Figure~\ref{fig:24Dond} are defined as
$$\theta_{\gamma}^{l_i,m_j} = -\gamma_{l,m} dt 2^{l_i}2^{m_j},$$  where $l_i$ and $m_j$ denote the respective positions of the target qubit in register of normal mode $l$ and the controlling qubit in register of normal mode $m$. The $\mathbf{R_x}$ gates in Figure~\ref{fig:24Doffd} have phase angles specified by $$\theta_{\mu}^{l_i,m_j} = \mu_{l,m} dt 2^{l_i}2^{m_i},$$ 
with $l_i$ and $m_j$ indicating the controlling qubit positions in their respective $l$ and $m$ normal mode registers.
As detailed in these figures, each bilinear on-diagonal term maps to $n^2$ two-qubit gates, and each off-diagonal term necessitates $n^2$ controlled-controlled-rotation gates.



To remain consistent with gate depth analyses of the 4D model, each three-qubit gate needs to be further translated into a sequence of two-qubit operations. Following the decomposition scheme outlined in~\cite{nielsen2010}, one $\mathbf{C^2R_x}$ gate can be implemented using 5 two-qubit gates. Accordingly, each term in the $G_4$ set contributes $5n^2$ gates to the total circuit depth.
Summing over all terms in the 24D model Hamiltonian, the $U_{S1}$ and $U_{S2}$ blocks together contributes $2((d+d_{G_2})n^2+d_{G_1}n+4)$ gates, while the $U_{c}$ block adds $5d_{G_4}n^2+d_{G_3}n$.


In the 24D simulation, the time evolution operators for the potential term in the Hamiltonian is notably more computationally demanding than the kinetic part. A more efficient operator-splitting scheme thus begins with a half time step of kinetic operators, followed by a full time step of potential operators, and concludes with another half time step of kinetic operators, represented as:
\begin{equation}
    e^{-iHdt}\approx  e^{-\frac{iKdt}{2}}U_\text{QFT}e^{-iVdt}U^{-1}_\text{QFT}e^{-\frac{iKdt}{2}},
\end{equation} 
instead of the previous Eq.~\ref{eq:1}. For the 4D simulations we were able to verify that this reordering has an insignificant impact on the accuracy of the results. The gate count per time step thus includes contributions from two kinetic propagation operators, one potential propagation operator, and two QFT subroutines, giving $(5d+5d_{G_4}+2d_{G_2})n^2 + (2d+2d_{G_1}+d_{G_3})n + 8$ gates. In addition to this, another two QFT processes are required at the initial and final time points to prepare the registers in the appropriate position or momentum spaces. In Table~\ref{table QPE24D}, we report the full gate depth evaluation associated with full 24D simulations.

For runs where $n=4$ qubits are used for each normal mode register and employing $n_t = 512$ time steps, they result in circuit depths of around $2.8\times10^6$ for both statistical and canonical phase estimation approaches.
For production runs with $n=5$ and $n_t= 1024$, this increases to $8.6\times10^6$.


\section{Conclusions}

Classical simulation methods for investigating the photodynamics of molecular systems with more than a handful of degrees of freedom rely on sophisticated numerical techniques to approximate the exponential complexity of coherent quantum dynamics using a polynomial number of parameters, and quickly become limited in the size of system that can be treated reliably. Quantum simulation methods are not limited in this way. In particular, through the SO-QFT grid-based approach the exponentially large quantum state can be stored using a linear number of qubits and the state can be time-evolved according to the Hamiltonian with a polynomial number of quantum gates. The SO-QFT quantum computational approach for investigating molecular photodynamics is therefore a promising direction toward quantum advantage in scientific computing.

This work presents and assesses an end-to-end SO-QFT quantum algorithm for simulating the vibronic dynamics of photoexcited pyrazine, encompassing initial state preparation, time evolution and signal processing through measurement. Through a full resource assay, using realistic algorithmic parameters for the temporal and spatial resolutions required to obtain scientifically meaningful results, we have determined the qubit budget and gate depth that would be required for this approach once fault tolerant devices of sufficient scale become available. We chose pyrazine as a case study since it has been studied extensively and is well understood, but the algorithm extends naturally to general treatments of the photodynamics through conical intersections of more complicated molecular systems~\cite{Oliver2012}. We find that the time and space resolutions and extents required in practice for the quantum algorithm are commensurate with those used for the classical analogues. 



The number of qubits required scales linearly with the number of degrees of freedom and logarithmically with the number of time steps: $d n + 1 + \log_2 n_t$. 
The depth of the deepest quantum circuit required for the calculation scales only quadratically with the size of the maximum qubit register used to encode an individual normal mode, linearly with the number of normal modes, and linearly with the number of Trotter steps in the time evolution, i.e., asymptotically $\mathcal{O}(dn^2n_t)$ plus overhead. 
For a 24D simulation of pyrazine using a vibronic Hamiltonian complete to second order, exploratory calculations can be performed using 4 qubit normal mode registers, corresponding to 16 grid points, and 512 time steps spanning the 264~fs dynamics. This requires 106 qubits and $2.8\times10^6$ sequential gates. Production calculations with 5 qubit normal modes registers and 1024 time steps requires 131 qubits and $8.6\times10^6$ sequential gates. We find that the number of measurements required to obtain an accurate spectrum is around $2\times10^5$.

The unfavourable scaling with the number of normal modes arises partly from the global control imposed during measurement, which suppresses the natural concurrency available in SO-QFT propagation. In the absence of such control, the kinetic and QFT blocks acting on independent registers can be executed in parallel, avoiding a direct increase of circuit depth with system dimensionality. Under ancilla control, however, their internal operations must be applied sequentially, resulting in a substantial depth overhead. Our future work will therefore explore control-free alternatives~\cite{Yang2024,brien2021,polla2023} as a promising route to restore parallelism and mitigate this restriction. 


The framework and cost analyses developed in this work provide a solid foundation for quantum algorithms aimed at simulating vibronic dynamics of complex molecular systems. Although the 24D pyrazine example is tractable using the best-in-class classical algorithms for simulating quantum dynamics, the addition of just a few more atoms increases the dimensionality to above 30, which is currently out of reach. 
Because the gate depths of individual algorithmic fragments are explicitly reported, the scaling to higher-dimensional problems beyond pyrazine can be directly extrapolated by enumerating fragment occurrences in the target Hamiltonian and summing their contributions. 
The resulting cost estimates will provide lower bounds on the resources required and serve as a quantitative target for next-generation hardware. 
Meanwhile, we note that parameterization of Hamiltonians would remain a nontrivial task for quantum implementation. 
In this context, practical scalability will also depend on future advances in electronic-structure methods capable of providing accurate vibronic Hamiltonians.

\section{Author Contributions}
Xiaoning Feng performed all simulations and calculations. Xiaoning Feng and David P. Tew conceived the project, designed the research and carried out the analysis. Hans Hon Sang Chan provided supportive feedback through discussions.


\section{Conflicts of Interest}
The authors declare that they have no competing interests.

\section{Data Availability}
All data supporting the findings of this study are present within the paper. The code used in this work has been deposited on Zenodo and is openly available at \url{https://doi.org/10.5281/zenodo.17432222}.


\bibliographystyle{apsrev4-1.bst}
\bibliography{references.bib}

\begin{thebibliography}{103}%
\makeatletter
\providecommand \@ifxundefined [1]{%
 \@ifx{#1\undefined}
}%
\providecommand \@ifnum [1]{%
 \ifnum #1\expandafter \@firstoftwo
 \else \expandafter \@secondoftwo
 \fi
}%
\providecommand \@ifx [1]{%
 \ifx #1\expandafter \@firstoftwo
 \else \expandafter \@secondoftwo
 \fi
}%
\providecommand \natexlab [1]{#1}%
\providecommand \enquote  [1]{``#1''}%
\providecommand \bibnamefont  [1]{#1}%
\providecommand \bibfnamefont [1]{#1}%
\providecommand \citenamefont [1]{#1}%
\providecommand \href@noop [0]{\@secondoftwo}%
\providecommand \href [0]{\begingroup \@sanitize@url \@href}%
\providecommand \@href[1]{\@@startlink{#1}\@@href}%
\providecommand \@@href[1]{\endgroup#1\@@endlink}%
\providecommand \@sanitize@url [0]{\catcode `\\12\catcode `\$12\catcode
  `\&12\catcode `\#12\catcode `\^12\catcode `\_12\catcode `\%12\relax}%
\providecommand \@@startlink[1]{}%
\providecommand \@@endlink[0]{}%
\providecommand \url  [0]{\begingroup\@sanitize@url \@url }%
\providecommand \@url [1]{\endgroup\@href {#1}{\urlprefix }}%
\providecommand \urlprefix  [0]{URL }%
\providecommand \Eprint [0]{\href }%
\providecommand \doibase [0]{http://dx.doi.org/}%
\providecommand \selectlanguage [0]{\@gobble}%
\providecommand \bibinfo  [0]{\@secondoftwo}%
\providecommand \bibfield  [0]{\@secondoftwo}%
\providecommand \translation [1]{[#1]}%
\providecommand \BibitemOpen [0]{}%
\providecommand \bibitemStop [0]{}%
\providecommand \bibitemNoStop [0]{.\EOS\space}%
\providecommand \EOS [0]{\spacefactor3000\relax}%
\providecommand \BibitemShut  [1]{\csname bibitem#1\endcsname}%
\let\auto@bib@innerbib\@empty
\bibitem [{\citenamefont {Montanaro}(2016)}]{Montanaro2016}%
  \BibitemOpen
  \bibfield  {author} {\bibinfo {author} {\bibfnamefont {A.}~\bibnamefont
  {Montanaro}},\ }\href {\doibase 10.1038/npjqi.2015.23} {\bibfield  {journal}
  {\bibinfo  {journal} {npj Quantum Information}\ }\textbf {\bibinfo {volume}
  {2}},\ \bibinfo {pages} {15023} (\bibinfo {year} {2016})}\BibitemShut
  {NoStop}%
\bibitem [{\citenamefont {Brown}\ \emph {et~al.}(2010)\citenamefont {Brown},
  \citenamefont {Munro},\ and\ \citenamefont {Kendon}}]{brown2010}%
  \BibitemOpen
  \bibfield  {author} {\bibinfo {author} {\bibfnamefont {K.~L.}\ \bibnamefont
  {Brown}}, \bibinfo {author} {\bibfnamefont {W.~J.}\ \bibnamefont {Munro}}, \
  and\ \bibinfo {author} {\bibfnamefont {V.~M.}\ \bibnamefont {Kendon}},\
  }\href {\doibase 10.3390/e12112268} {\bibfield  {journal} {\bibinfo
  {journal} {Entropy}\ }\textbf {\bibinfo {volume} {12}},\ \bibinfo {pages}
  {2268} (\bibinfo {year} {2010})}\BibitemShut {NoStop}%
\bibitem [{\citenamefont {Zalka}(1998)}]{Zalka1998}%
  \BibitemOpen
  \bibfield  {author} {\bibinfo {author} {\bibfnamefont {C.}~\bibnamefont
  {Zalka}},\ }\href {\doibase 10.1098/rspa.1998.0162} {\bibfield  {journal}
  {\bibinfo  {journal} {Proceedings of the Royal Society of London. Series A:
  Mathematical, Physical and Engineering Sciences}\ }\textbf {\bibinfo {volume}
  {454}},\ \bibinfo {pages} {313} (\bibinfo {year} {1998})}\BibitemShut
  {NoStop}%
\bibitem [{\citenamefont {Shen}\ \emph {et~al.}(2023)\citenamefont {Shen},
  \citenamefont {Klymko}, \citenamefont {Sud}, \citenamefont {Williams-Young},
  \citenamefont {Jong},\ and\ \citenamefont {Tubman}}]{Shen2023}%
  \BibitemOpen
  \bibfield  {author} {\bibinfo {author} {\bibfnamefont {Y.}~\bibnamefont
  {Shen}}, \bibinfo {author} {\bibfnamefont {K.}~\bibnamefont {Klymko}},
  \bibinfo {author} {\bibfnamefont {J.}~\bibnamefont {Sud}}, \bibinfo {author}
  {\bibfnamefont {D.~B.}\ \bibnamefont {Williams-Young}}, \bibinfo {author}
  {\bibfnamefont {W.~A.~d.}\ \bibnamefont {Jong}}, \ and\ \bibinfo {author}
  {\bibfnamefont {N.~M.}\ \bibnamefont {Tubman}},\ }\href {\doibase
  10.22331/q-2023-07-25-1066} {\bibfield  {journal} {\bibinfo  {journal}
  {Quantum}\ }\textbf {\bibinfo {volume} {7}},\ \bibinfo {pages} {1066}
  (\bibinfo {year} {2023})}\BibitemShut {NoStop}%
\bibitem [{\citenamefont {Ollitrault}\ \emph {et~al.}(2023)\citenamefont
  {Ollitrault}, \citenamefont {Jandura}, \citenamefont {Miessen}, \citenamefont
  {Burghardt}, \citenamefont {Martinazzo}, \citenamefont {Tacchino},\ and\
  \citenamefont {Tavernelli}}]{Pauline2023}%
  \BibitemOpen
  \bibfield  {author} {\bibinfo {author} {\bibfnamefont {P.~J.}\ \bibnamefont
  {Ollitrault}}, \bibinfo {author} {\bibfnamefont {S.}~\bibnamefont {Jandura}},
  \bibinfo {author} {\bibfnamefont {A.}~\bibnamefont {Miessen}}, \bibinfo
  {author} {\bibfnamefont {I.}~\bibnamefont {Burghardt}}, \bibinfo {author}
  {\bibfnamefont {R.}~\bibnamefont {Martinazzo}}, \bibinfo {author}
  {\bibfnamefont {F.}~\bibnamefont {Tacchino}}, \ and\ \bibinfo {author}
  {\bibfnamefont {I.}~\bibnamefont {Tavernelli}},\ }\href {\doibase
  10.22331/q-2023-10-12-1139} {\bibfield  {journal} {\bibinfo  {journal}
  {Quantum}\ }\textbf {\bibinfo {volume} {7}},\ \bibinfo {pages} {1139}
  (\bibinfo {year} {2023})}\BibitemShut {NoStop}%
\bibitem [{\citenamefont {Klymko}\ \emph {et~al.}(2022)\citenamefont {Klymko},
  \citenamefont {Mejuto-Zaera}, \citenamefont {Cotton}, \citenamefont
  {Wudarski}, \citenamefont {Urbanek}, \citenamefont {Hait}, \citenamefont
  {Head-Gordon}, \citenamefont {Whaley}, \citenamefont {Moussa}, \citenamefont
  {Wiebe}, \citenamefont {de~Jong},\ and\ \citenamefont
  {Tubman}}]{Katherine2022}%
  \BibitemOpen
  \bibfield  {author} {\bibinfo {author} {\bibfnamefont {K.}~\bibnamefont
  {Klymko}}, \bibinfo {author} {\bibfnamefont {C.}~\bibnamefont
  {Mejuto-Zaera}}, \bibinfo {author} {\bibfnamefont {S.~J.}\ \bibnamefont
  {Cotton}}, \bibinfo {author} {\bibfnamefont {F.}~\bibnamefont {Wudarski}},
  \bibinfo {author} {\bibfnamefont {M.}~\bibnamefont {Urbanek}}, \bibinfo
  {author} {\bibfnamefont {D.}~\bibnamefont {Hait}}, \bibinfo {author}
  {\bibfnamefont {M.}~\bibnamefont {Head-Gordon}}, \bibinfo {author}
  {\bibfnamefont {K.~B.}\ \bibnamefont {Whaley}}, \bibinfo {author}
  {\bibfnamefont {J.}~\bibnamefont {Moussa}}, \bibinfo {author} {\bibfnamefont
  {N.}~\bibnamefont {Wiebe}}, \bibinfo {author} {\bibfnamefont {W.~A.}\
  \bibnamefont {de~Jong}}, \ and\ \bibinfo {author} {\bibfnamefont {N.~M.}\
  \bibnamefont {Tubman}},\ }\href {\doibase 10.1103/PRXQuantum.3.020323}
  {\bibfield  {journal} {\bibinfo  {journal} {PRX Quantum}\ }\textbf {\bibinfo
  {volume} {3}},\ \bibinfo {pages} {020323} (\bibinfo {year}
  {2022})}\BibitemShut {NoStop}%
\bibitem [{\citenamefont {Nishi}\ \emph {et~al.}(2023)\citenamefont {Nishi},
  \citenamefont {Hamada}, \citenamefont {Nishiya}, \citenamefont {Kosugi},\
  and\ \citenamefont {Matsushita}}]{Nishi2023}%
  \BibitemOpen
  \bibfield  {author} {\bibinfo {author} {\bibfnamefont {H.}~\bibnamefont
  {Nishi}}, \bibinfo {author} {\bibfnamefont {K.}~\bibnamefont {Hamada}},
  \bibinfo {author} {\bibfnamefont {Y.}~\bibnamefont {Nishiya}}, \bibinfo
  {author} {\bibfnamefont {T.}~\bibnamefont {Kosugi}}, \ and\ \bibinfo {author}
  {\bibfnamefont {Y.-i.}\ \bibnamefont {Matsushita}},\ }\href
  {http://dx.doi.org/10.1103/PhysRevResearch.5.043048} {\bibfield  {journal}
  {\bibinfo  {journal} {Physical Review Research}\ }\textbf {\bibinfo {volume}
  {5}} (\bibinfo {year} {2023})}\BibitemShut {NoStop}%
\bibitem [{\citenamefont {Barison}\ \emph {et~al.}(2021)\citenamefont
  {Barison}, \citenamefont {Vicentini},\ and\ \citenamefont
  {Carleo}}]{Barison2021}%
  \BibitemOpen
  \bibfield  {author} {\bibinfo {author} {\bibfnamefont {S.}~\bibnamefont
  {Barison}}, \bibinfo {author} {\bibfnamefont {F.}~\bibnamefont {Vicentini}},
  \ and\ \bibinfo {author} {\bibfnamefont {G.}~\bibnamefont {Carleo}},\ }\href
  {\doibase 10.22331/q-2021-07-28-512} {\bibfield  {journal} {\bibinfo
  {journal} {Quantum}\ }\textbf {\bibinfo {volume} {5}},\ \bibinfo {pages}
  {512} (\bibinfo {year} {2021})}\BibitemShut {NoStop}%
\bibitem [{\citenamefont {Abrams}\ and\ \citenamefont
  {Lloyd}(1997)}]{Abrams1997}%
  \BibitemOpen
  \bibfield  {author} {\bibinfo {author} {\bibfnamefont {D.~S.}\ \bibnamefont
  {Abrams}}\ and\ \bibinfo {author} {\bibfnamefont {S.}~\bibnamefont {Lloyd}},\
  }\href {\doibase 10.1103/physrevlett.79.2586} {\bibfield  {journal} {\bibinfo
   {journal} {Physical Review Letters}\ }\textbf {\bibinfo {volume} {79}},\
  \bibinfo {pages} {2586} (\bibinfo {year} {1997})}\BibitemShut {NoStop}%
\bibitem [{\citenamefont {Hastings}\ \emph {et~al.}(2014)\citenamefont
  {Hastings}, \citenamefont {Wecker}, \citenamefont {Bauer},\ and\
  \citenamefont {Troyer}}]{hastings2014}%
  \BibitemOpen
  \bibfield  {author} {\bibinfo {author} {\bibfnamefont {M.~B.}\ \bibnamefont
  {Hastings}}, \bibinfo {author} {\bibfnamefont {D.}~\bibnamefont {Wecker}},
  \bibinfo {author} {\bibfnamefont {B.}~\bibnamefont {Bauer}}, \ and\ \bibinfo
  {author} {\bibfnamefont {M.}~\bibnamefont {Troyer}},\ }\href
  {https://arxiv.org/abs/1403.1539} {\enquote {\bibinfo {title} {Improving
  quantum algorithms for quantum chemistry},}\ } (\bibinfo {year}
  {2014})\BibitemShut {NoStop}%
\bibitem [{\citenamefont {Jordan}\ \emph {et~al.}(2012)\citenamefont {Jordan},
  \citenamefont {Lee},\ and\ \citenamefont {Preskill}}]{Jordan2012}%
  \BibitemOpen
  \bibfield  {author} {\bibinfo {author} {\bibfnamefont {S.~P.}\ \bibnamefont
  {Jordan}}, \bibinfo {author} {\bibfnamefont {K.~S.~M.}\ \bibnamefont {Lee}},
  \ and\ \bibinfo {author} {\bibfnamefont {J.}~\bibnamefont {Preskill}},\
  }\href {\doibase 10.1126/science.1217069} {\bibfield  {journal} {\bibinfo
  {journal} {Science}\ }\textbf {\bibinfo {volume} {336}},\ \bibinfo {pages}
  {1130} (\bibinfo {year} {2012})}\BibitemShut {NoStop}%
\bibitem [{\citenamefont {Georgescu}\ \emph {et~al.}(2014)\citenamefont
  {Georgescu}, \citenamefont {Ashhab},\ and\ \citenamefont
  {Nori}}]{george2014}%
  \BibitemOpen
  \bibfield  {author} {\bibinfo {author} {\bibfnamefont {I.~M.}\ \bibnamefont
  {Georgescu}}, \bibinfo {author} {\bibfnamefont {S.}~\bibnamefont {Ashhab}}, \
  and\ \bibinfo {author} {\bibfnamefont {F.}~\bibnamefont {Nori}},\ }\href
  {\doibase 10.1103/RevModPhys.86.153} {\bibfield  {journal} {\bibinfo
  {journal} {Reviews of Modern Physics}\ }\textbf {\bibinfo {volume} {86}},\
  \bibinfo {pages} {153} (\bibinfo {year} {2014})}\BibitemShut {NoStop}%
\bibitem [{\citenamefont {Berry}\ \emph {et~al.}(2015)\citenamefont {Berry},
  \citenamefont {Childs},\ and\ \citenamefont {Kothari}}]{Berry_2015}%
  \BibitemOpen
  \bibfield  {author} {\bibinfo {author} {\bibfnamefont {D.~W.}\ \bibnamefont
  {Berry}}, \bibinfo {author} {\bibfnamefont {A.~M.}\ \bibnamefont {Childs}}, \
  and\ \bibinfo {author} {\bibfnamefont {R.}~\bibnamefont {Kothari}},\ }in\
  \href {\doibase 10.1109/focs.2015.54} {\emph {\bibinfo {booktitle} {2015 IEEE
  56th Annual Symposium on Foundations of Computer Science}}}\ (\bibinfo
  {publisher} {IEEE},\ \bibinfo {year} {2015})\ p.\ \bibinfo {pages}
  {792–809}\BibitemShut {NoStop}%
\bibitem [{\citenamefont {Zhang}\ \emph {et~al.}(2022)\citenamefont {Zhang},
  \citenamefont {Li},\ and\ \citenamefont {Yuan}}]{Zhang2022}%
  \BibitemOpen
  \bibfield  {author} {\bibinfo {author} {\bibfnamefont {X.-M.}\ \bibnamefont
  {Zhang}}, \bibinfo {author} {\bibfnamefont {T.}~\bibnamefont {Li}}, \ and\
  \bibinfo {author} {\bibfnamefont {X.}~\bibnamefont {Yuan}},\ }\href
  {http://dx.doi.org/10.1103/PhysRevLett.129.230504} {\bibfield  {journal}
  {\bibinfo  {journal} {Physical Review Letters}\ }\textbf {\bibinfo {volume}
  {129}} (\bibinfo {year} {2022})}\BibitemShut {NoStop}%
\bibitem [{\citenamefont {Kökcü}\ \emph {et~al.}(2022)\citenamefont
  {Kökcü}, \citenamefont {Camps}, \citenamefont {Bassman~Oftelie},
  \citenamefont {Freericks}, \citenamefont {de~Jong}, \citenamefont
  {Van~Beeumen},\ and\ \citenamefont {Kemper}}]{Kokcu2022}%
  \BibitemOpen
  \bibfield  {author} {\bibinfo {author} {\bibfnamefont {E.}~\bibnamefont
  {Kökcü}}, \bibinfo {author} {\bibfnamefont {D.}~\bibnamefont {Camps}},
  \bibinfo {author} {\bibfnamefont {L.}~\bibnamefont {Bassman~Oftelie}},
  \bibinfo {author} {\bibfnamefont {J.~K.}\ \bibnamefont {Freericks}}, \bibinfo
  {author} {\bibfnamefont {W.~A.}\ \bibnamefont {de~Jong}}, \bibinfo {author}
  {\bibfnamefont {R.}~\bibnamefont {Van~Beeumen}}, \ and\ \bibinfo {author}
  {\bibfnamefont {A.~F.}\ \bibnamefont {Kemper}},\ }\href
  {http://dx.doi.org/10.1103/PhysRevA.105.032420} {\bibfield  {journal}
  {\bibinfo  {journal} {Physical Review A}\ }\textbf {\bibinfo {volume} {105}}
  (\bibinfo {year} {2022})}\BibitemShut {NoStop}%
\bibitem [{\citenamefont {Wada}\ \emph {et~al.}(2022)\citenamefont {Wada},
  \citenamefont {Raymond}, \citenamefont {Ohnishi}, \citenamefont {Kaminishi},
  \citenamefont {Sugawara}, \citenamefont {Yamamoto},\ and\ \citenamefont
  {Watanabe}}]{Wada2022}%
  \BibitemOpen
  \bibfield  {author} {\bibinfo {author} {\bibfnamefont {K.}~\bibnamefont
  {Wada}}, \bibinfo {author} {\bibfnamefont {R.}~\bibnamefont {Raymond}},
  \bibinfo {author} {\bibfnamefont {Y.-y.}\ \bibnamefont {Ohnishi}}, \bibinfo
  {author} {\bibfnamefont {E.}~\bibnamefont {Kaminishi}}, \bibinfo {author}
  {\bibfnamefont {M.}~\bibnamefont {Sugawara}}, \bibinfo {author}
  {\bibfnamefont {N.}~\bibnamefont {Yamamoto}}, \ and\ \bibinfo {author}
  {\bibfnamefont {H.~C.}\ \bibnamefont {Watanabe}},\ }\href
  {http://dx.doi.org/10.1103/PhysRevA.105.062421} {\bibfield  {journal}
  {\bibinfo  {journal} {Physical Review A}\ }\textbf {\bibinfo {volume} {105}}
  (\bibinfo {year} {2022})}\BibitemShut {NoStop}%
\bibitem [{\citenamefont {Chan}\ \emph {et~al.}(2023)\citenamefont {Chan},
  \citenamefont {Meister}, \citenamefont {Jones}, \citenamefont {Tew},\ and\
  \citenamefont {Benjamin}}]{Chan2023}%
  \BibitemOpen
  \bibfield  {author} {\bibinfo {author} {\bibfnamefont {H.~H.~S.}\
  \bibnamefont {Chan}}, \bibinfo {author} {\bibfnamefont {R.}~\bibnamefont
  {Meister}}, \bibinfo {author} {\bibfnamefont {T.}~\bibnamefont {Jones}},
  \bibinfo {author} {\bibfnamefont {D.~P.}\ \bibnamefont {Tew}}, \ and\
  \bibinfo {author} {\bibfnamefont {S.~C.}\ \bibnamefont {Benjamin}},\ }\href
  {\doibase 10.1126/sciadv.abo7484} {\bibfield  {journal} {\bibinfo  {journal}
  {Science Advances}\ }\textbf {\bibinfo {volume} {9}} (\bibinfo {year}
  {2023}),\ 10.1126/sciadv.abo7484}\BibitemShut {NoStop}%
\bibitem [{\citenamefont {Astrakhantsev}\ \emph {et~al.}(2023)\citenamefont
  {Astrakhantsev}, \citenamefont {Lin}, \citenamefont {Pollmann},\ and\
  \citenamefont {Smith}}]{Astrakhantsev2023}%
  \BibitemOpen
  \bibfield  {author} {\bibinfo {author} {\bibfnamefont {N.}~\bibnamefont
  {Astrakhantsev}}, \bibinfo {author} {\bibfnamefont {S.-H.}\ \bibnamefont
  {Lin}}, \bibinfo {author} {\bibfnamefont {F.}~\bibnamefont {Pollmann}}, \
  and\ \bibinfo {author} {\bibfnamefont {A.}~\bibnamefont {Smith}},\ }\href
  {http://dx.doi.org/10.1103/PhysRevResearch.5.033187} {\bibfield  {journal}
  {\bibinfo  {journal} {Physical Review Research}\ }\textbf {\bibinfo {volume}
  {5}} (\bibinfo {year} {2023})}\BibitemShut {NoStop}%
\bibitem [{\citenamefont {Tepaske}\ \emph {et~al.}(2023)\citenamefont
  {Tepaske}, \citenamefont {Hahn},\ and\ \citenamefont {Luitz}}]{Tepaske2023}%
  \BibitemOpen
  \bibfield  {author} {\bibinfo {author} {\bibfnamefont {M.~S.~J.}\
  \bibnamefont {Tepaske}}, \bibinfo {author} {\bibfnamefont {D.}~\bibnamefont
  {Hahn}}, \ and\ \bibinfo {author} {\bibfnamefont {D.~J.}\ \bibnamefont
  {Luitz}},\ }\href {http://dx.doi.org/10.21468/SciPostPhys.14.4.073}
  {\bibfield  {journal} {\bibinfo  {journal} {SciPost Physics}\ }\textbf
  {\bibinfo {volume} {14}} (\bibinfo {year} {2023})}\BibitemShut {NoStop}%
\bibitem [{\citenamefont {Haah}\ \emph {et~al.}(2023)\citenamefont {Haah},
  \citenamefont {Hastings}, \citenamefont {Kothari},\ and\ \citenamefont
  {Low}}]{Haah2023}%
  \BibitemOpen
  \bibfield  {author} {\bibinfo {author} {\bibfnamefont {J.}~\bibnamefont
  {Haah}}, \bibinfo {author} {\bibfnamefont {M.~B.}\ \bibnamefont {Hastings}},
  \bibinfo {author} {\bibfnamefont {R.}~\bibnamefont {Kothari}}, \ and\
  \bibinfo {author} {\bibfnamefont {G.~H.}\ \bibnamefont {Low}},\ }\href
  {\doibase 10.1137/18M1231511} {\bibfield  {journal} {\bibinfo  {journal}
  {SIAM Journal on Computing}\ }\textbf {\bibinfo {volume} {52}},\ \bibinfo
  {pages} {FOCS18} (\bibinfo {year} {2023})}\BibitemShut {NoStop}%
\bibitem [{\citenamefont {Yuan}\ and\ \citenamefont {Zhang}(2023)}]{Yuan2023}%
  \BibitemOpen
  \bibfield  {author} {\bibinfo {author} {\bibfnamefont {P.}~\bibnamefont
  {Yuan}}\ and\ \bibinfo {author} {\bibfnamefont {S.}~\bibnamefont {Zhang}},\
  }\href {\doibase 10.22331/q-2023-03-20-956} {\bibfield  {journal} {\bibinfo
  {journal} {Quantum}\ }\textbf {\bibinfo {volume} {7}},\ \bibinfo {pages}
  {956} (\bibinfo {year} {2023})}\BibitemShut {NoStop}%
\bibitem [{\citenamefont {Breda}\ \emph {et~al.}(2006)\citenamefont {Breda},
  \citenamefont {Reva}, \citenamefont {Lapinski}, \citenamefont {Nowak},\ and\
  \citenamefont {Fausto}}]{BREDA2006}%
  \BibitemOpen
  \bibfield  {author} {\bibinfo {author} {\bibfnamefont {S.}~\bibnamefont
  {Breda}}, \bibinfo {author} {\bibfnamefont {I.}~\bibnamefont {Reva}},
  \bibinfo {author} {\bibfnamefont {L.}~\bibnamefont {Lapinski}}, \bibinfo
  {author} {\bibfnamefont {M.}~\bibnamefont {Nowak}}, \ and\ \bibinfo {author}
  {\bibfnamefont {R.}~\bibnamefont {Fausto}},\ }\href {\doibase
  https://doi.org/10.1016/j.molstruc.2005.09.010} {\bibfield  {journal}
  {\bibinfo  {journal} {Journal of Molecular Structure}\ }\textbf {\bibinfo
  {volume} {786}},\ \bibinfo {pages} {193} (\bibinfo {year}
  {2006})}\BibitemShut {NoStop}%
\bibitem [{\citenamefont {Woywod}\ \emph {et~al.}(1994)\citenamefont {Woywod},
  \citenamefont {Domcke}, \citenamefont {Sobolewski},\ and\ \citenamefont
  {Werner}}]{Woywod1994}%
  \BibitemOpen
  \bibfield  {author} {\bibinfo {author} {\bibfnamefont {C.}~\bibnamefont
  {Woywod}}, \bibinfo {author} {\bibfnamefont {W.}~\bibnamefont {Domcke}},
  \bibinfo {author} {\bibfnamefont {A.~L.}\ \bibnamefont {Sobolewski}}, \ and\
  \bibinfo {author} {\bibfnamefont {H.}~\bibnamefont {Werner}},\ }\href
  {\doibase 10.1063/1.466618} {\bibfield  {journal} {\bibinfo  {journal} {The
  Journal of Chemical Physics}\ }\textbf {\bibinfo {volume} {100}},\ \bibinfo
  {pages} {1400} (\bibinfo {year} {1994})}\BibitemShut {NoStop}%
\bibitem [{\citenamefont {He}\ \emph {et~al.}(2009)\citenamefont {He},
  \citenamefont {Zhu}, \citenamefont {Chin},\ and\ \citenamefont
  {Lin}}]{HE2009}%
  \BibitemOpen
  \bibfield  {author} {\bibinfo {author} {\bibfnamefont {R.}~\bibnamefont
  {He}}, \bibinfo {author} {\bibfnamefont {C.}~\bibnamefont {Zhu}}, \bibinfo
  {author} {\bibfnamefont {C.-H.}\ \bibnamefont {Chin}}, \ and\ \bibinfo
  {author} {\bibfnamefont {S.~H.}\ \bibnamefont {Lin}},\ }\href {\doibase
  10.1016/j.cplett.2009.05.043} {\bibfield  {journal} {\bibinfo  {journal}
  {Chemical Physics Letters}\ }\textbf {\bibinfo {volume} {476}},\ \bibinfo
  {pages} {19} (\bibinfo {year} {2009})}\BibitemShut {NoStop}%
\bibitem [{\citenamefont {Krempl}\ \emph {et~al.}(1994)\citenamefont {Krempl},
  \citenamefont {Winterstetter}, \citenamefont {Plöhn},\ and\ \citenamefont
  {Domcke}}]{Krempl1994}%
  \BibitemOpen
  \bibfield  {author} {\bibinfo {author} {\bibfnamefont {S.}~\bibnamefont
  {Krempl}}, \bibinfo {author} {\bibfnamefont {M.}~\bibnamefont
  {Winterstetter}}, \bibinfo {author} {\bibfnamefont {H.}~\bibnamefont
  {Plöhn}}, \ and\ \bibinfo {author} {\bibfnamefont {W.}~\bibnamefont
  {Domcke}},\ }\href {\doibase 10.1063/1.467253} {\bibfield  {journal}
  {\bibinfo  {journal} {The Journal of Chemical Physics}\ }\textbf {\bibinfo
  {volume} {100}},\ \bibinfo {pages} {926} (\bibinfo {year}
  {1994})}\BibitemShut {NoStop}%
\bibitem [{\citenamefont {Krempl}\ \emph {et~al.}(1995)\citenamefont {Krempl},
  \citenamefont {Winterstetter},\ and\ \citenamefont {Domcke}}]{Krempl1995}%
  \BibitemOpen
  \bibfield  {author} {\bibinfo {author} {\bibfnamefont {S.}~\bibnamefont
  {Krempl}}, \bibinfo {author} {\bibfnamefont {M.}~\bibnamefont
  {Winterstetter}}, \ and\ \bibinfo {author} {\bibfnamefont {W.}~\bibnamefont
  {Domcke}},\ }\href {\doibase 10.1063/1.469364} {\bibfield  {journal}
  {\bibinfo  {journal} {The Journal of Chemical Physics}\ }\textbf {\bibinfo
  {volume} {102}},\ \bibinfo {pages} {6499} (\bibinfo {year}
  {1995})}\BibitemShut {NoStop}%
\bibitem [{\citenamefont {Worth}\ \emph {et~al.}(1996)\citenamefont {Worth},
  \citenamefont {Meyer},\ and\ \citenamefont {Cederbaum}}]{Worth1996}%
  \BibitemOpen
  \bibfield  {author} {\bibinfo {author} {\bibfnamefont {G.~A.}\ \bibnamefont
  {Worth}}, \bibinfo {author} {\bibfnamefont {H.}~\bibnamefont {Meyer}}, \ and\
  \bibinfo {author} {\bibfnamefont {L.~S.}\ \bibnamefont {Cederbaum}},\ }\href
  {\doibase 10.1063/1.472327} {\bibfield  {journal} {\bibinfo  {journal} {The
  Journal of Chemical Physics}\ }\textbf {\bibinfo {volume} {105}},\ \bibinfo
  {pages} {4412} (\bibinfo {year} {1996})}\BibitemShut {NoStop}%
\bibitem [{\citenamefont {Raab}\ \emph {et~al.}(1999)\citenamefont {Raab},
  \citenamefont {Worth}, \citenamefont {Meyer},\ and\ \citenamefont
  {Cederbaum}}]{Raab1999}%
  \BibitemOpen
  \bibfield  {author} {\bibinfo {author} {\bibfnamefont {A.}~\bibnamefont
  {Raab}}, \bibinfo {author} {\bibfnamefont {G.~A.}\ \bibnamefont {Worth}},
  \bibinfo {author} {\bibfnamefont {H.~D.}\ \bibnamefont {Meyer}}, \ and\
  \bibinfo {author} {\bibfnamefont {L.~S.}\ \bibnamefont {Cederbaum}},\ }\href
  {\doibase 10.1063/1.478061} {\bibfield  {journal} {\bibinfo  {journal} {The
  Journal of Chemical Physics}\ }\textbf {\bibinfo {volume} {110}},\ \bibinfo
  {pages} {936} (\bibinfo {year} {1999})}\BibitemShut {NoStop}%
\bibitem [{\citenamefont {Sala}\ \emph {et~al.}(2014)\citenamefont {Sala},
  \citenamefont {Saab}, \citenamefont {Lasorne}, \citenamefont {Gatti},\ and\
  \citenamefont {Guérin}}]{Sala2014}%
  \BibitemOpen
  \bibfield  {author} {\bibinfo {author} {\bibfnamefont {M.}~\bibnamefont
  {Sala}}, \bibinfo {author} {\bibfnamefont {M.}~\bibnamefont {Saab}}, \bibinfo
  {author} {\bibfnamefont {B.}~\bibnamefont {Lasorne}}, \bibinfo {author}
  {\bibfnamefont {F.}~\bibnamefont {Gatti}}, \ and\ \bibinfo {author}
  {\bibfnamefont {S.}~\bibnamefont {Guérin}},\ }\href {\doibase
  10.1063/1.4875736} {\bibfield  {journal} {\bibinfo  {journal} {The Journal of
  Chemical Physics}\ }\textbf {\bibinfo {volume} {140}},\ \bibinfo {pages}
  {194309} (\bibinfo {year} {2014})}\BibitemShut {NoStop}%
\bibitem [{\citenamefont {Thoss}\ \emph {et~al.}(2000)\citenamefont {Thoss},
  \citenamefont {Miller},\ and\ \citenamefont {Stock}}]{Thoss2000}%
  \BibitemOpen
  \bibfield  {author} {\bibinfo {author} {\bibfnamefont {M.}~\bibnamefont
  {Thoss}}, \bibinfo {author} {\bibfnamefont {W.~H.}\ \bibnamefont {Miller}}, \
  and\ \bibinfo {author} {\bibfnamefont {G.}~\bibnamefont {Stock}},\ }\href
  {\doibase 10.1063/1.481668} {\bibfield  {journal} {\bibinfo  {journal} {The
  Journal of Chemical Physics}\ }\textbf {\bibinfo {volume} {112}},\ \bibinfo
  {pages} {10282} (\bibinfo {year} {2000})}\BibitemShut {NoStop}%
\bibitem [{\citenamefont {Puzari}\ \emph {et~al.}(2005)\citenamefont {Puzari},
  \citenamefont {Swathi}, \citenamefont {Sarkar},\ and\ \citenamefont
  {Adhikari}}]{Puzari2005}%
  \BibitemOpen
  \bibfield  {author} {\bibinfo {author} {\bibfnamefont {P.}~\bibnamefont
  {Puzari}}, \bibinfo {author} {\bibfnamefont {R.~S.}\ \bibnamefont {Swathi}},
  \bibinfo {author} {\bibfnamefont {B.}~\bibnamefont {Sarkar}}, \ and\ \bibinfo
  {author} {\bibfnamefont {S.}~\bibnamefont {Adhikari}},\ }\href {\doibase
  10.1063/1.2050647} {\bibfield  {journal} {\bibinfo  {journal} {The Journal of
  Chemical Physics}\ }\textbf {\bibinfo {volume} {123}},\ \bibinfo {pages}
  {134317} (\bibinfo {year} {2005})}\BibitemShut {NoStop}%
\bibitem [{\citenamefont {Chen}\ and\ \citenamefont
  {Batista}(2006)}]{Chen2006}%
  \BibitemOpen
  \bibfield  {author} {\bibinfo {author} {\bibfnamefont {X.}~\bibnamefont
  {Chen}}\ and\ \bibinfo {author} {\bibfnamefont {V.~S.}\ \bibnamefont
  {Batista}},\ }\href {https://doi.org/10.1063/1.2356477} {\bibfield  {journal}
  {\bibinfo  {journal} {The Journal of Chemical Physics}\ }\textbf {\bibinfo
  {volume} {125}} (\bibinfo {year} {2006})}\BibitemShut {NoStop}%
\bibitem [{\citenamefont {Saller}\ and\ \citenamefont
  {Habershon}(2015)}]{Saller2015}%
  \BibitemOpen
  \bibfield  {author} {\bibinfo {author} {\bibfnamefont {M.~A.~C.}\
  \bibnamefont {Saller}}\ and\ \bibinfo {author} {\bibfnamefont
  {S.}~\bibnamefont {Habershon}},\ }\href {\doibase 10.1021/ct500657f}
  {\bibfield  {journal} {\bibinfo  {journal} {Journal of Chemical Theory and
  Computation}\ }\textbf {\bibinfo {volume} {11}},\ \bibinfo {pages} {8}
  (\bibinfo {year} {2015})}\BibitemShut {NoStop}%
\bibitem [{\citenamefont {Saller}\ and\ \citenamefont
  {Habershon}(2017)}]{Saller2017}%
  \BibitemOpen
  \bibfield  {author} {\bibinfo {author} {\bibfnamefont {M.~A.~C.}\
  \bibnamefont {Saller}}\ and\ \bibinfo {author} {\bibfnamefont
  {S.}~\bibnamefont {Habershon}},\ }\href {\doibase 10.1021/acs.jctc.7b00021}
  {\bibfield  {journal} {\bibinfo  {journal} {Journal of Chemical Theory and
  Computation}\ }\textbf {\bibinfo {volume} {13}},\ \bibinfo {pages} {3085}
  (\bibinfo {year} {2017})}\BibitemShut {NoStop}%
\bibitem [{\citenamefont {Shor}(1994)}]{Shor1994}%
  \BibitemOpen
  \bibfield  {author} {\bibinfo {author} {\bibfnamefont {P.}~\bibnamefont
  {Shor}},\ }in\ \href {\doibase 10.1109/SFCS.1994.365700} {\emph {\bibinfo
  {booktitle} {Proceedings 35th Annual Symposium on Foundations of Computer
  Science}}}\ (\bibinfo {year} {1994})\ pp.\ \bibinfo {pages}
  {124--134}\BibitemShut {NoStop}%
\bibitem [{\citenamefont {Oliveira}\ \emph {et~al.}(2007)\citenamefont
  {Oliveira}, \citenamefont {Bonagamba}, \citenamefont {Sarthour},
  \citenamefont {Freitas},\ and\ \citenamefont {deAzevedo}}]{OLIVEIRA2007}%
  \BibitemOpen
  \bibfield  {author} {\bibinfo {author} {\bibfnamefont {I.~S.}\ \bibnamefont
  {Oliveira}}, \bibinfo {author} {\bibfnamefont {T.~J.}\ \bibnamefont
  {Bonagamba}}, \bibinfo {author} {\bibfnamefont {R.~S.}\ \bibnamefont
  {Sarthour}}, \bibinfo {author} {\bibfnamefont {J.~C.}\ \bibnamefont
  {Freitas}}, \ and\ \bibinfo {author} {\bibfnamefont {E.~R.}\ \bibnamefont
  {deAzevedo}},\ }in\ \href {\doibase
  https://doi.org/10.1016/B978-044452782-0/50007-5} {\emph {\bibinfo
  {booktitle} {NMR Quantum Information Processing}}}\ (\bibinfo  {publisher}
  {Elsevier Science B.V.},\ \bibinfo {address} {Amsterdam},\ \bibinfo {year}
  {2007})\ pp.\ \bibinfo {pages} {183--205}\BibitemShut {NoStop}%
\bibitem [{\citenamefont {Band}\ and\ \citenamefont
  {Avishai}(2013)}]{BAND2013}%
  \BibitemOpen
  \bibfield  {author} {\bibinfo {author} {\bibfnamefont {Y.~B.}\ \bibnamefont
  {Band}}\ and\ \bibinfo {author} {\bibfnamefont {Y.}~\bibnamefont {Avishai}},\
  }in\ \href {\doibase https://doi.org/10.1016/B978-0-444-53786-7.00005-8}
  {\emph {\bibinfo {booktitle} {Quantum Mechanics with Applications to
  Nanotechnology and Information Science}}}\ (\bibinfo  {publisher} {Academic
  Press},\ \bibinfo {address} {Amsterdam},\ \bibinfo {year} {2013})\ pp.\
  \bibinfo {pages} {193--258}\BibitemShut {NoStop}%
\bibitem [{\citenamefont {Trotter}(1959)}]{Trotter1959}%
  \BibitemOpen
  \bibfield  {author} {\bibinfo {author} {\bibfnamefont {H.~F.}\ \bibnamefont
  {Trotter}},\ }\href {http://www.jstor.org/stable/2033649} {\bibfield
  {journal} {\bibinfo  {journal} {Proceedings of the American Mathematical
  Society}\ }\textbf {\bibinfo {volume} {10}},\ \bibinfo {pages} {545}
  (\bibinfo {year} {1959})}\BibitemShut {NoStop}%
\bibitem [{\citenamefont {{Suzuki}}(1976)}]{Suzuki1976}%
  \BibitemOpen
  \bibfield  {author} {\bibinfo {author} {\bibfnamefont {M.}~\bibnamefont
  {{Suzuki}}},\ }\href {\doibase 10.1007/BF01609348} {\bibfield  {journal}
  {\bibinfo  {journal} {Communications in Mathematical Physics}\ }\textbf
  {\bibinfo {volume} {51}},\ \bibinfo {pages} {183} (\bibinfo {year}
  {1976})}\BibitemShut {NoStop}%
\bibitem [{\citenamefont {Suzuki}(1990)}]{SUZUKI1990}%
  \BibitemOpen
  \bibfield  {author} {\bibinfo {author} {\bibfnamefont {M.}~\bibnamefont
  {Suzuki}},\ }\href {\doibase https://doi.org/10.1016/0375-9601(90)90962-N}
  {\bibfield  {journal} {\bibinfo  {journal} {Physics Letters A}\ }\textbf
  {\bibinfo {volume} {146}},\ \bibinfo {pages} {319} (\bibinfo {year}
  {1990})}\BibitemShut {NoStop}%
\bibitem [{\citenamefont {{Suzuki}}(1991)}]{Suzuki1991}%
  \BibitemOpen
  \bibfield  {author} {\bibinfo {author} {\bibfnamefont {M.}~\bibnamefont
  {{Suzuki}}},\ }\href {\doibase 10.1063/1.529425} {\bibfield  {journal}
  {\bibinfo  {journal} {Journal of Mathematical Physics}\ }\textbf {\bibinfo
  {volume} {32}},\ \bibinfo {pages} {400} (\bibinfo {year} {1991})}\BibitemShut
  {NoStop}%
\bibitem [{\citenamefont {Bandrauk}\ and\ \citenamefont
  {Shen}(1991)}]{Andre1991}%
  \BibitemOpen
  \bibfield  {author} {\bibinfo {author} {\bibfnamefont {A.~D.}\ \bibnamefont
  {Bandrauk}}\ and\ \bibinfo {author} {\bibfnamefont {H.}~\bibnamefont
  {Shen}},\ }\href {\doibase https://doi.org/10.1016/0009-2614(91)90232-X}
  {\bibfield  {journal} {\bibinfo  {journal} {Chemical Physics Letters}\
  }\textbf {\bibinfo {volume} {176}},\ \bibinfo {pages} {428} (\bibinfo {year}
  {1991})}\BibitemShut {NoStop}%
\bibitem [{\citenamefont {Bandrauk}\ and\ \citenamefont
  {Shen}(1992)}]{Andre1992}%
  \BibitemOpen
  \bibfield  {author} {\bibinfo {author} {\bibfnamefont {A.~D.}\ \bibnamefont
  {Bandrauk}}\ and\ \bibinfo {author} {\bibfnamefont {H.}~\bibnamefont
  {Shen}},\ }\href {\doibase 10.1139/v92-078} {\bibfield  {journal} {\bibinfo
  {journal} {Canadian Journal of Chemistry}\ }\textbf {\bibinfo {volume}
  {70}},\ \bibinfo {pages} {555} (\bibinfo {year} {1992})}\BibitemShut
  {NoStop}%
\bibitem [{\citenamefont {Hatano}\ and\ \citenamefont
  {Suzuki}(2005)}]{Hatano2005}%
  \BibitemOpen
  \bibfield  {author} {\bibinfo {author} {\bibfnamefont {N.}~\bibnamefont
  {Hatano}}\ and\ \bibinfo {author} {\bibfnamefont {M.}~\bibnamefont
  {Suzuki}},\ }in\ \href {\doibase 10.1007/11526216_2} {\emph {\bibinfo
  {booktitle} {Quantum Annealing and Other Optimization Methods}}}\ (\bibinfo
  {publisher} {Springer Berlin Heidelberg},\ \bibinfo {year} {2005})\ pp.\
  \bibinfo {pages} {37--68}\BibitemShut {NoStop}%
\bibitem [{\citenamefont {Tannor}(2007)}]{tannor2007}%
  \BibitemOpen
  \bibfield  {author} {\bibinfo {author} {\bibfnamefont {D.}~\bibnamefont
  {Tannor}},\ }\href {https://books.google.co.uk/books?id=t7m08j3Wi9YC} {\emph
  {\bibinfo {title} {Introduction to Quantum Mechanics}}}\ (\bibinfo
  {publisher} {University Science Books},\ \bibinfo {year} {2007})\BibitemShut
  {NoStop}%
\bibitem [{\citenamefont {Childs}\ \emph {et~al.}(2021)\citenamefont {Childs},
  \citenamefont {Su}, \citenamefont {Tran}, \citenamefont {Wiebe},\ and\
  \citenamefont {Zhu}}]{Childs2021}%
  \BibitemOpen
  \bibfield  {author} {\bibinfo {author} {\bibfnamefont {A.~M.}\ \bibnamefont
  {Childs}}, \bibinfo {author} {\bibfnamefont {Y.}~\bibnamefont {Su}}, \bibinfo
  {author} {\bibfnamefont {M.~C.}\ \bibnamefont {Tran}}, \bibinfo {author}
  {\bibfnamefont {N.}~\bibnamefont {Wiebe}}, \ and\ \bibinfo {author}
  {\bibfnamefont {S.}~\bibnamefont {Zhu}},\ }\href
  {http://dx.doi.org/10.1103/PhysRevX.11.011020} {\bibfield  {journal}
  {\bibinfo  {journal} {Physical Review X}\ }\textbf {\bibinfo {volume} {11}}
  (\bibinfo {year} {2021})}\BibitemShut {NoStop}%
\bibitem [{\citenamefont {Roulet}\ and\ \citenamefont
  {Vaníček}(2021)}]{roulet2021}%
  \BibitemOpen
  \bibfield  {author} {\bibinfo {author} {\bibfnamefont {J.}~\bibnamefont
  {Roulet}}\ and\ \bibinfo {author} {\bibfnamefont {J.}~\bibnamefont
  {Vaníček}},\ }\href {https://doi.org/10.1063/5.0071153} {\bibfield
  {journal} {\bibinfo  {journal} {The Journal of Chemical Physics}\ }\textbf
  {\bibinfo {volume} {155}} (\bibinfo {year} {2021})}\BibitemShut {NoStop}%
\bibitem [{\citenamefont {Kassal}\ \emph {et~al.}(2008)\citenamefont {Kassal},
  \citenamefont {Jordan}, \citenamefont {Love}, \citenamefont {Mohseni},\ and\
  \citenamefont {Aspuru-Guzik}}]{Kassal2008}%
  \BibitemOpen
  \bibfield  {author} {\bibinfo {author} {\bibfnamefont {I.}~\bibnamefont
  {Kassal}}, \bibinfo {author} {\bibfnamefont {S.~P.}\ \bibnamefont {Jordan}},
  \bibinfo {author} {\bibfnamefont {P.~J.}\ \bibnamefont {Love}}, \bibinfo
  {author} {\bibfnamefont {M.}~\bibnamefont {Mohseni}}, \ and\ \bibinfo
  {author} {\bibfnamefont {A.}~\bibnamefont {Aspuru-Guzik}},\ }\href {\doibase
  10.1073/pnas.0808245105} {\bibfield  {journal} {\bibinfo  {journal}
  {Proceedings of the National Academy of Sciences}\ }\textbf {\bibinfo
  {volume} {105}},\ \bibinfo {pages} {18681} (\bibinfo {year}
  {2008})}\BibitemShut {NoStop}%
\bibitem [{\citenamefont {Jones}\ \emph {et~al.}(2019)\citenamefont {Jones},
  \citenamefont {O'Brien}, \citenamefont {White}, \citenamefont {Campbell},\
  and\ \citenamefont {Clark}}]{jones2019}%
  \BibitemOpen
  \bibfield  {author} {\bibinfo {author} {\bibfnamefont {B.~D.~M.}\
  \bibnamefont {Jones}}, \bibinfo {author} {\bibfnamefont {G.~O.}\ \bibnamefont
  {O'Brien}}, \bibinfo {author} {\bibfnamefont {D.~R.}\ \bibnamefont {White}},
  \bibinfo {author} {\bibfnamefont {E.~T.}\ \bibnamefont {Campbell}}, \ and\
  \bibinfo {author} {\bibfnamefont {J.~A.}\ \bibnamefont {Clark}},\ }\href
  {https://arxiv.org/abs/1904.01336} {\enquote {\bibinfo {title} {Optimising
  trotter-suzuki decompositions for quantum simulation using evolutionary
  strategies},}\ } (\bibinfo {year} {2019})\BibitemShut {NoStop}%
\bibitem [{\citenamefont {Kołaczek}\ \emph {et~al.}(2019)\citenamefont
  {Kołaczek}, \citenamefont {Spisak},\ and\ \citenamefont
  {Wołoszyn}}]{Damian2019}%
  \BibitemOpen
  \bibfield  {author} {\bibinfo {author} {\bibfnamefont {D.}~\bibnamefont
  {Kołaczek}}, \bibinfo {author} {\bibfnamefont {B.~J.}\ \bibnamefont
  {Spisak}}, \ and\ \bibinfo {author} {\bibfnamefont {M.}~\bibnamefont
  {Wołoszyn}},\ }\href {\doibase doi:10.2478/amcs-2019-0032} {\bibfield
  {journal} {\bibinfo  {journal} {International Journal of Applied Mathematics
  and Computer Science}\ }\textbf {\bibinfo {volume} {29}},\ \bibinfo {pages}
  {439} (\bibinfo {year} {2019})}\BibitemShut {NoStop}%
\bibitem [{\citenamefont {Navickas}\ \emph {et~al.}(2025)\citenamefont
  {Navickas}, \citenamefont {MacDonell}, \citenamefont {Valahu}, \citenamefont
  {Olaya-Agudelo}, \citenamefont {Scuccimarra}, \citenamefont {Millican},
  \citenamefont {Matsos}, \citenamefont {Nourse}, \citenamefont {Rao},
  \citenamefont {Biercuk}, \citenamefont {Hempel}, \citenamefont {Kassal},\
  and\ \citenamefont {Tan}}]{Navickas2025}%
  \BibitemOpen
  \bibfield  {author} {\bibinfo {author} {\bibfnamefont {T.}~\bibnamefont
  {Navickas}}, \bibinfo {author} {\bibfnamefont {R.~J.}\ \bibnamefont
  {MacDonell}}, \bibinfo {author} {\bibfnamefont {C.~H.}\ \bibnamefont
  {Valahu}}, \bibinfo {author} {\bibfnamefont {V.~C.}\ \bibnamefont
  {Olaya-Agudelo}}, \bibinfo {author} {\bibfnamefont {F.}~\bibnamefont
  {Scuccimarra}}, \bibinfo {author} {\bibfnamefont {M.~J.}\ \bibnamefont
  {Millican}}, \bibinfo {author} {\bibfnamefont {V.~G.}\ \bibnamefont
  {Matsos}}, \bibinfo {author} {\bibfnamefont {H.~L.}\ \bibnamefont {Nourse}},
  \bibinfo {author} {\bibfnamefont {A.~D.}\ \bibnamefont {Rao}}, \bibinfo
  {author} {\bibfnamefont {M.~J.}\ \bibnamefont {Biercuk}}, \bibinfo {author}
  {\bibfnamefont {C.}~\bibnamefont {Hempel}}, \bibinfo {author} {\bibfnamefont
  {I.}~\bibnamefont {Kassal}}, \ and\ \bibinfo {author} {\bibfnamefont {T.~R.}\
  \bibnamefont {Tan}},\ }\href {\doibase 10.1021/jacs.5c03336} {\bibfield
  {journal} {\bibinfo  {journal} {Journal of the American Chemical Society}\
  }\textbf {\bibinfo {volume} {147}},\ \bibinfo {pages} {23566–23573}
  (\bibinfo {year} {2025})}\BibitemShut {NoStop}%
\bibitem [{\citenamefont {MacDonell}\ \emph {et~al.}(2021)\citenamefont
  {MacDonell}, \citenamefont {Dickerson}, \citenamefont {Birch}, \citenamefont
  {Kumar}, \citenamefont {Edmunds}, \citenamefont {Biercuk}, \citenamefont
  {Hempel},\ and\ \citenamefont {Kassal}}]{Ryan2021}%
  \BibitemOpen
  \bibfield  {author} {\bibinfo {author} {\bibfnamefont {R.~J.}\ \bibnamefont
  {MacDonell}}, \bibinfo {author} {\bibfnamefont {C.~E.}\ \bibnamefont
  {Dickerson}}, \bibinfo {author} {\bibfnamefont {C.~J.~T.}\ \bibnamefont
  {Birch}}, \bibinfo {author} {\bibfnamefont {A.}~\bibnamefont {Kumar}},
  \bibinfo {author} {\bibfnamefont {C.~L.}\ \bibnamefont {Edmunds}}, \bibinfo
  {author} {\bibfnamefont {M.~J.}\ \bibnamefont {Biercuk}}, \bibinfo {author}
  {\bibfnamefont {C.}~\bibnamefont {Hempel}}, \ and\ \bibinfo {author}
  {\bibfnamefont {I.}~\bibnamefont {Kassal}},\ }\href {\doibase
  10.1039/D1SC02142G} {\bibfield  {journal} {\bibinfo  {journal} {Chem. Sci.}\
  }\textbf {\bibinfo {volume} {12}},\ \bibinfo {pages} {9794} (\bibinfo {year}
  {2021})}\BibitemShut {NoStop}%
\bibitem [{\citenamefont {Schneider}\ and\ \citenamefont
  {Domcke}(1988)}]{SCHNEIDER1988}%
  \BibitemOpen
  \bibfield  {author} {\bibinfo {author} {\bibfnamefont {R.}~\bibnamefont
  {Schneider}}\ and\ \bibinfo {author} {\bibfnamefont {W.}~\bibnamefont
  {Domcke}},\ }\href {\doibase https://doi.org/10.1016/0009-2614(88)80034-4}
  {\bibfield  {journal} {\bibinfo  {journal} {Chemical Physics Letters}\
  }\textbf {\bibinfo {volume} {150}},\ \bibinfo {pages} {235} (\bibinfo {year}
  {1988})}\BibitemShut {NoStop}%
\bibitem [{\citenamefont {Kanno}\ \emph {et~al.}(2015)\citenamefont {Kanno},
  \citenamefont {Ito}, \citenamefont {Shimakura}, \citenamefont {Koseki},
  \citenamefont {Kono},\ and\ \citenamefont {Fujimura}}]{Kanno2015}%
  \BibitemOpen
  \bibfield  {author} {\bibinfo {author} {\bibfnamefont {M.}~\bibnamefont
  {Kanno}}, \bibinfo {author} {\bibfnamefont {Y.}~\bibnamefont {Ito}}, \bibinfo
  {author} {\bibfnamefont {N.}~\bibnamefont {Shimakura}}, \bibinfo {author}
  {\bibfnamefont {S.}~\bibnamefont {Koseki}}, \bibinfo {author} {\bibfnamefont
  {H.}~\bibnamefont {Kono}}, \ and\ \bibinfo {author} {\bibfnamefont
  {Y.}~\bibnamefont {Fujimura}},\ }\href {\doibase 10.1039/C4CP04807E}
  {\bibfield  {journal} {\bibinfo  {journal} {Physical Chemistry Chemical
  Physics}\ }\textbf {\bibinfo {volume} {17}},\ \bibinfo {pages} {2012}
  (\bibinfo {year} {2015})}\BibitemShut {NoStop}%
\bibitem [{\citenamefont {Schile}\ and\ \citenamefont
  {Limmer}(2019)}]{Schile2019}%
  \BibitemOpen
  \bibfield  {author} {\bibinfo {author} {\bibfnamefont {A.~J.}\ \bibnamefont
  {Schile}}\ and\ \bibinfo {author} {\bibfnamefont {D.~T.}\ \bibnamefont
  {Limmer}},\ }\href {\doibase 10.1063/1.5106379} {\bibfield  {journal}
  {\bibinfo  {journal} {The Journal of Chemical Physics}\ }\textbf {\bibinfo
  {volume} {151}},\ \bibinfo {pages} {014106} (\bibinfo {year}
  {2019})}\BibitemShut {NoStop}%
\bibitem [{\citenamefont {Gu}\ and\ \citenamefont {Mukamel}(2020)}]{Gu2020}%
  \BibitemOpen
  \bibfield  {author} {\bibinfo {author} {\bibfnamefont {B.}~\bibnamefont
  {Gu}}\ and\ \bibinfo {author} {\bibfnamefont {S.}~\bibnamefont {Mukamel}},\
  }\href {\doibase 10.1021/acs.jpclett.0c00381} {\bibfield  {journal} {\bibinfo
   {journal} {The Journal of Physical Chemistry Letters}\ }\textbf {\bibinfo
  {volume} {11}},\ \bibinfo {pages} {5555} (\bibinfo {year}
  {2020})}\BibitemShut {NoStop}%
\bibitem [{\citenamefont {Neville}\ \emph {et~al.}(2022)\citenamefont
  {Neville}, \citenamefont {Stolow},\ and\ \citenamefont
  {Schuurman}}]{Neville2022}%
  \BibitemOpen
  \bibfield  {author} {\bibinfo {author} {\bibfnamefont {S.~P.}\ \bibnamefont
  {Neville}}, \bibinfo {author} {\bibfnamefont {A.}~\bibnamefont {Stolow}}, \
  and\ \bibinfo {author} {\bibfnamefont {M.~S.}\ \bibnamefont {Schuurman}},\
  }\href {\doibase 10.1088/1361-6455/ac5460} {\bibfield  {journal} {\bibinfo
  {journal} {Journal of Physics B: Atomic, Molecular and Optical Physics}\
  }\textbf {\bibinfo {volume} {55}},\ \bibinfo {pages} {044004} (\bibinfo
  {year} {2022})}\BibitemShut {NoStop}%
\bibitem [{\citenamefont {Köuppel}\ \emph {et~al.}(1984)\citenamefont
  {Köuppel}, \citenamefont {Domcke},\ and\ \citenamefont
  {Cederbaum}}]{Köuppel1984}%
  \BibitemOpen
  \bibfield  {author} {\bibinfo {author} {\bibfnamefont {H.}~\bibnamefont
  {Köuppel}}, \bibinfo {author} {\bibfnamefont {W.}~\bibnamefont {Domcke}}, \
  and\ \bibinfo {author} {\bibfnamefont {L.~S.}\ \bibnamefont {Cederbaum}},\
  }in\ \href {\doibase 10.1002/9780470142813.ch2} {\emph {\bibinfo {booktitle}
  {Advances in Chemical Physics}}}\ (\bibinfo  {publisher} {John Wiley \&
  Sons},\ \bibinfo {year} {1984})\ pp.\ \bibinfo {pages} {59--246}\BibitemShut
  {NoStop}%
\bibitem [{\citenamefont {Innes}\ \emph {et~al.}(1988)\citenamefont {Innes},
  \citenamefont {Ross},\ and\ \citenamefont {Moomaw}}]{Innes1988}%
  \BibitemOpen
  \bibfield  {author} {\bibinfo {author} {\bibfnamefont {K.}~\bibnamefont
  {Innes}}, \bibinfo {author} {\bibfnamefont {I.}~\bibnamefont {Ross}}, \ and\
  \bibinfo {author} {\bibfnamefont {W.~R.}\ \bibnamefont {Moomaw}},\ }\href
  {\doibase https://doi.org/10.1016/0022-2852(88)90343-8} {\bibfield  {journal}
  {\bibinfo  {journal} {Journal of Molecular Spectroscopy}\ }\textbf {\bibinfo
  {volume} {132}},\ \bibinfo {pages} {492} (\bibinfo {year}
  {1988})}\BibitemShut {NoStop}%
\bibitem [{\citenamefont {Zeng}\ \emph {et~al.}(2025)\citenamefont {Zeng},
  \citenamefont {Sun}, \citenamefont {Jiang},\ and\ \citenamefont
  {Zhao}}]{Zeng2025}%
  \BibitemOpen
  \bibfield  {author} {\bibinfo {author} {\bibfnamefont {P.}~\bibnamefont
  {Zeng}}, \bibinfo {author} {\bibfnamefont {J.}~\bibnamefont {Sun}}, \bibinfo
  {author} {\bibfnamefont {L.}~\bibnamefont {Jiang}}, \ and\ \bibinfo {author}
  {\bibfnamefont {Q.}~\bibnamefont {Zhao}},\ }\href
  {http://dx.doi.org/10.1103/PRXQuantum.6.010359} {\bibfield  {journal}
  {\bibinfo  {journal} {PRX Quantum}\ }\textbf {\bibinfo {volume} {6}}
  (\bibinfo {year} {2025})}\BibitemShut {NoStop}%
\bibitem [{\citenamefont {Childs}\ \emph {et~al.}(2022)\citenamefont {Childs},
  \citenamefont {Leng}, \citenamefont {Li}, \citenamefont {Liu},\ and\
  \citenamefont {Zhang}}]{Childs2022}%
  \BibitemOpen
  \bibfield  {author} {\bibinfo {author} {\bibfnamefont {A.~M.}\ \bibnamefont
  {Childs}}, \bibinfo {author} {\bibfnamefont {J.}~\bibnamefont {Leng}},
  \bibinfo {author} {\bibfnamefont {T.}~\bibnamefont {Li}}, \bibinfo {author}
  {\bibfnamefont {J.-P.}\ \bibnamefont {Liu}}, \ and\ \bibinfo {author}
  {\bibfnamefont {C.}~\bibnamefont {Zhang}},\ }\href {\doibase
  10.22331/q-2022-11-17-860} {\bibfield  {journal} {\bibinfo  {journal}
  {{Quantum}}\ }\textbf {\bibinfo {volume} {6}},\ \bibinfo {pages} {860}
  (\bibinfo {year} {2022})}\BibitemShut {NoStop}%
\bibitem [{\citenamefont {Berry}\ \emph {et~al.}(2024)\citenamefont {Berry},
  \citenamefont {Rubin}, \citenamefont {Elnabawy}, \citenamefont {Ahlers},
  \citenamefont {DePrince}, \citenamefont {Lee}, \citenamefont {Gogolin},\ and\
  \citenamefont {Babbush}}]{Berry2024}%
  \BibitemOpen
  \bibfield  {author} {\bibinfo {author} {\bibfnamefont {D.~W.}\ \bibnamefont
  {Berry}}, \bibinfo {author} {\bibfnamefont {N.~C.}\ \bibnamefont {Rubin}},
  \bibinfo {author} {\bibfnamefont {A.~O.}\ \bibnamefont {Elnabawy}}, \bibinfo
  {author} {\bibfnamefont {G.}~\bibnamefont {Ahlers}}, \bibinfo {author}
  {\bibfnamefont {A.~E.}\ \bibnamefont {DePrince}}, \bibinfo {author}
  {\bibfnamefont {J.}~\bibnamefont {Lee}}, \bibinfo {author} {\bibfnamefont
  {C.}~\bibnamefont {Gogolin}}, \ and\ \bibinfo {author} {\bibfnamefont
  {R.}~\bibnamefont {Babbush}},\ }\href {\doibase 10.1038/s41534-024-00896-9}
  {\bibfield  {journal} {\bibinfo  {journal} {npj Quantum Information}\
  }\textbf {\bibinfo {volume} {10}},\ \bibinfo {pages} {130} (\bibinfo {year}
  {2024})}\BibitemShut {NoStop}%
\bibitem [{\citenamefont {An}\ \emph {et~al.}(2023)\citenamefont {An},
  \citenamefont {Liu},\ and\ \citenamefont {Lin}}]{An2023}%
  \BibitemOpen
  \bibfield  {author} {\bibinfo {author} {\bibfnamefont {D.}~\bibnamefont
  {An}}, \bibinfo {author} {\bibfnamefont {J.-P.}\ \bibnamefont {Liu}}, \ and\
  \bibinfo {author} {\bibfnamefont {L.}~\bibnamefont {Lin}},\ }\href
  {http://dx.doi.org/10.1103/PhysRevLett.131.150603} {\bibfield  {journal}
  {\bibinfo  {journal} {Physical Review Letters}\ }\textbf {\bibinfo {volume}
  {131}} (\bibinfo {year} {2023})}\BibitemShut {NoStop}%
\bibitem [{\citenamefont {Su}\ \emph {et~al.}(2021)\citenamefont {Su},
  \citenamefont {Berry}, \citenamefont {Wiebe}, \citenamefont {Rubin},\ and\
  \citenamefont {Babbush}}]{Berry2021}%
  \BibitemOpen
  \bibfield  {author} {\bibinfo {author} {\bibfnamefont {Y.}~\bibnamefont
  {Su}}, \bibinfo {author} {\bibfnamefont {D.~W.}\ \bibnamefont {Berry}},
  \bibinfo {author} {\bibfnamefont {N.}~\bibnamefont {Wiebe}}, \bibinfo
  {author} {\bibfnamefont {N.}~\bibnamefont {Rubin}}, \ and\ \bibinfo {author}
  {\bibfnamefont {R.}~\bibnamefont {Babbush}},\ }\href {\doibase
  10.1103/PRXQuantum.2.040332} {\bibfield  {journal} {\bibinfo  {journal} {PRX
  Quantum}\ }\textbf {\bibinfo {volume} {2}},\ \bibinfo {pages} {040332}
  (\bibinfo {year} {2021})}\BibitemShut {NoStop}%
\bibitem [{\citenamefont {Mukhopadhyay}\ \emph {et~al.}(2024)\citenamefont
  {Mukhopadhyay}, \citenamefont {Stetina},\ and\ \citenamefont
  {Wiebe}}]{Mukhop2024}%
  \BibitemOpen
  \bibfield  {author} {\bibinfo {author} {\bibfnamefont {P.}~\bibnamefont
  {Mukhopadhyay}}, \bibinfo {author} {\bibfnamefont {T.~F.}\ \bibnamefont
  {Stetina}}, \ and\ \bibinfo {author} {\bibfnamefont {N.}~\bibnamefont
  {Wiebe}},\ }\href {http://dx.doi.org/10.1103/PRXQuantum.5.010345} {\bibfield
  {journal} {\bibinfo  {journal} {PRX Quantum}\ }\textbf {\bibinfo {volume}
  {5}} (\bibinfo {year} {2024})}\BibitemShut {NoStop}%
\bibitem [{\citenamefont {Low}\ and\ \citenamefont {Chuang}(2019)}]{Low2019}%
  \BibitemOpen
  \bibfield  {author} {\bibinfo {author} {\bibfnamefont {G.~H.}\ \bibnamefont
  {Low}}\ and\ \bibinfo {author} {\bibfnamefont {I.~L.}\ \bibnamefont
  {Chuang}},\ }\href {\doibase 10.22331/q-2019-07-12-163} {\bibfield  {journal}
  {\bibinfo  {journal} {Quantum}\ }\textbf {\bibinfo {volume} {3}},\ \bibinfo
  {pages} {163} (\bibinfo {year} {2019})}\BibitemShut {NoStop}%
\bibitem [{\citenamefont {von Burg}\ \emph {et~al.}(2021)\citenamefont {von
  Burg}, \citenamefont {Low}, \citenamefont {Häner}, \citenamefont {Steiger},
  \citenamefont {Reiher}, \citenamefont {Roetteler},\ and\ \citenamefont
  {Troyer}}]{vonBurg2021}%
  \BibitemOpen
  \bibfield  {author} {\bibinfo {author} {\bibfnamefont {V.}~\bibnamefont {von
  Burg}}, \bibinfo {author} {\bibfnamefont {G.~H.}\ \bibnamefont {Low}},
  \bibinfo {author} {\bibfnamefont {T.}~\bibnamefont {Häner}}, \bibinfo
  {author} {\bibfnamefont {D.~S.}\ \bibnamefont {Steiger}}, \bibinfo {author}
  {\bibfnamefont {M.}~\bibnamefont {Reiher}}, \bibinfo {author} {\bibfnamefont
  {M.}~\bibnamefont {Roetteler}}, \ and\ \bibinfo {author} {\bibfnamefont
  {M.}~\bibnamefont {Troyer}},\ }\href
  {http://dx.doi.org/10.1103/PhysRevResearch.3.033055} {\bibfield  {journal}
  {\bibinfo  {journal} {Physical Review Research}\ }\textbf {\bibinfo {volume}
  {3}} (\bibinfo {year} {2021})}\BibitemShut {NoStop}%
\bibitem [{\citenamefont {Feit}\ \emph {et~al.}(1982)\citenamefont {Feit},
  \citenamefont {Fleck},\ and\ \citenamefont {Steiger}}]{FEIT1982}%
  \BibitemOpen
  \bibfield  {author} {\bibinfo {author} {\bibfnamefont {M.}~\bibnamefont
  {Feit}}, \bibinfo {author} {\bibfnamefont {J.}~\bibnamefont {Fleck}}, \ and\
  \bibinfo {author} {\bibfnamefont {A.}~\bibnamefont {Steiger}},\ }\href
  {\doibase https://doi.org/10.1016/0021-9991(82)90091-2} {\bibfield  {journal}
  {\bibinfo  {journal} {Journal of Computational Physics}\ }\textbf {\bibinfo
  {volume} {47}},\ \bibinfo {pages} {412} (\bibinfo {year} {1982})}\BibitemShut
  {NoStop}%
\bibitem [{\citenamefont {Kosloff}(1988)}]{kosloff1988}%
  \BibitemOpen
  \bibfield  {author} {\bibinfo {author} {\bibfnamefont {R.}~\bibnamefont
  {Kosloff}},\ }\href {\doibase 10.1021/j100319a003} {\bibfield  {journal}
  {\bibinfo  {journal} {The Journal of Physical Chemistry}\ }\textbf {\bibinfo
  {volume} {92}},\ \bibinfo {pages} {2087{\textendash}2100} (\bibinfo {year}
  {1988})}\BibitemShut {NoStop}%
\bibitem [{\citenamefont {Benenti}\ and\ \citenamefont
  {Strini}(2008)}]{Benenti2008}%
  \BibitemOpen
  \bibfield  {author} {\bibinfo {author} {\bibfnamefont {G.}~\bibnamefont
  {Benenti}}\ and\ \bibinfo {author} {\bibfnamefont {G.}~\bibnamefont
  {Strini}},\ }\href {\doibase 10.1119/1.2894532} {\bibfield  {journal}
  {\bibinfo  {journal} {American Journal of Physics}\ }\textbf {\bibinfo
  {volume} {76}},\ \bibinfo {pages} {657} (\bibinfo {year} {2008})}\BibitemShut
  {NoStop}%
\bibitem [{\citenamefont {Franck}\ and\ \citenamefont
  {Dymond}(1926)}]{Franck1926}%
  \BibitemOpen
  \bibfield  {author} {\bibinfo {author} {\bibfnamefont {J.}~\bibnamefont
  {Franck}}\ and\ \bibinfo {author} {\bibfnamefont {E.~G.}\ \bibnamefont
  {Dymond}},\ }\href {\doibase 10.1039/TF9262100536} {\bibfield  {journal}
  {\bibinfo  {journal} {Trans. Faraday Soc.}\ }\textbf {\bibinfo {volume}
  {21}},\ \bibinfo {pages} {536} (\bibinfo {year} {1926})}\BibitemShut
  {NoStop}%
\bibitem [{\citenamefont {Condon}(1926)}]{Condon1926}%
  \BibitemOpen
  \bibfield  {author} {\bibinfo {author} {\bibfnamefont {E.}~\bibnamefont
  {Condon}},\ }\href {\doibase 10.1103/PhysRev.28.1182} {\bibfield  {journal}
  {\bibinfo  {journal} {Physical Review}\ }\textbf {\bibinfo {volume} {28}},\
  \bibinfo {pages} {1182} (\bibinfo {year} {1926})}\BibitemShut {NoStop}%
\bibitem [{\citenamefont {Birge}(1926)}]{Birge1926}%
  \BibitemOpen
  \bibfield  {author} {\bibinfo {author} {\bibfnamefont {R.~T.}\ \bibnamefont
  {Birge}},\ }\href {\doibase 10.1103/PhysRev.28.1157} {\bibfield  {journal}
  {\bibinfo  {journal} {Physical Review}\ }\textbf {\bibinfo {volume} {28}},\
  \bibinfo {pages} {1157} (\bibinfo {year} {1926})}\BibitemShut {NoStop}%
\bibitem [{\citenamefont {Luis}\ \emph {et~al.}(2004)\citenamefont {Luis},
  \citenamefont {Bishop},\ and\ \citenamefont {Kirtman}}]{Luis2004}%
  \BibitemOpen
  \bibfield  {author} {\bibinfo {author} {\bibfnamefont {J.~M.}\ \bibnamefont
  {Luis}}, \bibinfo {author} {\bibfnamefont {D.~M.}\ \bibnamefont {Bishop}}, \
  and\ \bibinfo {author} {\bibfnamefont {B.}~\bibnamefont {Kirtman}},\ }\href
  {\doibase 10.1063/1.1630566} {\bibfield  {journal} {\bibinfo  {journal} {The
  Journal of Chemical Physics}\ }\textbf {\bibinfo {volume} {120}},\ \bibinfo
  {pages} {813} (\bibinfo {year} {2004})}\BibitemShut {NoStop}%
\bibitem [{\citenamefont {{International Union of Pure and Applied
  Chemistry}}(2025)}]{IUPAC2025}%
  \BibitemOpen
  \bibfield  {author} {\bibinfo {author} {\bibnamefont {{International Union of
  Pure and Applied Chemistry}}},\ }\href
  {https://doi.org/10.1351/goldbook.F02510} {\enquote {\bibinfo {title}
  {Franck--condon principle},}\ }\bibinfo {howpublished} {IUPAC Gold Book}
  (\bibinfo {year} {2025})\BibitemShut {NoStop}%
\bibitem [{\citenamefont {Kundu}\ \emph {et~al.}(2022)\citenamefont {Kundu},
  \citenamefont {Roy}, \citenamefont {Fleming},\ and\ \citenamefont
  {Makri}}]{Kundu2022}%
  \BibitemOpen
  \bibfield  {author} {\bibinfo {author} {\bibfnamefont {S.}~\bibnamefont
  {Kundu}}, \bibinfo {author} {\bibfnamefont {P.~P.}\ \bibnamefont {Roy}},
  \bibinfo {author} {\bibfnamefont {G.~R.}\ \bibnamefont {Fleming}}, \ and\
  \bibinfo {author} {\bibfnamefont {N.}~\bibnamefont {Makri}},\ }\href
  {\doibase 10.1021/acs.jpcb.2c00846} {\bibfield  {journal} {\bibinfo
  {journal} {The Journal of Physical Chemistry B}\ }\textbf {\bibinfo {volume}
  {126}},\ \bibinfo {pages} {2899} (\bibinfo {year} {2022})}\BibitemShut
  {NoStop}%
\bibitem [{\citenamefont {M{\"o}tt{\"o}nen}\ \emph {et~al.}(2004)\citenamefont
  {M{\"o}tt{\"o}nen}, \citenamefont {Vartiainen}, \citenamefont {Bergholm},\
  and\ \citenamefont {Salomaa}}]{Mttnen2004}%
  \BibitemOpen
  \bibfield  {author} {\bibinfo {author} {\bibfnamefont {M.}~\bibnamefont
  {M{\"o}tt{\"o}nen}}, \bibinfo {author} {\bibfnamefont {J.~J.}\ \bibnamefont
  {Vartiainen}}, \bibinfo {author} {\bibfnamefont {V.}~\bibnamefont
  {Bergholm}}, \ and\ \bibinfo {author} {\bibfnamefont {M.~M.}\ \bibnamefont
  {Salomaa}},\ }\href {\doibase 10.26421/QIC5.6-5} {\bibfield  {journal}
  {\bibinfo  {journal} {Quantum Inf. Comput.}\ }\textbf {\bibinfo {volume}
  {5}},\ \bibinfo {pages} {467} (\bibinfo {year} {2004})}\BibitemShut {NoStop}%
\bibitem [{\citenamefont {Iaconis}\ \emph {et~al.}(2024)\citenamefont
  {Iaconis}, \citenamefont {Johri},\ and\ \citenamefont {Zhu}}]{Iaconis_2024}%
  \BibitemOpen
  \bibfield  {author} {\bibinfo {author} {\bibfnamefont {J.}~\bibnamefont
  {Iaconis}}, \bibinfo {author} {\bibfnamefont {S.}~\bibnamefont {Johri}}, \
  and\ \bibinfo {author} {\bibfnamefont {E.~Y.}\ \bibnamefont {Zhu}},\ }\href
  {http://dx.doi.org/10.1038/s41534-024-00805-0} {\bibfield  {journal}
  {\bibinfo  {journal} {npj Quantum Information}\ }\textbf {\bibinfo {volume}
  {10}} (\bibinfo {year} {2024})}\BibitemShut {NoStop}%
\bibitem [{\citenamefont {Moosa}\ \emph {et~al.}(2023)\citenamefont {Moosa},
  \citenamefont {Watts}, \citenamefont {Chen}, \citenamefont {Sarma},\ and\
  \citenamefont {McMahon}}]{moosa2023}%
  \BibitemOpen
  \bibfield  {author} {\bibinfo {author} {\bibfnamefont {M.}~\bibnamefont
  {Moosa}}, \bibinfo {author} {\bibfnamefont {T.~W.}\ \bibnamefont {Watts}},
  \bibinfo {author} {\bibfnamefont {Y.}~\bibnamefont {Chen}}, \bibinfo {author}
  {\bibfnamefont {A.}~\bibnamefont {Sarma}}, \ and\ \bibinfo {author}
  {\bibfnamefont {P.~L.}\ \bibnamefont {McMahon}},\ }\href {\doibase
  10.1088/2058-9565/acfc62} {\bibfield  {journal} {\bibinfo  {journal} {Quantum
  Science and Technology}\ }\textbf {\bibinfo {volume} {9}},\ \bibinfo {pages}
  {015002} (\bibinfo {year} {2023})}\BibitemShut {NoStop}%
\bibitem [{\citenamefont {Rattew}\ \emph {et~al.}(2021)\citenamefont {Rattew},
  \citenamefont {Sun}, \citenamefont {Minssen},\ and\ \citenamefont
  {Pistoia}}]{Rattew_2021}%
  \BibitemOpen
  \bibfield  {author} {\bibinfo {author} {\bibfnamefont {A.~G.}\ \bibnamefont
  {Rattew}}, \bibinfo {author} {\bibfnamefont {Y.}~\bibnamefont {Sun}},
  \bibinfo {author} {\bibfnamefont {P.}~\bibnamefont {Minssen}}, \ and\
  \bibinfo {author} {\bibfnamefont {M.}~\bibnamefont {Pistoia}},\ }\href
  {\doibase 10.22331/q-2021-12-23-609} {\bibfield  {journal} {\bibinfo
  {journal} {Quantum}\ }\textbf {\bibinfo {volume} {5}},\ \bibinfo {pages}
  {609} (\bibinfo {year} {2021})}\BibitemShut {NoStop}%
\bibitem [{\citenamefont {{Qiskit contributors}}(2023)}]{Qiskit}%
  \BibitemOpen
  \bibfield  {author} {\bibinfo {author} {\bibnamefont {{Qiskit
  contributors}}},\ }\href {\doibase 10.5281/zenodo.2573505} {\enquote
  {\bibinfo {title} {Qiskit: An open-source framework for quantum computing},}\
  } (\bibinfo {year} {2023})\BibitemShut {NoStop}%
\bibitem [{\citenamefont {Ollitrault}\ \emph {et~al.}(2020)\citenamefont
  {Ollitrault}, \citenamefont {Mazzola},\ and\ \citenamefont
  {Tavernelli}}]{Pauline2020}%
  \BibitemOpen
  \bibfield  {author} {\bibinfo {author} {\bibfnamefont {P.~J.}\ \bibnamefont
  {Ollitrault}}, \bibinfo {author} {\bibfnamefont {G.}~\bibnamefont {Mazzola}},
  \ and\ \bibinfo {author} {\bibfnamefont {I.}~\bibnamefont {Tavernelli}},\
  }\href {\doibase 10.1103/PhysRevLett.125.260511} {\bibfield  {journal}
  {\bibinfo  {journal} {Physical Review Letters}\ }\textbf {\bibinfo {volume}
  {125}},\ \bibinfo {pages} {260511} (\bibinfo {year} {2020})}\BibitemShut
  {NoStop}%
\bibitem [{\citenamefont {Jensen}\ \emph {et~al.}(2023)\citenamefont {Jensen},
  \citenamefont {Johnson},\ and\ \citenamefont {Kunitsa}}]{jensen2023}%
  \BibitemOpen
  \bibfield  {author} {\bibinfo {author} {\bibfnamefont {P.~W.~K.}\
  \bibnamefont {Jensen}}, \bibinfo {author} {\bibfnamefont {P.~D.}\
  \bibnamefont {Johnson}}, \ and\ \bibinfo {author} {\bibfnamefont {A.~A.}\
  \bibnamefont {Kunitsa}},\ }\href {\doibase 10.1103/PhysRevA.108.022422}
  {\bibfield  {journal} {\bibinfo  {journal} {Phys. Rev. A}\ }\textbf {\bibinfo
  {volume} {108}},\ \bibinfo {pages} {022422} (\bibinfo {year}
  {2023})}\BibitemShut {NoStop}%
\bibitem [{\citenamefont {Aharonov}\ \emph {et~al.}(2006)\citenamefont
  {Aharonov}, \citenamefont {Jones},\ and\ \citenamefont
  {Landau}}]{aharonov2006}%
  \BibitemOpen
  \bibfield  {author} {\bibinfo {author} {\bibfnamefont {D.}~\bibnamefont
  {Aharonov}}, \bibinfo {author} {\bibfnamefont {V.}~\bibnamefont {Jones}}, \
  and\ \bibinfo {author} {\bibfnamefont {Z.}~\bibnamefont {Landau}},\ }\href
  {https://arxiv.org/abs/quant-ph/0511096} {\enquote {\bibinfo {title} {A
  polynomial quantum algorithm for approximating the jones polynomial},}\ }
  (\bibinfo {year} {2006})\BibitemShut {NoStop}%
\bibitem [{\citenamefont {Cleve}\ \emph {et~al.}(1998)\citenamefont {Cleve},
  \citenamefont {Ekert}, \citenamefont {Macchiavello},\ and\ \citenamefont
  {Mosca}}]{Cleve1998}%
  \BibitemOpen
  \bibfield  {author} {\bibinfo {author} {\bibfnamefont {R.}~\bibnamefont
  {Cleve}}, \bibinfo {author} {\bibfnamefont {A.}~\bibnamefont {Ekert}},
  \bibinfo {author} {\bibfnamefont {C.}~\bibnamefont {Macchiavello}}, \ and\
  \bibinfo {author} {\bibfnamefont {M.}~\bibnamefont {Mosca}},\ }\href
  {\doibase 10.1098/rspa.1998.0164} {\bibfield  {journal} {\bibinfo  {journal}
  {Proceedings of the Royal Society of London. Series A: Mathematical, Physical
  and Engineering Sciences}\ }\textbf {\bibinfo {volume} {454}},\ \bibinfo
  {pages} {339} (\bibinfo {year} {1998})}\BibitemShut {NoStop}%
\bibitem [{\citenamefont {MacDonell}\ \emph {et~al.}(2023)\citenamefont
  {MacDonell}, \citenamefont {Navickas}, \citenamefont {Wohlers-Reichel},
  \citenamefont {Valahu}, \citenamefont {Rao}, \citenamefont {Millican},
  \citenamefont {Currington}, \citenamefont {Biercuk}, \citenamefont {Tan},
  \citenamefont {Hempel},\ and\ \citenamefont {Kassal}}]{Ryan2023}%
  \BibitemOpen
  \bibfield  {author} {\bibinfo {author} {\bibfnamefont {R.~J.}\ \bibnamefont
  {MacDonell}}, \bibinfo {author} {\bibfnamefont {T.}~\bibnamefont {Navickas}},
  \bibinfo {author} {\bibfnamefont {T.~F.}\ \bibnamefont {Wohlers-Reichel}},
  \bibinfo {author} {\bibfnamefont {C.~H.}\ \bibnamefont {Valahu}}, \bibinfo
  {author} {\bibfnamefont {A.~D.}\ \bibnamefont {Rao}}, \bibinfo {author}
  {\bibfnamefont {M.~J.}\ \bibnamefont {Millican}}, \bibinfo {author}
  {\bibfnamefont {M.~A.}\ \bibnamefont {Currington}}, \bibinfo {author}
  {\bibfnamefont {M.~J.}\ \bibnamefont {Biercuk}}, \bibinfo {author}
  {\bibfnamefont {T.~R.}\ \bibnamefont {Tan}}, \bibinfo {author} {\bibfnamefont
  {C.}~\bibnamefont {Hempel}}, \ and\ \bibinfo {author} {\bibfnamefont
  {I.}~\bibnamefont {Kassal}},\ }\href {\doibase 10.1039/D3SC02453A} {\bibfield
   {journal} {\bibinfo  {journal} {Chem. Sci.}\ }\textbf {\bibinfo {volume}
  {14}},\ \bibinfo {pages} {9439} (\bibinfo {year} {2023})}\BibitemShut
  {NoStop}%
\bibitem [{\citenamefont {Del~Re}\ \emph {et~al.}(2024)\citenamefont {Del~Re},
  \citenamefont {Rost}, \citenamefont {Foss-Feig}, \citenamefont {Kemper},\
  and\ \citenamefont {Freericks}}]{lorenzo2024}%
  \BibitemOpen
  \bibfield  {author} {\bibinfo {author} {\bibfnamefont {L.}~\bibnamefont
  {Del~Re}}, \bibinfo {author} {\bibfnamefont {B.}~\bibnamefont {Rost}},
  \bibinfo {author} {\bibfnamefont {M.}~\bibnamefont {Foss-Feig}}, \bibinfo
  {author} {\bibfnamefont {A.~F.}\ \bibnamefont {Kemper}}, \ and\ \bibinfo
  {author} {\bibfnamefont {J.~K.}\ \bibnamefont {Freericks}},\ }\href {\doibase
  10.1103/PhysRevLett.132.100601} {\bibfield  {journal} {\bibinfo  {journal}
  {Phys. Rev. Lett.}\ }\textbf {\bibinfo {volume} {132}},\ \bibinfo {pages}
  {100601} (\bibinfo {year} {2024})}\BibitemShut {NoStop}%
\bibitem [{\citenamefont {Li}\ \emph {et~al.}(2024)\citenamefont {Li},
  \citenamefont {Dulal}, \citenamefont {Ohorodnikov}, \citenamefont {Wang},\
  and\ \citenamefont {Ding}}]{dantong2024}%
  \BibitemOpen
  \bibfield  {author} {\bibinfo {author} {\bibfnamefont {D.}~\bibnamefont
  {Li}}, \bibinfo {author} {\bibfnamefont {D.}~\bibnamefont {Dulal}}, \bibinfo
  {author} {\bibfnamefont {M.}~\bibnamefont {Ohorodnikov}}, \bibinfo {author}
  {\bibfnamefont {H.}~\bibnamefont {Wang}}, \ and\ \bibinfo {author}
  {\bibfnamefont {Y.}~\bibnamefont {Ding}},\ }\href
  {https://arxiv.org/abs/2408.05406} {\enquote {\bibinfo {title} {Efficient
  quantum gradient and higher-order derivative estimation via generalized
  hadamard test},}\ } (\bibinfo {year} {2024})\BibitemShut {NoStop}%
\bibitem [{\citenamefont {Wan}\ \emph {et~al.}(2022)\citenamefont {Wan},
  \citenamefont {Berta},\ and\ \citenamefont {Campbell}}]{Wan2022}%
  \BibitemOpen
  \bibfield  {author} {\bibinfo {author} {\bibfnamefont {K.}~\bibnamefont
  {Wan}}, \bibinfo {author} {\bibfnamefont {M.}~\bibnamefont {Berta}}, \ and\
  \bibinfo {author} {\bibfnamefont {E.~T.}\ \bibnamefont {Campbell}},\ }\href
  {http://dx.doi.org/10.1103/PhysRevLett.129.030503} {\bibfield  {journal}
  {\bibinfo  {journal} {Physical Review Letters}\ }\textbf {\bibinfo {volume}
  {129}} (\bibinfo {year} {2022})}\BibitemShut {NoStop}%
\bibitem [{\citenamefont {Lin}\ and\ \citenamefont {Tong}(2022)}]{Lin2022}%
  \BibitemOpen
  \bibfield  {author} {\bibinfo {author} {\bibfnamefont {L.}~\bibnamefont
  {Lin}}\ and\ \bibinfo {author} {\bibfnamefont {Y.}~\bibnamefont {Tong}},\
  }\href {http://dx.doi.org/10.1103/PRXQuantum.3.010318} {\bibfield  {journal}
  {\bibinfo  {journal} {PRX Quantum}\ }\textbf {\bibinfo {volume} {3}}
  (\bibinfo {year} {2022})}\BibitemShut {NoStop}%
\bibitem [{\citenamefont {Blunt}\ \emph {et~al.}(2023)\citenamefont {Blunt},
  \citenamefont {Caune}, \citenamefont {Izsák}, \citenamefont {Campbell},\
  and\ \citenamefont {Holzmann}}]{Blunt2023}%
  \BibitemOpen
  \bibfield  {author} {\bibinfo {author} {\bibfnamefont {N.~S.}\ \bibnamefont
  {Blunt}}, \bibinfo {author} {\bibfnamefont {L.}~\bibnamefont {Caune}},
  \bibinfo {author} {\bibfnamefont {R.}~\bibnamefont {Izsák}}, \bibinfo
  {author} {\bibfnamefont {E.~T.}\ \bibnamefont {Campbell}}, \ and\ \bibinfo
  {author} {\bibfnamefont {N.}~\bibnamefont {Holzmann}},\ }\href
  {http://dx.doi.org/10.1103/PRXQuantum.4.040341} {\bibfield  {journal}
  {\bibinfo  {journal} {PRX Quantum}\ }\textbf {\bibinfo {volume} {4}}
  (\bibinfo {year} {2023})}\BibitemShut {NoStop}%
\bibitem [{\citenamefont {Kitaev}(1995)}]{kitaev1995}%
  \BibitemOpen
  \bibfield  {author} {\bibinfo {author} {\bibfnamefont {A.~Y.}\ \bibnamefont
  {Kitaev}},\ }\href {https://arxiv.org/abs/quant-ph/9511026} {\enquote
  {\bibinfo {title} {Quantum measurements and the abelian stabilizer
  problem},}\ } (\bibinfo {year} {1995})\BibitemShut {NoStop}%
\bibitem [{\citenamefont {Nielsen}\ and\ \citenamefont
  {Chuang}(2010)}]{nielsen2010}%
  \BibitemOpen
  \bibfield  {author} {\bibinfo {author} {\bibfnamefont {M.~A.}\ \bibnamefont
  {Nielsen}}\ and\ \bibinfo {author} {\bibfnamefont {I.~L.}\ \bibnamefont
  {Chuang}},\ }\href {\doibase 10.1017/CBO9780511976667} {\emph {\bibinfo
  {title} {Quantum Computation and Quantum Information: 10th Anniversary
  Edition}}}\ (\bibinfo  {publisher} {Cambridge University Press},\ \bibinfo
  {year} {2010})\BibitemShut {NoStop}%
\bibitem [{\citenamefont {Dobšíček}(2008)}]{miroslav2008}%
  \BibitemOpen
  \bibfield  {author} {\bibinfo {author} {\bibfnamefont {M.}~\bibnamefont
  {Dobšíček}},\ }\href {https://arxiv.org/abs/0803.0909} {\enquote {\bibinfo
  {title} {Quantum computing, phase estimation and applications},}\ } (\bibinfo
  {year} {2008})\BibitemShut {NoStop}%
\bibitem [{\citenamefont {Cao}\ \emph {et~al.}(2019)\citenamefont {Cao},
  \citenamefont {Romero}, \citenamefont {Olson}, \citenamefont {Degroote},
  \citenamefont {Johnson}, \citenamefont {Kieferová}, \citenamefont
  {Kivlichan}, \citenamefont {Menke}, \citenamefont {Peropadre}, \citenamefont
  {Sawaya}, \citenamefont {Sim}, \citenamefont {Veis},\ and\ \citenamefont
  {Aspuru-Guzik}}]{Cao2019}%
  \BibitemOpen
  \bibfield  {author} {\bibinfo {author} {\bibfnamefont {Y.}~\bibnamefont
  {Cao}}, \bibinfo {author} {\bibfnamefont {J.}~\bibnamefont {Romero}},
  \bibinfo {author} {\bibfnamefont {J.~P.}\ \bibnamefont {Olson}}, \bibinfo
  {author} {\bibfnamefont {M.}~\bibnamefont {Degroote}}, \bibinfo {author}
  {\bibfnamefont {P.~D.}\ \bibnamefont {Johnson}}, \bibinfo {author}
  {\bibfnamefont {M.}~\bibnamefont {Kieferová}}, \bibinfo {author}
  {\bibfnamefont {I.~D.}\ \bibnamefont {Kivlichan}}, \bibinfo {author}
  {\bibfnamefont {T.}~\bibnamefont {Menke}}, \bibinfo {author} {\bibfnamefont
  {B.}~\bibnamefont {Peropadre}}, \bibinfo {author} {\bibfnamefont {N.~P.~D.}\
  \bibnamefont {Sawaya}}, \bibinfo {author} {\bibfnamefont {S.}~\bibnamefont
  {Sim}}, \bibinfo {author} {\bibfnamefont {L.}~\bibnamefont {Veis}}, \ and\
  \bibinfo {author} {\bibfnamefont {A.}~\bibnamefont {Aspuru-Guzik}},\ }\href
  {\doibase 10.1021/acs.chemrev.8b00803} {\bibfield  {journal} {\bibinfo
  {journal} {Chemical Reviews}\ }\textbf {\bibinfo {volume} {119}},\ \bibinfo
  {pages} {10856–10915} (\bibinfo {year} {2019})}\BibitemShut {NoStop}%
\bibitem [{\citenamefont {O’Brien}\ \emph {et~al.}(2019)\citenamefont
  {O’Brien}, \citenamefont {Tarasinski},\ and\ \citenamefont
  {Terhal}}]{OBrien2019}%
  \BibitemOpen
  \bibfield  {author} {\bibinfo {author} {\bibfnamefont {T.~E.}\ \bibnamefont
  {O’Brien}}, \bibinfo {author} {\bibfnamefont {B.}~\bibnamefont
  {Tarasinski}}, \ and\ \bibinfo {author} {\bibfnamefont {B.~M.}\ \bibnamefont
  {Terhal}},\ }\href {\doibase 10.1088/1367-2630/aafb8e} {\bibfield  {journal}
  {\bibinfo  {journal} {New Journal of Physics}\ }\textbf {\bibinfo {volume}
  {21}},\ \bibinfo {pages} {023022} (\bibinfo {year} {2019})}\BibitemShut
  {NoStop}%
\bibitem [{\citenamefont {Ni}\ \emph {et~al.}(2023)\citenamefont {Ni},
  \citenamefont {Li},\ and\ \citenamefont {Ying}}]{Ni2023}%
  \BibitemOpen
  \bibfield  {author} {\bibinfo {author} {\bibfnamefont {H.}~\bibnamefont
  {Ni}}, \bibinfo {author} {\bibfnamefont {H.}~\bibnamefont {Li}}, \ and\
  \bibinfo {author} {\bibfnamefont {L.}~\bibnamefont {Ying}},\ }\href {\doibase
  10.22331/q-2023-11-06-1165} {\bibfield  {journal} {\bibinfo  {journal}
  {Quantum}\ }\textbf {\bibinfo {volume} {7}},\ \bibinfo {pages} {1165}
  (\bibinfo {year} {2023})}\BibitemShut {NoStop}%
\bibitem [{\citenamefont {Yamazaki}\ \emph {et~al.}(1983)\citenamefont
  {Yamazaki}, \citenamefont {Murao}, \citenamefont {Yamanaka},\ and\
  \citenamefont {Yoshihara}}]{Yamazaki1983}%
  \BibitemOpen
  \bibfield  {author} {\bibinfo {author} {\bibfnamefont {I.}~\bibnamefont
  {Yamazaki}}, \bibinfo {author} {\bibfnamefont {T.}~\bibnamefont {Murao}},
  \bibinfo {author} {\bibfnamefont {T.}~\bibnamefont {Yamanaka}}, \ and\
  \bibinfo {author} {\bibfnamefont {K.}~\bibnamefont {Yoshihara}},\ }\href
  {\doibase 10.1039/DC9837500395} {\bibfield  {journal} {\bibinfo  {journal}
  {Faraday Discuss. Chem. Soc.}\ }\textbf {\bibinfo {volume} {75}},\ \bibinfo
  {pages} {395} (\bibinfo {year} {1983})}\BibitemShut {NoStop}%
\bibitem [{\citenamefont {Greene}\ and\ \citenamefont
  {Batista}(2017)}]{Greene2017}%
  \BibitemOpen
  \bibfield  {author} {\bibinfo {author} {\bibfnamefont {S.~M.}\ \bibnamefont
  {Greene}}\ and\ \bibinfo {author} {\bibfnamefont {V.~S.}\ \bibnamefont
  {Batista}},\ }\href {\doibase 10.1021/acs.jctc.7b00608} {\bibfield  {journal}
  {\bibinfo  {journal} {Journal of Chemical Theory and Computation}\ }\textbf
  {\bibinfo {volume} {13}},\ \bibinfo {pages} {4034} (\bibinfo {year}
  {2017})}\BibitemShut {NoStop}%
\bibitem [{\citenamefont {Oliver}\ \emph {et~al.}(2012)\citenamefont {Oliver},
  \citenamefont {King}, \citenamefont {Tew}, \citenamefont {Dixon},\ and\
  \citenamefont {Ashfold}}]{Oliver2012}%
  \BibitemOpen
  \bibfield  {author} {\bibinfo {author} {\bibfnamefont {T.~A.~A.}\
  \bibnamefont {Oliver}}, \bibinfo {author} {\bibfnamefont {G.~A.}\
  \bibnamefont {King}}, \bibinfo {author} {\bibfnamefont {D.~P.}\ \bibnamefont
  {Tew}}, \bibinfo {author} {\bibfnamefont {R.~N.}\ \bibnamefont {Dixon}}, \
  and\ \bibinfo {author} {\bibfnamefont {M.~N.~R.}\ \bibnamefont {Ashfold}},\
  }\href {\doibase 10.1021/jp308804d} {\bibfield  {journal} {\bibinfo
  {journal} {The Journal of Physical Chemistry A}\ }\textbf {\bibinfo {volume}
  {116}},\ \bibinfo {pages} {12444–12459} (\bibinfo {year}
  {2012})}\BibitemShut {NoStop}%
\bibitem [{\citenamefont {Yang}\ \emph {et~al.}(2024)\citenamefont {Yang},
  \citenamefont {Christianen}, \citenamefont {Bañuls}, \citenamefont {Wild},\
  and\ \citenamefont {Cirac}}]{Yang2024}%
  \BibitemOpen
  \bibfield  {author} {\bibinfo {author} {\bibfnamefont {Y.}~\bibnamefont
  {Yang}}, \bibinfo {author} {\bibfnamefont {A.}~\bibnamefont {Christianen}},
  \bibinfo {author} {\bibfnamefont {M.~C.}\ \bibnamefont {Bañuls}}, \bibinfo
  {author} {\bibfnamefont {D.~S.}\ \bibnamefont {Wild}}, \ and\ \bibinfo
  {author} {\bibfnamefont {J.~I.}\ \bibnamefont {Cirac}},\ }\href
  {http://dx.doi.org/10.1103/PhysRevLett.132.220601} {\bibfield  {journal}
  {\bibinfo  {journal} {Physical Review Letters}\ }\textbf {\bibinfo {volume}
  {132}} (\bibinfo {year} {2024})}\BibitemShut {NoStop}%
\bibitem [{\citenamefont {O'Brien}\ \emph {et~al.}(2021)\citenamefont
  {O'Brien}, \citenamefont {Polla}, \citenamefont {Rubin}, \citenamefont
  {Huggins}, \citenamefont {McArdle}, \citenamefont {Boixo}, \citenamefont
  {McClean},\ and\ \citenamefont {Babbush}}]{brien2021}%
  \BibitemOpen
  \bibfield  {author} {\bibinfo {author} {\bibfnamefont {T.~E.}\ \bibnamefont
  {O'Brien}}, \bibinfo {author} {\bibfnamefont {S.}~\bibnamefont {Polla}},
  \bibinfo {author} {\bibfnamefont {N.~C.}\ \bibnamefont {Rubin}}, \bibinfo
  {author} {\bibfnamefont {W.~J.}\ \bibnamefont {Huggins}}, \bibinfo {author}
  {\bibfnamefont {S.}~\bibnamefont {McArdle}}, \bibinfo {author} {\bibfnamefont
  {S.}~\bibnamefont {Boixo}}, \bibinfo {author} {\bibfnamefont {J.~R.}\
  \bibnamefont {McClean}}, \ and\ \bibinfo {author} {\bibfnamefont
  {R.}~\bibnamefont {Babbush}},\ }\href {\doibase 10.1103/PRXQuantum.2.020317}
  {\bibfield  {journal} {\bibinfo  {journal} {PRX Quantum}\ }\textbf {\bibinfo
  {volume} {2}},\ \bibinfo {pages} {020317} (\bibinfo {year}
  {2021})}\BibitemShut {NoStop}%
\bibitem [{\citenamefont {Polla}\ \emph {et~al.}(2023)\citenamefont {Polla},
  \citenamefont {Anselmetti},\ and\ \citenamefont {O'Brien}}]{polla2023}%
  \BibitemOpen
  \bibfield  {author} {\bibinfo {author} {\bibfnamefont {S.}~\bibnamefont
  {Polla}}, \bibinfo {author} {\bibfnamefont {G.-L.~R.}\ \bibnamefont
  {Anselmetti}}, \ and\ \bibinfo {author} {\bibfnamefont {T.~E.}\ \bibnamefont
  {O'Brien}},\ }\href {\doibase 10.1103/PhysRevA.108.012403} {\bibfield
  {journal} {\bibinfo  {journal} {Phys. Rev. A}\ }\textbf {\bibinfo {volume}
  {108}},\ \bibinfo {pages} {012403} (\bibinfo {year} {2023})}\BibitemShut
  {NoStop}%
\end{thebibliography}%

\clearpage 

\appendix

\section{Circuits of State Preparation}
\label{App state prep}

\begin{figure*}[!htbp]
\flushright
\scalebox{0.9}{
\Qcircuit @C=1em @R=1.5em {
\lstick{\ket{k_{0}}} & \qw &\qw &\qw & \qw& \qw &\qw &\qw & \ctrl{3}& \qw &\qw &\qw & \qw& \qw &\qw &\qw & \ctrl{3}& \qw\\
\lstick{\ket{k_{1}}} & \qw &\qw &\qw & \ctrl{2}& \qw &\qw &\qw & \qw& \qw &\qw &\qw & \ctrl{2}& \qw &\qw &\qw & \qw& \qw\\
\lstick{\ket{k_{2}}} &\qw & \ctrl{1}& \qw &\qw &\qw & \ctrl{1}& \qw &\qw &\qw & \ctrl{1}& \qw &\qw &\qw & \ctrl{1}&\qw & \qw& \qw \\
\lstick{\ket{k_{3}}}   & \gate{R_{y}(\theta_{0,0}^{(y)})} & \targ & \gate{R_{y}(\theta_{0,1}^{(y)})} & \targ& \gate{R_{y}(\theta_{0,2}^{(y)})} & \targ& \gate{R_{y}(\theta_{0,3}^{(y)})} & \targ& \gate{R_{y}(\theta_{0,4}^{(y)})} & \targ& \gate{R_{y}(\theta_{0,5}^{(y)})} & \targ& \gate{R_{y}(\theta_{0,6}^{(y)})} & \targ& \gate{R_{y}(\theta_{0,7}^{(y)})} & \targ& \qw
}}
\caption{\justifying Quantum circuit of the uniformly 3-controlled rotations.}
\label{fig:p3}
\end{figure*}
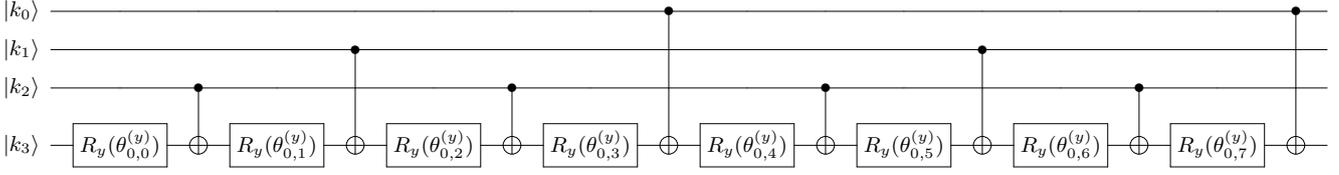

In this work, we select the uniformly controlled rotations as the approach of loading the initial Gaussian states on quantum computers. In our current settings, the amplitudes are all real and positive, and hence only $\mathbf{R_{y}}$ and $\mathbf{CNOT}$ gates are needed to load the target wavefunction, as suggested in~\cite{moosa2023,Mttnen2004}. For each 4-qubit register, the corresponding quantum circuit consists of 4 uniformly $j$-controlled rotation sequences, with $j={0,1,2,3}$, as shown in Figure~\ref{fig:p0}, \ref{fig:p1}, \ref{fig:p2} and \ref{fig:p3}. The total gate count in this section is $2^{n+1}-3=29$.


\begin{figure}[!htbp]
\centering
\scalebox{1.1}{
\Qcircuit @C=1em @R=1.5em {
\lstick{\ket{k_{0}}} & \gate{R_{y}(\theta_{3,0}^{(y)})} &\qw   \\
}}
\caption{\justifying Quantum circuits of the uniformly 0-controlled rotations.}
\label{fig:p0}
\end{figure}

\begin{figure}[!htbp]
\centering
\scalebox{1.1}{
\Qcircuit @C=1em @R=1.5em {
\lstick{\ket{k_{0}}} & \qw &\ctrl{1} & \qw &\ctrl{1}& \qw  \\
\lstick{\ket{k_{1}}} & \gate{R_{y}(\theta_{2,0}^{(y)})}      &\targ  &  \gate{R_{y}(\theta_{2,1}^{(y)})} & \targ& \qw    \\
}}
\caption{\justifying Quantum circuit of the uniformly 1-controlled rotations.}
\label{fig:p1}
\end{figure}

\begin{figure}[!htbp]
\flushright
\scalebox{0.8}{
\Qcircuit @C=1em @R=1.5em {
\lstick{\ket{k_{0}}} & \qw &\qw & \qw& \ctrl{2}& \qw &\qw & \qw& \ctrl{2}& \qw   \\
\lstick{\ket{k_{1}}} & \qw &\ctrl{1} & \qw&\qw & \qw & \ctrl{1} & \qw&\qw& \qw \\
\lstick{\ket{k_{2}}} & \gate{R_{y}(\theta_{1,0}^{(y)})} & \targ & \gate{R_{y}(\theta_{1,1}^{(y)})}  & \targ&\gate{R_{y}(\theta_{1,2}^{(y)})}&\targ& \gate{R_{y}(\theta_{1,3}^{(y)})}  & \targ  & \qw\\
}}
\caption{\justifying Quantum circuit of the uniformly 2-controlled rotations.}
\label{fig:p2}
\end{figure}

To actually perform the circuits, the rotation angle of the $k^\text{th}$ $\mathbf{R_{y}}$ gate in the $j$-controlled rotations are determined by~\cite{moosa2023,Mttnen2004}
\begin{equation}
\begin{split}
    \theta_{3-j,k}^{(y)} = \sum_{w=0}^{2^{j}-1}M_{kw}\alpha_{3-j,w}^{(y)},
\end{split}   
\label{eq:12}
\end{equation}
where $j={0,1,2,3}$ indicates the target qubit $\ket{k_{j}}$ of the corresponding $\mathbf{R_{y}}$ gates. 

The coefficient $M_{kw} = 2^{-j}(-1)^{b_{w}g_{k}}$ contains the binary code $b_{w}$ and binary reflected Gray code $g_{k}$ of the integer $w$ and $k$. By denoting $i=3-j$, the $\alpha_{3-j,w}$ is calculated from wavefunction amplitudes according to~\cite{moosa2023,Mttnen2004}
\begin{equation}
\begin{split}
    \alpha_{i,w}^{(y)} = 2\arcsin \sqrt{\frac{\sum_{l=0}^{2^{i}-1}|\psi_{(2w+1)2^{i}+l}|^{2}}{\sum_{l=0}^{2^{i+1}-1}|\psi_{w2^{i+1}+l}|^{2}}}.
\end{split}   
\label{eq:14}
\end{equation}

\begin{figure}[!htbp]
    \centering
    \includegraphics[scale=0.18]{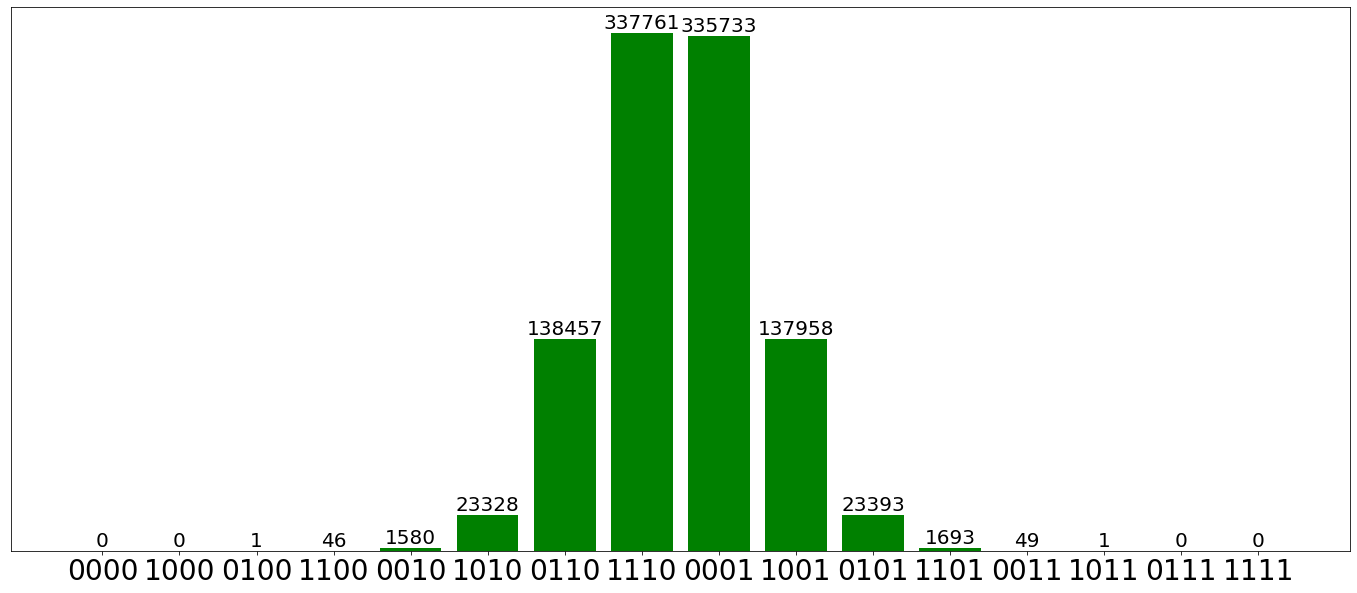}
    \includegraphics[scale=0.18]{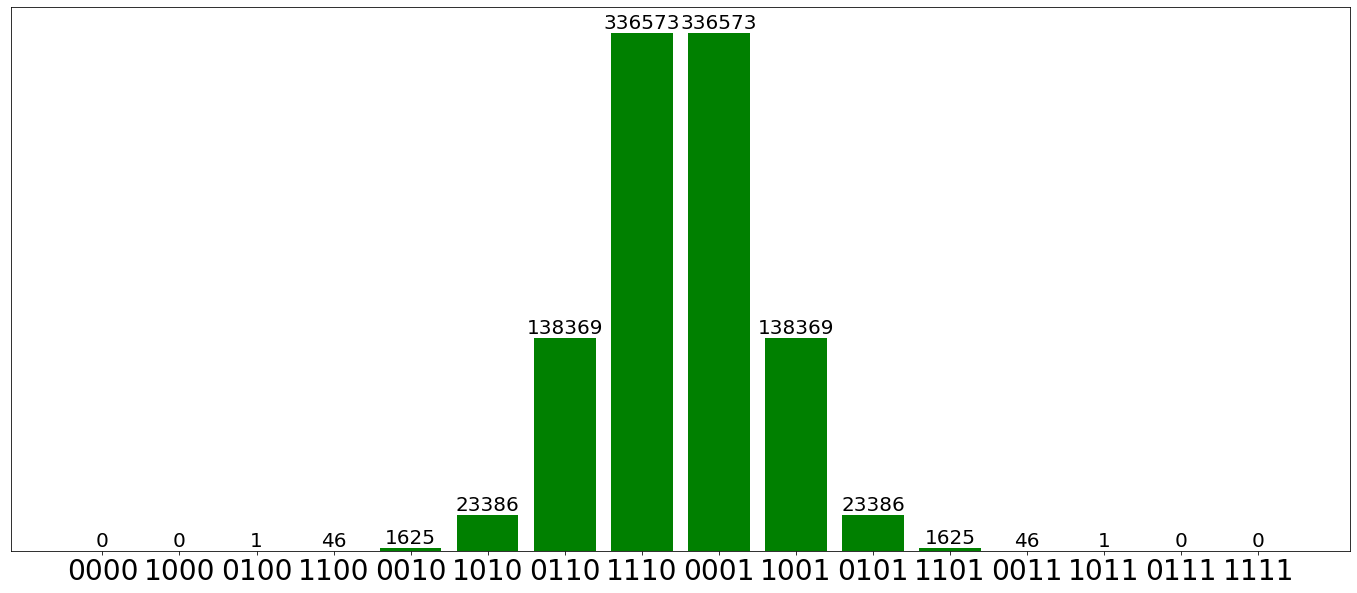}
    \caption{\justifying (TOP) Emulated random measurement outcomes of the state preparation circuit within $1\times10^{6}$ shots from Qiskit. (BOTTOM) The analytical probability distribution of the target wavefunction.}
    \label{fig:16}
\end{figure}

Taking one of the tuning modes, namely $\nu_{6a}$, as an illustrative example, Figure~\ref{fig:16} presents the measurement outcomes from $1\times10^{6}$ shots performed on Qiskit. The numerical bar chart (top) aligns exceptionally well with the analytical probability distribution of the target wavefunction (bottom). This strong agreement proves the effectiveness and precision of encoding the desired initial harmonic state using uniformly controlled rotations on a four-qubit quantum register. 

\section{Circuits of Time Evolution}
\label{App evolve}

Here we briefly illustrate quantum circuits applicable to the time evolution operators in the 4D model simulation, taking one $n$-qubit normal mode as an example. Specifically, $\mathbf{U_{1}}$ gates are frequently used when encoding $U_\text{diag}$ and $U_{K}$. This single-qubit rotation gate applies a phase operation around the $Z$-axis of the Bloch sphere, with $\mathbf{U_{1}}(\theta)(\alpha\ket{0}+\beta\ket{1})=\alpha\ket{0}+e^{i\theta}\beta\ket{1}$.

\begin{figure}[!htbp]
\centering
\scalebox{1.1}{
\Qcircuit @C=1em @R=1.5em {
\lstick{\ket{k_{0}}} & \gate{U_{1}(\theta_{0})} &\gate{X} & \gate{U_{1}(\theta_{0})} & \gate{X}& \qw  & \qw   \\
\lstick{\ket{k_{1}}} & \qw      &\qw  & \qw & \qw&\qw  & \qw   \\
\lstick{\ket{k_{2}}} & \qw      & \qw            & \qw      & \qw&\qw   & \qw  \\
\lstick{\vdots} & & \vdots & &   & &  \\
\lstick{\ket{k_{n-1}}}   & \qw & \qw & \qw & \qw& \qw  &  \qw 
}}
\caption{\justifying Quantum circuit of the time evolution operator incorporating $0^\text{th}$ order polynomial terms in the exponent.}
\label{fig:2}
\end{figure}
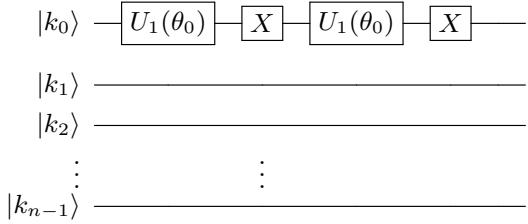

The $0^\text{th}$-order term in $V_\text{diag}$ builds the energy gap between $S_{1}$ and $S_{2}$ states by 4 single-qubit gates. Figure~\ref{fig:2} shows a repetitive sequence of $\mathbf{U_{1}}(\theta_0)$ gates and Pauli $\mathbf{X}$ gates utilised to apply the constant phase 
$$e^{i\theta_{0}}=e^{-\frac{i(-\Delta)dt}{2}} \text{, with } \theta_{0} = -\frac{-\Delta dt}{2},$$ onto the $S_{1}$ state (replace $-\Delta$ by $\Delta$ for the $S_{2}$ state).
As the $\mathbf{U_{1}}(\theta_0)$ gate only rotates the phase of state $\ket{1}$, a following Pauli $\mathbf{X}$ gate is required to invert the original $\ket{0}$ to $\ket{1}$. Then another $\mathbf{U_{1}}(\theta_0)$ gate is applied on this new $\ket{1}$. The final Pauli $\mathbf{X}$ gate restores everything. 

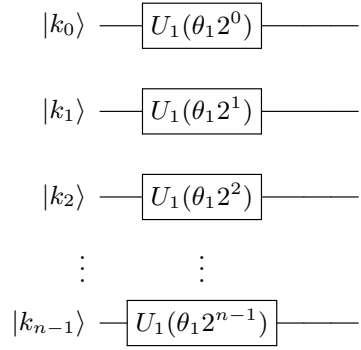
\begin{figure}[!htbp]
\centering
\scalebox{1.1}{
\Qcircuit @C=1em @R=1.5em {
\lstick{\ket{k_{0}}} & \gate{U_{1}(\theta_{1}2^{0})} &\qw & \qw  &  \qw   \\
\lstick{\ket{k_{1}}} & \gate{U_{1}(\theta_{1}2^{1})}      &\qw  &\qw     & \qw   \\
\lstick{\ket{k_{2}}} & \gate{U_{1}(\theta_{1}2^{2})} & \qw &\qw &    \qw  \\
\lstick{\vdots} &     \vdots & & & \\
\lstick{\ket{k_{n-1}}}   & \gate{U_{1}(\theta_{1}2^{n-1})} & \qw &  \qw  &   \qw 
}}
\caption{\justifying Quantum circuit of the time evolution operator incorporating $1^\text{st}$ order polynomial terms in the exponent.}
\label{fig:3}
\end{figure}

The interpretation of the $1^\text{st}$-order intra-state coupling term (with $\kappa$ parameters in Eq.~\ref{eq:hamil}) solely requires $\mathbf{U_{1}}$ gates. Re-writing the coordinate $Q$ in binary expansion $$Q = k_{0}2^{0} + k_{1}2^{1} + \cdots + k_{n-1}2^{n-1}$$ indicates a $2^{i}$ term in the rotation angle of the $\mathbf{U_{1}}$ gate on the $k_{i}$ qubit, as presented in Figure~\ref{fig:3}. 
These individual $\mathbf{U_{1}}(\theta_1 2^i)$ gates can, in principle, be applied in parallel to encode $$e^{-\frac{i\kappa^{(1)}Qdt}{2}} \text{, with }\theta_{1}=-\frac{\kappa^{(1)}dt}{2},$$ for the $U_{S1}$ block (replace $\kappa^{(1)}$ by $\kappa^{(2)}$ for the $U_{S2}$ block). However, because either the $U_{S1}$ or $U_{S2}$ block is controlled by the electronic state degree of freedom as a whole, these single-qubit $\mathbf{U_{1}}$ gates, though acting on different qubits, cannot be executed concurrently, giving an $n$-gate rather than a one-gate contribution to the circuit depth.


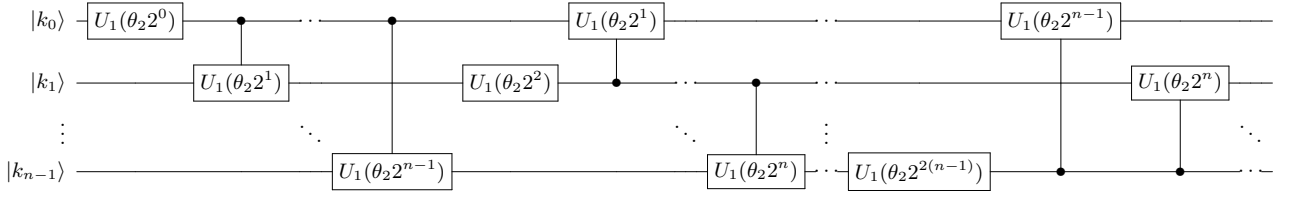
\begin{figure*}[!htbp]
\centering
\scalebox{0.88}{
\Qcircuit @C=0.5em @R=1.2em {
\lstick{\ket{k_{0}}} & \gate{U_{1}(\theta_{2}2^{0})} & \ctrl{1}  &\qw& \cdots & &\ctrl{3} & \qw & \gate{U_{1}(\theta_{2}2^{1})}  & \qw &\qw&\qw&\qw &\qw& \cdots & & \qw&\gate{U_{1}(\theta_{2}2^{n-1})} &\qw &\qw& \qw&\qw&\qw   \\
\lstick{\ket{k_{1}}} & \qw & \gate{U_{1}(\theta_{2}2^{1})} & \qw & \qw&\qw &  \qw&\gate{U_{1}(\theta_{2}2^{2})} & \ctrl{-1} &\qw &\cdots &  & \ctrl{2} & \qw & \cdots & & \qw& \qw& \gate{U_{1}(\theta_{2}2^{n})} &\qw& \qw&\qw&\qw \\
\lstick{\vdots} & & &   & \ddots &  &  & && &\ddots  &   & && \vdots &&&&&&\ddots\\
\lstick{\ket{k_{n-1}}}   & \qw  & \qw & \qw & \qw & \qw & \gate{U_{1}(\theta_{2}2^{n-1})}& \qw & \qw &\qw & \qw  &\qw&\gate{U_{1}(\theta_{2}2^{n})}&\qw&\cdots& &\gate{U_{1}(\theta_{2}2^{2(n-1)})} &\ctrl{-3} &\ctrl{-2}&\qw&\cdots&&\qw  
}}
\caption{\justifying Quantum circuit of the time evolution operator incorporating $2^{\text{nd}}$ order polynomial terms in the exponent.}
\label{fig:4}
\end{figure*}

Both the $K$ term in momentum space and the vibrational frequency term in $V_\text{diag}$ are quadratic polynomials. Taking $V_\text{diag}$ as an example, we re-calculate $Q^{2}$ in binary expansion: $$Q^{2} = (\sum_{i=0}^{n-1}k_{i}2^{i})(\sum_{j=0}^{n-1}k_{j}2^{j}).$$ Each component in this expression translates to regarding $i$ as the order of the controlling qubit and $j$ as the order of the target qubit for one $\mathbf{U_{1}}$ gate (or vice versa), leading to elimination of the controlling dot when $i=j$.
By allocating the corresponding $2^{i+j}$ in the rotation angle, $n^2$ $\mathbf{U_{1}}(\theta_2 2^{i+j})$ gates in Figure~\ref{fig:4} collectively apply 
$$e^{-\frac{i\omega Q^{2}dt}{4}} \text{, with }\theta_{2}=-\frac{\omega}{2}\frac{dt}{2},$$
to the corresponding normal mode $Q$.


\begin{figure}[!htbp]
\flushright
\scalebox{1.05}{
\Qcircuit @C=1em @R=1.5em {
\lstick{\ket{q}} & \gate{R_{x}(\theta_c 2^0)} & \gate{R_{x}(\theta_c 2^1)}& \qw &\cdots & & \gate{R_{x}(\theta_c 2^{n-1})} & \qw \\
\lstick{\ket{k_{0}}} & \ctrl{-1} &\qw  & \qw &\cdots & & \qw & \qw  \\
\lstick{\ket{k_{1}}} & \qw      & \ctrl{-2}   & \qw &\cdots  &  & \qw &\qw  \\
\lstick{\vdots} & & &   & \ddots &   \\
\lstick{\ket{k_{n-1}}}   & \qw & \qw  & \qw & \cdots&  & \ctrl{-4}& \qw  
}}
\caption{\justifying Quantum circuit of the time evolution operator incorporating off-diagonal coupling terms in the exponent.}
\label{fig:5}
\end{figure}
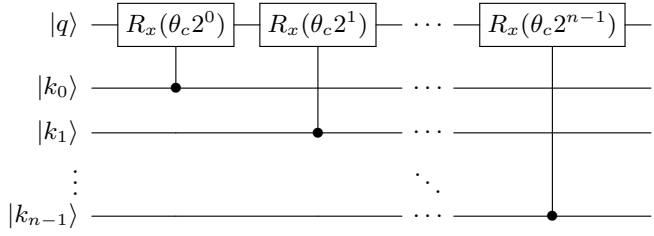

To incorporate the linear vibronic coupling term $V_\text{c}$, successive $\mathbf{R_{x}}(\theta_\text{c}2^i)$ gates targeting on the 1-qubit register representing the electronic state are employed to recover the $U_\text{c}$ block
$$e^{-\frac{i\lambda Q_{10a}dt }{2} \mathbf{X}} \text{, with }\theta_\text{c} = \lambda dt,$$ 
as indicated in Figure~\ref{fig:5}. Here the off-diagonal matrix, implemented by the Pauli $\mathbf{X}$ operation, is naturally included in the $\mathbf{R_{x}}$ gate as
$$\mathbf{R_{x}}(\theta)=e^{-i\frac{\theta}{2}\mathbf{X}}.$$
All these $\mathbf{R_{x}}$ gates are controlled by the register representing $Q_{10a}$ normal mode, with a $2^{i}$ term specified in the rotation angle of the $\mathbf{R_{x_{i}}}$ gate controlled by the $k_{i}$ qubit of the $Q_{10a}$ register.

\section{Circuits of Quantum Fourier Transform}
\label{App qft}
The Quantum Fourier Transform (QFT) serves as a fundamental operation in quantum computing, executing the discrete Fourier Transform for conversion between the computational and Fourier bases. Its circuit consists of two essential components: $n$ Hadamard gates $\mathbf{H}$ acting on single qubits, and $\frac{n\times(n-1)}{2}$ controlled rotation gates $\mathbf{R}$ applying phase shifts conditional on the control qubit being in the $|1\rangle$ state~\cite{Benenti2008}. 

\begin{figure}[!htbp]
\centering
\scalebox{0.56}{
\Qcircuit @C=0.8em @R=1.5em {
\lstick{\ket{k_{n-1}}} & \gate{H} & \gate{R_{2}} & \gate{R_{3}} & \qw &\cdots & & \gate{R_{n}} & \qw &\qw &\qw &\qw& \qw &\qw &\qw &\qw & \qw & \qw & \qw & \qw & \qw   & \qw&\qw& \qw&\qw \\
\lstick{\ket{k_{n-2}}} & \qw      & \ctrl{-1}      & \qw & \qw & \qw   & \qw & \qw &\gate{H} & \gate{R_{2}}& \qw &\cdots &  & \gate{R_{n-1}} & \qw &\qw & \qw &\qw &\qw &\qw & \qw & \qw & \qw& \qw &\qw \\
\lstick{\ket{k_{n-3}}} & \qw      & \qw            & \ctrl{-2}      & \qw & \qw   & \qw & \qw &\qw &\ctrl{-1}&\qw&\qw&\qw& \qw& \gate{H} & \gate{R_{2}}&\qw&\cdots& & \gate{R_{n-2}} & \qw & \qw & \qw&\qw&\qw\\
\lstick{\vdots} & & & &   & \ddots &  &  &  & & &\ddots  &  &  & & & & \ddots \\
\lstick{\ket{k_{0}}}   & \qw & \qw & \qw & \qw & \qw & \qw & \ctrl{-4}& \qw & \qw &\qw & \qw  & \qw &\ctrl{-3}&\qw&\qw&\qw&\qw&\qw&\ctrl{-2}&\qw&\cdots& &\gate{H} &\qw 
}}
\caption{\justifying Illustrative QFT circuit for an $n$-qubit register.}
\label{fig:14}
\end{figure}
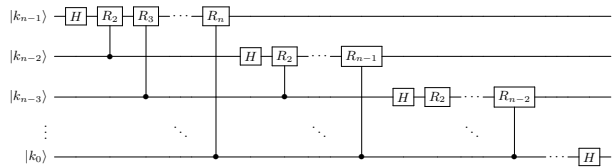

Figure~\ref{fig:14} outlines the standard QFT circuit with a reversed qubit order for clarity, where the least significant qubit is positioned at the bottom and the most significant at the top. While the figure does not explicitly include the SWAP gates required to restore the correct qubit ordering, these operations contribute an additional $\frac{n}{2}$ gates for even $n$ (or $\frac{n-1}{2}$ in the case of odd $n$). The inverse QFT circuit follows a straightforward reversal of operations, applying $\mathbf{R}^{-1}$ from right to left. Crucially, the QFT exhibits a quadratic complexity of $\mathcal{O}(n^{2})$, vastly outperforming the classical FFT, whose computational burden scales as $\mathcal{O}(n2^{n})$.

\section{Hadamard Test}
\label{App hadamard}

\begin{figure}[!htbp]
\centering
\scalebox{1.1}{
\Qcircuit @C=1em @R=1.5em {
\lstick{\ket{0}} & \gate{H} & \ctrl{1}   & \gate{H}  & \meter  \\
\lstick{\ket{\psi}} & \qw {/}^{dn+1} & \gate{U} & \qw & \qw  
}}
\caption{\justifying Ancilla-assisted measurement of the real part of the autocorrelation at a specific time step.}
\label{fig:had}
\end{figure}
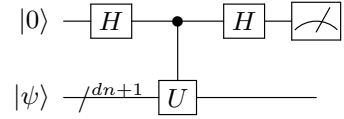

The measurement of the desired autocorrelation values at each time step is efficiently realised using the Hadamard test~\cite{Cleve1998} circuit illustrated in Figure~\ref{fig:had}. This circuit transforms the system $$\text{from } \ket{0}\ket{\psi} \text{ to } \frac{1}{2}[(I+U)\ket{0}+(I-U)\ket{1}]\ket{\psi},$$ employing an ancillary qubit to facilitate extraction of the real part of the autocorrelation. Specifically, the difference in measurement probabilities of the ancilla being found in $\ket{0}$ and $\ket{1}$ directly yields the real component:
\begin{equation}
\begin{split}
P_{\ket{\text{ancilla}}}(0)-P_{\ket{\text{ancilla}}}(1) 
&= \operatorname{\mathbb{R}e}\{ \bra{\psi(0)}U\ket{\psi(0)}\}
\\&= \operatorname{\mathbb{R}e}\{ \bra{\psi(0)}\ket{\psi(t)}\}.
\end{split}
\label{eq:real hada}
\end{equation}


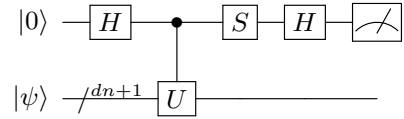
\begin{figure}[!htbp]
\centering
\scalebox{1.1}{
\Qcircuit @C=1em @R=1.5em {
\lstick{\ket{0}} & \gate{H} & \ctrl{1} & \gate{S}  & \gate{H}  & \meter  \\
\lstick{\ket{\psi}} & \qw {/}^{dn+1} & \gate{U} & \qw & \qw  & \qw
}}
\caption{\justifying Ancilla-assisted measurement of the imaginary part of the autocorrelation at a specific time step.}
\label{fig:imag hada}
\end{figure}

The complementary circuit for extracting the imaginary part of the autocorrelation is depicted in Figure~\ref{fig:imag hada}. With $\mathbf{S}\ket{0}=\ket{0}$ and $\mathbf{S}\ket{1}=i\ket{0}$, this circuit converts the system $$\text{from } \ket{0}\ket{\psi} \text{ to } \frac{1}{2}[(I+iU)\ket{0}+(I-iU)\ket{1}]\ket{\psi}.$$ As with the real component, the imaginary part is obtained by evaluating probability differences following:
\begin{equation}
\begin{split}
P_{\ket{\text{ancilla}}}(1)-P_{\ket{\text{ancilla}}}(0) 
&= \operatorname{\mathbb{I}m}\{ \bra{\psi(0)}U\ket{\psi(0)} \}
\\&= \operatorname{\mathbb{I}m}\{ \bra{\psi(0)}\ket{\psi(t)} \}.
\end{split}
\label{eq:imag hada}
\end{equation}


\section{Canonical Quantum Phase Estimation}
\label{General QPE}

\begin{figure}[!!htbp]
\flushright
\scalebox{1}{
\Qcircuit @C=0.8em @R=0.8em {
\lstick{\ket{0}} & \gate{H} & \qw & \qw & \qw  & \cdots & & \ctrl{4} &\multigate{3}{\text{QFT}^{-1}} & \meter  \\
\lstick{\vdots} & & & & & \vdots & & & \\
\lstick{\ket{0}} & \gate{H} & \qw & \ctrl{2}& \qw  & \cdots & & \qw &\ghost{QFT^{-1}} & \meter  \\
\lstick{\ket{0}} & \gate{H} & \ctrl{1} & \qw& \qw  & \cdots & & \qw &\ghost{QFT^{-1}} & \meter  \\
\lstick{\ket{\psi}} & \qw {/}^{dn+1} & \gate{U^{2^{0}}} & \gate{U^{2^{1}}} & \qw &\cdots&& \gate{U^{2^{m-1}}}& \qw
}}
\caption{\justifying Retrieval of spectrum signals using canonical QPE techniques, with $m$ qubits in the time register.}
\label{fig:APP QPE}
\end{figure}
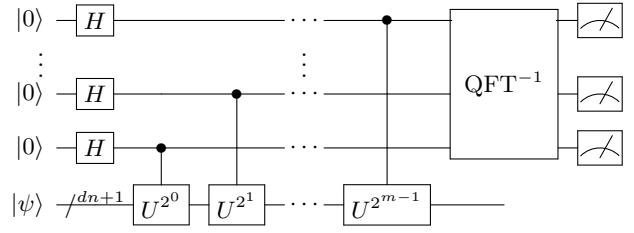

Here we briefly illustrate the application of canonical Quantum Phase Estimation (QPE) in retrieving spectrum signals. The time evolution in a single time step is denoted as $$U\ket{\psi} = e^{-iHdt}\ket{\psi}= e^{2\pi i \theta}\ket{\psi}.$$ $U^{2^{j}}$ indicates repeated operations to the system: $$U^{2^{j}}\ket{\psi} = e^{2^{j} 2\pi i \theta}\ket{\psi} \text{, with \( j \in [0, m) \)}.$$ The controlled $U^{2^{j}}$ architecture in Figure~\ref{fig:APP QPE} encapsulates $2^{j}$ time steps within each block. The structure of the time register is dictated by the total number of evolution steps, necessitating $m$ qubits to represent $2^{m}$ discrete time steps within the simulation.

The transformation of quantum states throughout the QPE process is schematically described as follows. The Hadamard gate, in conjunction with the controlled $U$ sequence, evolves the system 
$$\text{ from } \ket{0}^{\otimes m}\otimes\ket{\psi} \text{ to } \frac{1}{2^{m/2}}\sum_{j=0}^{2^{m}-1}e^{2\pi i \theta j}\ket{j}\otimes \ket{\psi},$$ 
with $j$ constrained to \( j \in [0, 2^{m}-1) \) (rather than the one \( j \in [0, m) \) mentioned above). Here $\ket{j}$ represents the binary encoding of index $j$. Upon the final application of the inverse QFT, measurements of the qubit states $\ket{\theta_{i}}$ (\( i \in [0, m) \)) within the time register enable recovery of the phase $\theta$, which resides in the interval $[0,1]$. The retrieved phase appears as a binary fraction, $0.\theta_{0}\theta_{1}\theta_{2}\ldots\theta_{m-1}$, from which the corresponding eigenvalues or spectral peak positions are further determined.

The precision of $\theta$ is intrinsically linked to the number of fractional digits, with error suppression achieved by increasing the quantity of qubits encoded within the time register. Given that the quantum system exists as a superposition of multiple eigenstates, each associated with a distinct energy level, repeated executions of the circuit generate measurement outcomes that manifest as peaks centred around the corresponding eigenvalues. The statistical distribution of these peaks, governed by the inherent vibronic properties of the pyrazine system, reflects the desired characteristic spectrum.

\end{document}